\documentclass[sigconf]{acmart}

\usepackage{xcolor}
\usepackage{graphicx}
\usepackage{subcaption}
\usepackage{epstopdf}
\usepackage{wrapfig}
\usepackage{hyperref}
\usepackage{array}
\usepackage{fixme}
\usepackage{amssymb}
\usepackage{tikz}
\usepackage{multirow}
\usepackage{booktabs} 

\usepackage{wrapfig}
\usepackage{colortbl}
\usepackage{algorithmic}

\usepackage{cleveref}
\crefformat{section}{(\S#2#1#3)}

\usepackage{epigraph}

\usepackage[vlined,linesnumbered]{algorithm2e}

\SetCommentSty{mycommfont}
\SetAlFnt{\sffamily}

\SetAlCapFnt{\normalfont\sffamily\small}

\makeatletter
\renewcommand{\algocf@Vline}[1]{
  \strut\par\nointerlineskip
  \algocf@push{\skiprule}
  \hbox{\bgroup\color{cyan}\vrule\egroup%
    \vtop{\algocf@push{\skiptext}
      \vtop{\algocf@addskiptotal #1}\bgroup\color{cyan}\Hlne\egroup}}\vskip\skiphlne
  \algocf@pop{\skiprule}
  \nointerlineskip}
\renewcommand{\algocf@Vsline}[1]{
  \strut\par\nointerlineskip
  \algocf@bblockcode%
  \algocf@push{\skiprule}
  \hbox{\bgroup\color{cyan}\vrule\egroup
    \vtop{\algocf@push{\skiptext}
      \vtop{\algocf@addskiptotal #1}}}
  \algocf@pop{\skiprule}
  \algocf@eblockcode%
}
\makeatother
\SetKwProg{Fn}{Function}{}{}

\newcommand{\doublerule}[1][.4pt]{%
  \noindent
    \makebox[0pt][l]{\rule[.7ex]{\linewidth}{#1}}%
      \rule[.3ex]{\linewidth}{#1}}

\fxsetup{
    status=draft,
    author={Todo},
    layout=inline,
    theme=color,
    silent=true,
    }
\definecolor{fxnote}{rgb}{0.8000,0.0000,0.0000}

\newcommand\ch[1]{\textcolor{black}{#1}}
\newcommand\chm[1]{\textcolor{black}{#1}}

\newcommand{\showDOI}[1]{\unskip}
\newcommand{\showURL}[1]{\unskip}
\newcolumntype{x}[1]{>{\raggedright}p{#1}} \newcommand{\tn}{\tabularnewline}

\graphicspath{{graphics/}}

\begin{document}

\copyrightyear{2018} 
\acmYear{2018} 
\fancyhead{}

\title{The Internals of the Data Calculator}

\author{Stratos Idreos, Kostas Zoumpatianos, Brian Hentschel, Michael S. Kester, Demi Guo}
\affiliation{
\institution{Harvard University}
}

\begin{abstract}
Data structures are critical in any data-driven scenario, but they are notoriously hard to design 
due to a massive design space and the dependence of performance on workload and hardware which evolve continuously.  
We present a design engine, the Data Calculator, 
which enables interactive and semi-automated design of data structures. 
It brings two innovations.
First, it offers a set of fine-grained design primitives that capture the first principles of data layout design: 
how data structure nodes lay data out, and how they are positioned relative to each other.
This allows for a structured description of the universe of possible data structure designs 
that can be synthesized as combinations of those primitives.
The second innovation is computation of performance using learned cost models.
These models are trained on diverse hardware and data profiles and capture the cost properties of fundamental data access primitives (e.g., random access). 
With these models, we synthesize the performance cost of complex operations on arbitrary data structure designs
without having to: 1) implement the data structure, 2) run the workload, or even 3) access the target hardware.
We demonstrate that the Data Calculator can assist data structure designers and researchers by accurately answering rich  what-if design questions on the order of a few seconds or minutes, i.e., computing 
how the performance (response time) of a given data structure design is impacted by variations in the: 
1) design, 2) hardware, 3) data, and 4) query workloads.
This makes it effortless to test numerous designs and ideas before embarking on lengthy implementation, deployment, and hardware acquisition steps.
We also demonstrate that the Data Calculator can synthesize entirely new designs, auto-complete partial designs, and detect suboptimal design choices. 

\vspace{.2cm}
\center{\emph{Let us calculate.}  ---Gottfried Leibniz}
   

\end{abstract}

\maketitle

\renewcommand{\thefootnote}{\fnsymbol{footnote}}
\footnotetext[1]{The name ``Calculator'' pays tribute to the early works that experimented with the concept of calculating complex objects from a small set of primitives \cite{Leibniz1666}.}
\renewcommand{\thefootnote}{\arabic{footnote}}

\section{From Manual to Interactive Design}
\label{sec:intro}

\begin{figure}[t]
    \includegraphics[width=0.6\columnwidth]{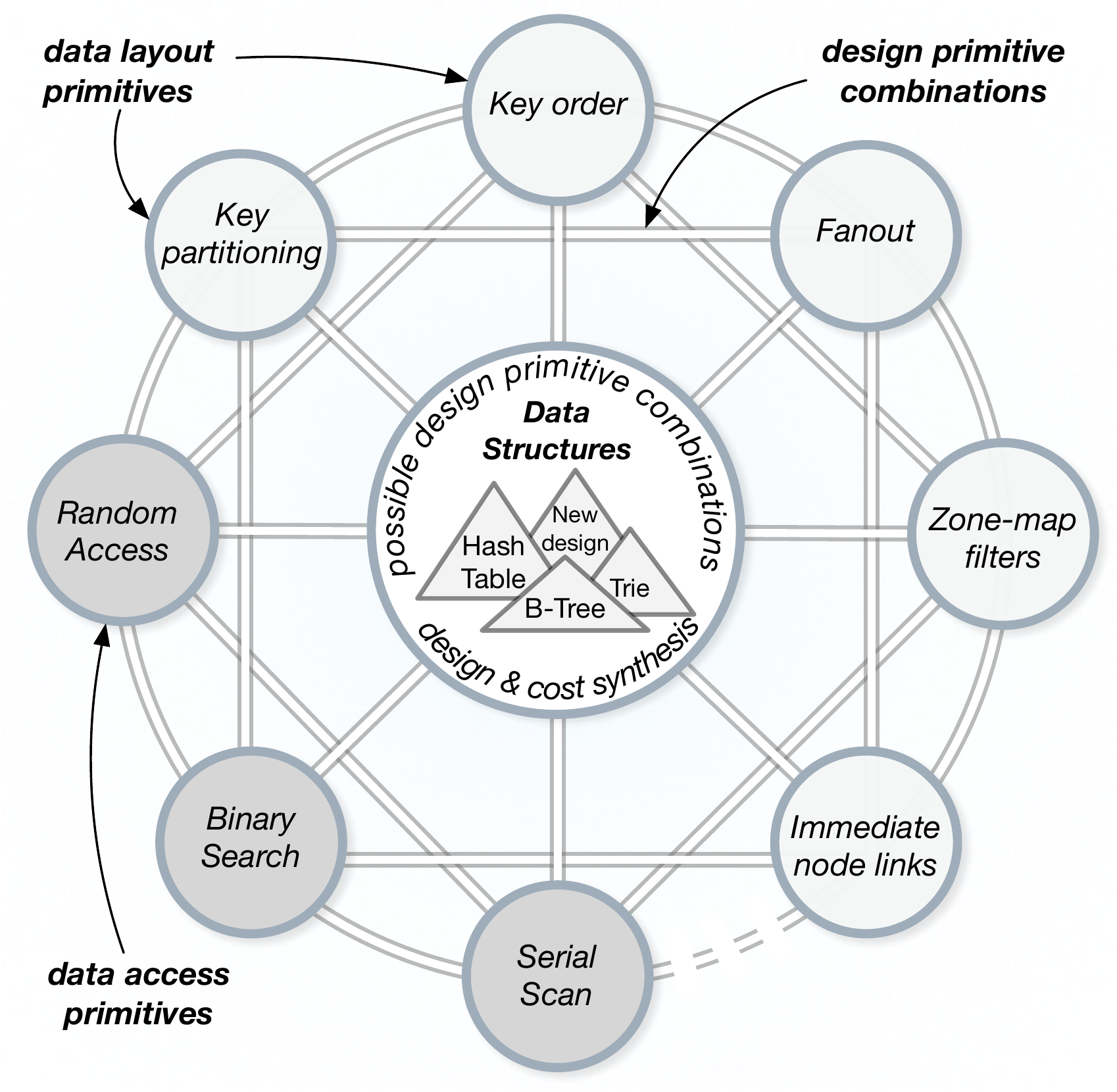}
    \vspace{-2ex}
    \caption{The concept of the Data Calculator: computing data access method designs as combinations of a small set of primitives. 
    (Drawing inspired by a figure in the Ph.D. thesis of Gottfried Leibniz who envisioned an engine that calculates physical laws
    from a small set of primitives \cite{Leibniz1666}.)}
    \label{figure:Leibniz}
    \vspace{1ex}
\end{figure}

\noindent{\textbf{The Importance of Data Structures.}} 
Data structures are at the core of any data-driven software, from relational database systems, NoSQL key-value stores, operating systems, compilers, HCI systems, and scientific data management to any ad-hoc program that deals with increasingly growing data. 
Any operation in any data-driven system/program goes through a data structure whenever it touches data. Any effort to rethink the design of a specific system or to add new functionality typically includes (or even begins by) rethinking how data should be stored and accessed \cite{Hellerstein1995, Yao1977, Yao1978, Lehman1986, Graefe2011, Athanassoulis2016, Abadi2013}. 
In this way, the design of data structures has been an active area of research since the onset of computer science 
 and there is an ever-growing need for alternative designs.
 This is fueled by 1) the continuous advent of new applications that require tailored storage and access patterns in both industry and science, and 2) new hardware that requires specific storage and access patterns to ensure longevity and maximum utilization.
Every year dozens of new data structure designs are published, e.g., 
more than fifty new designs appeared at ACM SIGMOD, PVLDB, EDBT and IEEE ICDE in 2017 according to data from DBLP.

\noindent{\textbf{A Vast and Complex Design Space.}}
A data structure design consists of 1) a data layout to describe how data is stored, and 2) algorithms that describe how its basic functionality (search, insert, etc.)
is achieved over the specific data layout. A data structure can be as simple as an array or arbitrarily complex using sophisticated combinations of hashing, range and radix partitioning, careful data placement, compression and encodings. 
Data structures may also be referred to as ``data containers'' or ``access methods'' (in which case the term ``structure'' applies only to the layout).
The data layout design itself may be further broken down into the base data layout and the indexing information which helps navigate the data, i.e., the leaves of a B+tree and its inner nodes, or buckets of a hash table and the hash-map.
We use the term data structure design throughout the paper to refer to the overall design of the data layout, indexing, and the algorithms together as a whole.

We define ``design'' as the set of all decisions that characterize the layout and algorithms of a data structure, e.g., 
``Should data nodes be sorted?'', ``Should they use pointers?'', and ``How should we scan them exactly?''. 
The number of possible \chm{valid} data structure designs explodes to $\gg10^{32}$
even if we limit the overall design to only two different kinds of nodes (e.g., as is the case for B+trees). If we allow every node to adopt different design decisions (e.g., based on access patterns), 
then the number of designs grows to $\gg10^{100}$.\footnote{For comparison, the estimated number of stars in the universe is $10^{24}$.} We explain how we derive these numbers in Section \ref{sec:design}.

\noindent{\textbf{The Problem: Human-Driven Design Only.}}
The design of data structures is a slow process, relying on the expertise and intuition of researchers and engineers who need to mentally navigate the 
vast design space. 
For example, consider the following design questions.


\begin{enumerate}
\item We need a data structure for a specific workload:
Should we strip down an existing complex data structure? 
Should we build off a simpler one?
Or should we design and build a new one from scratch?

\item We expect that the workload might shift (e.g., due to new application features):
How will performance change?
Should we redesign our core data structures? 

\item We add flash drives with more bandwidth and also add more system memory:
Should we change the layout of our B-tree nodes?
Should we change the size ratio in our LSM-tree?

\item We want to improve throughput:
How beneficial would it be to buy faster disks? more memory? or
should we invest the same budget in redesigning our core data structure? 
\end{enumerate}

This complexity leads to a slow design process and has severe cost side-effects \cite{Bernstein1987a,Cheung2015}. 
Time to market is of extreme importance, so new data structure design 
effectively stops when a design ``is due'' and only rarely when it ``is ready''.
Thus, the process of design extends beyond the initial design phase to periods of reconsidering the design given bugs or changes in the 
scenarios it should support.
Furthermore, this complexity makes it difficult to predict the impact of design choices, workloads, and hardware on performance.
We include two quotes from a systems architect with more than two decades of experience
 with relational systems and key-value stores.

\textit{(1) ``I know from experience that getting a new data structure into production takes years. Over several years, assumptions made about the workload and hardware are likely to change, and these changes threaten to reduce the benefit of a data structure. This risk of change makes it hard to commit to multi-year development efforts. We need to reduce the time it takes to get new data structures into production.''}

\textit{(2) ``Another problem is the limited ability we have to iterate. While some changes only require an online schema change, many require a dump and reload for a data service that might be running 24x7. The budget for such changes is limited. We can overcome the limited budget with tools that help us determine the changes most likely to be useful. Decisions today are frequently based on expert opinions, and these experts are in short supply.''}

\noindent{\textbf{Vision Step 1: Design Synthesis from First Principles.}}
We propose a move toward the new design paradigm captured in Figure \ref{figure:Leibniz}. 
Our intuition is that most designs (and even inventions) are about combining a small set of fundamental concepts in different ways or tunings. 
If we can describe the set of the first principles of data structure design, i.e., the core design principles out of which all data structures can be drawn, then we will have a structured way to express all possible designs we may invent, study, and employ as combinations of those principles. 
An analogy is the periodic table of elements in chemistry. 
It classifies elements based on their atomic number, electron configuration, and recurring chemical properties.
The structure of the table allows one to understand the elements and how they relate to each other but crucially it also enables arguing about the possible design space; more than one hundred years since the inception of the periodic table in the 18th century, we keep discovering elements that are predicted (synthesized) by the ``gaps'' in the table, accelerating science. 

\begin{figure*}[t]
    \center
   \includegraphics[width=\textwidth]{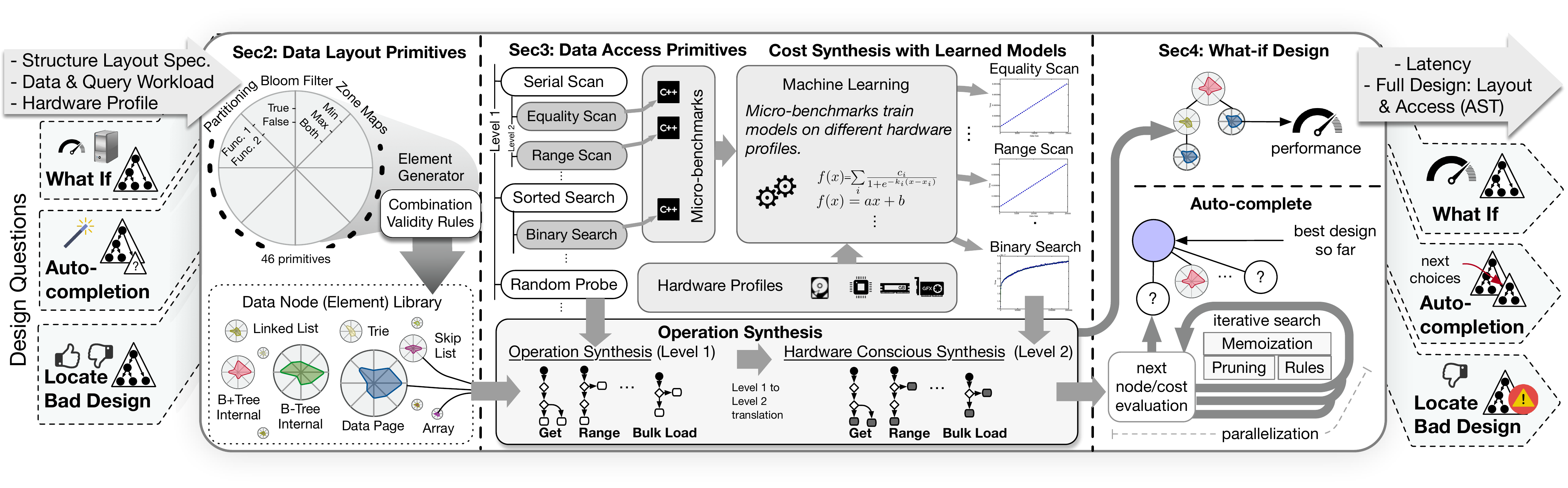}
    \vspace{-7.5ex}
    \caption{\chm{The architecture of the Data Calculator: From high-level layout specifications to performance cost calculation.}}
    \label{figure:architecture}
    \vspace{-2.5ex}    
\end{figure*} 

\noindent{\textbf{Our vision is to build the periodic table of data structures}} so we can express their massive design space. We take the first step in this paper, presenting a set of first principles that can synthesize orders of magnitude more data structure designs than what has been published in the literature. It captures basic hardware conscious layouts and read operations; future work includes extending the table for additional parts of the design space, such as updates, concurrency, compression, adaptivity, and security.

\noindent{\textbf{Vision Step 2: Cost Synthesis from Learned Models.}}
The second step in our vision is to accelerate and automate the design process. 
Key here, is being able to argue about the performance behavior of the massive number of designs so we can rank them.
Even with an intuition that a given design is an excellent choice, one has to implement the design, and test it on a given data and query workload and onto specific hardware. This process can take weeks at a time and has to be repeated when any part of the environment changes. Can we accelerate this process so we can quickly test alternative designs (or different combinations of hardware, data, and queries) on the order of a few seconds? 
If this is possible, then we can 1) accelerate design and research of new data structures, and 2) enable new kinds of adaptive systems that can decide core parts of their design, and the right hardware.

Arguing formally about the performance of diverse designs is a notoriously hard problem \cite{Manegold2002,Yao1975,Teorey1976,Cardenas1973,Yao1977,Zhou1999} especially as workload and hardware properties change; even if we can come up with a robust analytical model it may soon be obsolete \cite{Kester2017}. 
We take a hybrid route using a combination of analytical models, benchmarks, and machine learning for a small set of fundamental access primitives.
For example, all pointer based data structures need to perform random accesses as operations traverse their nodes. 
All data structures need to perform a write during an update operation, regardless of the exact update strategy. We synthesize the cost of complex operations out of models that describe those simpler more fundamental operations inspired by past work on generalized models \cite{Manegold2002,Yao1977}.
  In addition, our models start out as analytical models since we know how these primitives will likely behave. However, they are also trained across diverse hardware profiles by running benchmarks that isolate the behavior of those primitives. This way, we learn a set of coefficients for each model that capture the subtle performance details of diverse hardware settings.

\noindent{\textbf{The Data Calculator: Automated What-if Design.}}
\chm{We present a ``design engine'' -- the Data Calculator -- that can 
 compute the performance of arbitrary data structure designs as combinations of fundamental design primitives.}
It is an interactive tool that  
accelerates the process of design by turning it into an exploration process, improving the productivity of researchers and engineers; 
it is able to answer what-if data structure design questions to understand how the introduction of 
new design choices, workloads, and hardware affect the performance (latency) of an existing design. 
It currently supports read queries for basic hardware conscious layouts.
It allows users to give as input a high-level specification of the layout of a data structure (as a combination of primitives), 
in addition to workload, and hardware specifications.
\chm{The Data Calculator gives as output a calculation of the latency to run the input workload on the input hardware.} 
The architecture and components of the Data Calculator are captured in Figure \ref{figure:architecture} (from left to right): 
(1) a library of fine-grained data layout primitives that can be combined in arbitrary ways to describe data structure layouts; 
(2) a library of data access primitives that can be combined to generate designs of operations;
(3) an operation and cost synthesizer that computes the design of operations
and their latency for a given data structure layout specification, a workload
and a hardware profile,
and (4) a search component that can traverse part of the design space to supplement a partial data structure specification or inspect an existing one
with respect to both the layout and the access design choices.

\noindent{\textbf{Inspiration.}}
Our work is inspired by several lines of work across many fields of computer science.
John Ousterhout's project Magic in the area of 
computer architecture allows for quick verification of transistor designs so that engineers 
can easily test multiple designs \cite{Ousterhout1984}. 
Leland Wilkinson's ``grammar of graphics'' provides structure and formulation 
on the massive universe of possible graphics one can design \cite{Wilkinson2005}.
Mike Franklin's Ph.D. thesis explores the possible client-server architecture designs using 
caching based replication as the main design primitive and proposes a taxonomy that produced both published and unpublished (at the time) cache consistency algorithms. 
Joe Hellerstein's work on Generalized Search Indexes \cite{Hellerstein1995, Aoki1998, Aoki1999, Kornacker1997, Kornacker1999,Kornacker1998,Kornacker2003} makes it easy to design and test new data structures by providing templates that significantly minimize implementation time.
S. Bing Yao's work on generalized cost models \cite{Yao1977} for database organizations, and Stefan Manegold's work on generalized cost models tailored for the memory hierarchy \cite{ManegoldPhD} showed that it is possible to synthesize the costs of database operations from basic access patterns and based on hardware performance properties.  
Work on data representation synthesis in programming languages \cite{Schonberg1979,Schonberg1981,Cohen1993,Smaragdakis1997,Shacham2009,Hawkins2011,Hawkins2012,Loncaric2016,Steindorfer2016} enables selection and synthesis of representations out of
small sets of (3-5) existing data structures. 
The Data Calculator \cite{Calculator} can be seen as a step toward the Automatic Programmer 
challenge set by Jim Gray in his Turing award lecture \cite{Gray2000},
and as a step toward the ``calculus of data structures'' challenge set by 
Turing award winner Robert Tarjan \cite{Tarjan1978}: \textit{``What makes one data structure better than another for a certain application? The known results cry out for an underlying theory to explain them.''}

\textbf{Contributions.}
Our contributions are as follows: 
        \vspace{-3ex}    
\begin{enumerate}
  \item We introduce a set of data layout design primitives that capture the first principles of data layouts including hardware conscious designs that dictate the relative positioning of data structure nodes \cref{sec:design}. 
    \item We show how combinations of the design primitives can describe known data structure designs, including arrays, linked-lists, skip-lists, queues, hash-tables, binary trees and (Cache-conscious) b-trees, tries, MassTree, and FAST \cref{sec:design}.
    \item We show that in addition to known designs, the design primitives form a massive space of possible designs that has only been minimally explored in the literature \cref{sec:design}.
  \item We show how to synthesize the latency cost of basic operations (point and range queries, and bulk loading) of arbitrary data structure designs from a small set of access primitives. Access primitives represent fundamental ways to access data and come with learned cost models which are trained on diverse hardware to capture hardware properties \cref{sec:access}.
  \item We show how to use cost synthesis to interactively answer 
    complex what-if design questions, i.e., the impact of changes to design, workload, and hardware \cref{sec:search}. 
    \item We introduce a design synthesis algorithm that completes partial layout specifications given a workload and hardware input; it utilizes cost synthesis to rank designs \cref{sec:search}. 
  \item We demonstrate that the Data Calculator can accurately 
     compute the performance impact of design choices for state-of-the-art designs
    and diverse hardware \cref{sec:eval}.
  \item We demonstrate that the Data Calculator can accelerate the design process by answering rich design questions in a matter of seconds or minutes \cref{sec:eval}. 
\end{enumerate}

\section{Data Layout Primitives \newline and Structure Specifications}
\label{sec:design}
In this section, we discuss 
the library of data layout design primitives and how it enables the description of
a massive number of both known and unknown data structures.

\begin{figure*}
    \includegraphics[width=\textwidth]{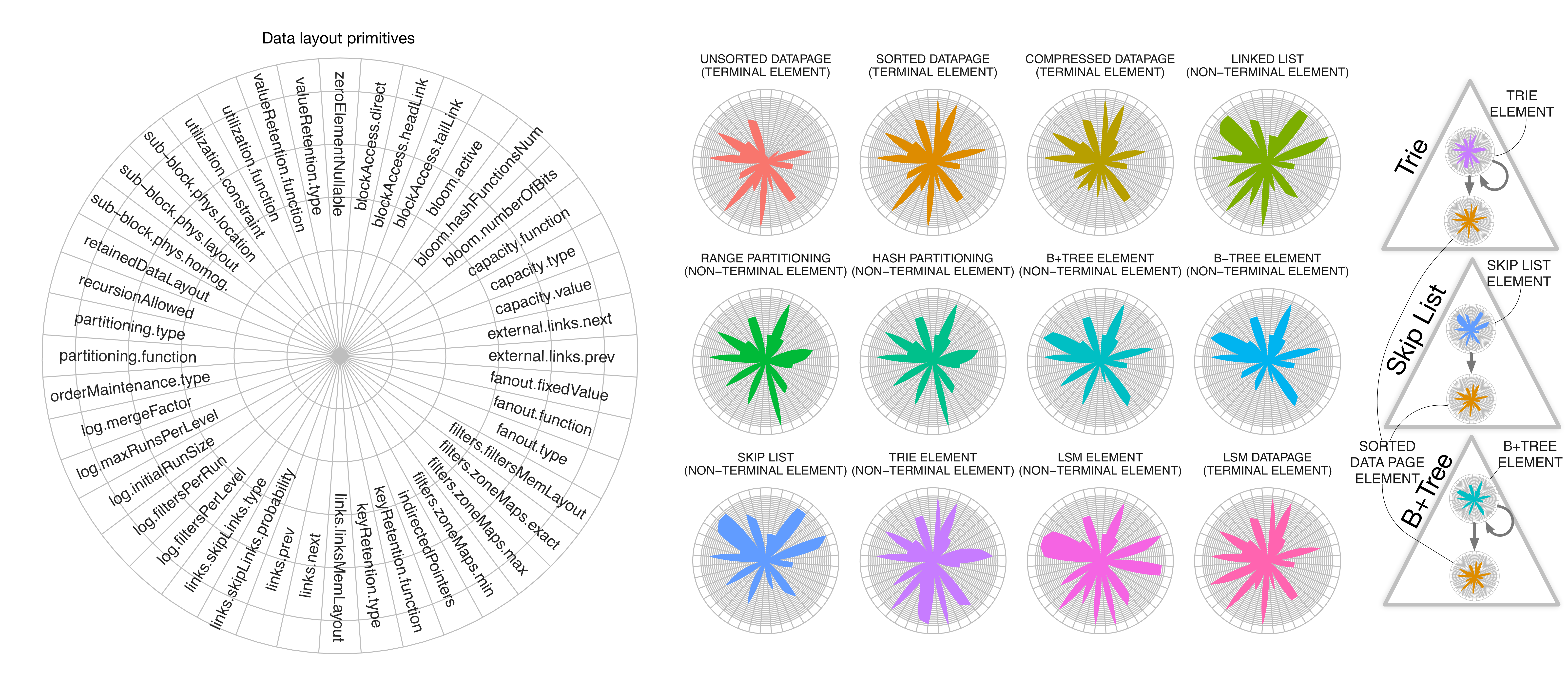}
    \vspace{-9ex}
    \caption{\ch{The data layout primitives and examples of synthesizing node layouts of state-of-the-art data structures.}}
    \vspace{-3ex}
    \label{figure:nodePrimitives}
\end{figure*}

\noindent{\textbf{Data Layout Primitives.}}
The Data Calculator contains a small set of design primitives 
that represent fundamental design choices when constructing a data structure layout.
Each primitive belongs to a class of primitives depending on the high-level design concept it refers to such as
node data organization, partitioning, node physical placement, and node metadata management.
Within each class, individual primitives define design choices and allow for alternative tunings.
The complete set of primitives we introduce in this paper is shown in Figure \ref{table:layoutprimitivestable} in the appendix;
they describe basic data layouts and cache conscious optimizations for reads.  
For example, ``Key Partitioning (none|sorted|k-ary|range|temporal)'' defines how data is laid out in a node.
Similarly, ``Key Retention (none|full|func)'' defines whether and how keys are included in a node.
In this way, in a B+tree all nodes use ``sorted'' for order maintenance, while internal nodes use ``none'' for key retention as they only store fences and pointers, and leaf nodes use ``full'' for key retention.

The logic we use to generate primitives is that each one should represent a fundamental design concept
that does not break down into more useful design choices 
(otherwise, there will be parts of the design space we cannot express).
Coming up with the set of primitives is a trial and error task to map the known space of design concepts
to an as clean and elegant set of primitives as possible.

Naturally, not all layout primitives can be combined. Most invalid relationships stem from the structure of the 
primitives, i.e., each primitive combines with every other standalone primitive.
Only a few pairs of primitive tunings do not combine which generates a small set of invalidation rules.
These are mentioned in Figure \ref{table:layoutprimitivestable}.

\noindent{\textbf{From Layout Primitives to Data Structures.}}
To describe complete data structures, we introduce the concept of $elements$.
An element is a full specification of a single data structure node; it defines the data 
and access methods used to access the node's data.
An element may be ``terminal'' or ``non-terminal''.
That is, an element may be describing a node that further partitions data to more nodes or not.
This is done with the ``fanout'' primitive whose value represents the maximum number of children that would be generated
when a node partitions data. Or it can be set to ``terminal'' in which case its value represents the capacity of a terminal node. 
A data structure specification contains one or more elements. It needs to have at least one terminal element,
and it may have zero or more non-terminal elements. 
Each element has a destination element (except terminal ones) and a source element (except the root). 
Recursive connections are allowed to the same element.

\noindent{\textbf{Examples.}}
A visualization of the primitives can be seen at the left side of Figure \ref{figure:nodePrimitives}.
It is a flat representation of the primitives shown in Figure \ref{table:layoutprimitivestable} which creates an entry for every primitive signature.
\ch{The radius depicts the domain of each primitive but different primitives may have different domains, visually
depicted via the multiple inner circles in the radar plots of Figure \ref{figure:nodePrimitives}.}
The small radar plots on the right side of Figure \ref{figure:nodePrimitives}
depict descriptions of nodes of known data structures as combinations of the base primitives.
Even visually it starts to become apparent that state-of-the-art designs
which are meant to handle different scenarios are ``synthesized from the same pool of design concepts''.
For example, using the non-terminal B+tree element and the terminal sorted data page element 
we can construct a full B+tree specification;
data is recursively broken down into internal nodes using the B+tree element until we reach
the leaf level, i.e., when partitions reach the terminal node size.
Figure \ref{figure:nodePrimitives} also depicts Trie and Skip-list specifications.
Figure \ref{table:layoutprimitivestable} provides complete specifications of Hash-table, Linked-list, B+tree, Cache-conscious B-tree, and FAST.

\noindent{\textbf{Elements ``Without Data''.}}
For flat data structures without an indexing layer, e.g., linked-lists and skip-lists, there need to be elements in the specification that describe the algorithm used to navigate the terminal nodes. Given that this algorithm is effectively a model, it does not rely on any data, and so such elements do not translate to actual nodes; they only affect algorithms that navigate across the terminal nodes. For example, a linked-list element in Figure \ref{table:layoutprimitivestable} describes that data is divided into nodes that can only be accessed via following the links that connect terminal nodes.
Similarly, one can create complex hierarchies of non-terminal elements that do not store any data but instead their job is to synthesize a collective model of how the keys should be distributed in the data structure, e.g., based on their value or other properties of the workload. These elements may lead to multiple hierarchies of both non-terminal nodes with data and terminal ones, synthesizing data structure designs that treat parts of the data differently. We will see such examples in the experimental analysis.  

\noindent{\textbf{Recursive Design Through Blocks.}}
A block is a logical portion of the data that we  
divide into smaller blocks to construct an instance of a data structure
specification.
The elements in a specification are the ``atoms'' with which we construct data structure instances
by applying them recursively onto blocks.
Initially, there is a single block of data, all data.  
Once all elements have been applied, the original block is broken down into a set of 
smaller blocks that correspond to the internal nodes (if any)
and the terminal nodes of the data structure.
Elements without data can be thought of as if they apply on a logical data block that represents part of the data
with a set of specific properties (i.e., all data if this is the first element) and partitions the data
with a particular logic into further logical blocks or physical nodes. 
This recursive construction is used when we test, cost, and search through multiple possible designs 
concurrently over the same data for a given workload and hardware as we will discuss in the next two sections,
but it is also helpful to visualize designs as if ``data is pushed through the design'' based on the elements and logical blocks.

\noindent{\textbf{Cache-Conscious Designs.}}
One critical aspect of data structure design is the relative positioning of its nodes, i.e., how ``far'' each node
is positioned with respect to its predecessors and successors in a query path. 
This aspect is critical to the overall cost of traversing a data structure.
The Data Calculator design space allows to dictate how nodes should be positioned explicitly:
each non-terminal element defines how its children are positioned physically with respect to each other and with respect 
to the current node.
For example, setting the layout primitive ``Sub-block physical layout'' to BFS tells the current node that its children are laid out sequentially. In addition, setting the layout primitive ``Sub-blocks homogeneous'' to true implies that all its children have the same layout (and therefore fixed width), and allows a parent node to access any of its children nodes directly with a single pointer and reference number. This, in turn, makes it possible to fit more data in internal nodes because only one pointer is needed and thus more fences can be stored within the same storage budget. Such primitives allow  specifying designs such as Cache Conscious B+tree \cite{Rao2000} (Figure \ref{table:layoutprimitivestable} provides the complete specification), but also the possibility of generalizing the optimizations made there to arbitrary structures.

Similarly, we can describe FAST \cite{Kim2010}. First, we set ``Sub-block physical location'' to inline, specifying that the children nodes are directly after the parent node physically. Second, we set that the children nodes are homogeneous, and finally, we set that the children have a sub-block layout of ``BFS Layer List (Page Size / Cache Line Size)''. Here, the BFS layer list specifies that on a higher level, we should have a BFS layout of sub-trees containing Page Size/Cache Line Size layers; however, inside of those sub-trees pages are laid out in BFS manner by a single level. The combination matches the combined Page Level blocking and Cache Line level blocking of FAST. Additionally, the Data Calculator realizes that all child node physical locations can be calculated via offsets, and so eliminates all pointers.  
Figure \ref{table:layoutprimitivestable} provides the complete specification.

\noindent{\textbf{Size of the Design Space.}}
To help with arguing about the possible design space we provide formal definitions of the various constructs.

\begin{definition}[Data Layout Primitive]
A primitive $p_i$ belongs to a domain of values $\mathcal{P}_i$ and describes a layout aspect of a data structure node.
\end{definition}

\begin{definition}[Data Structure Element]
A Data Structure Element $E$ is defined as a set of data layout primitives: $E=\{p_1, ..., p_n\} \in \mathcal{P}_i \times ... \times \mathcal{P}_n$, that uniquely identify it.
\end{definition}

Given a set of $Inv(\mathcal{P})$ invalid combinations,
the set of all possible elements $\mathcal{E}$, (i.e., node layouts) that can be designed
as distinct combinations of data layout primitives has the following cardinality.
\begin{equation}
\label{formula:primitives}
|\mathcal{E}| = \mathcal{P}_i \times ... \times \mathcal{P}_n - Inv(\mathcal{P})= \prod_{\forall P_i \in E} |\mathcal{P}_i| - Inv(\mathcal{P})
\end{equation}

\begin{definition}[Blocks]
Each non-terminal element $E \in \mathcal{E}$, applied on a set of data entries $D \in \mathcal{D}$, uses function
$B_E(D) = \{D_1, ..., D_f\}$ to divide $D$ into $f$ blocks such that $D_1 \cup ... \cup D_f = D$.
\end{definition}

A polymorphic design where every block may be described by a
different element leads to the following recursive formula
for the cardinality of all possible designs.
\begin{equation}
\label{formula:polymorphicRec}
c_{poly}(D) = |\mathcal{E}| + \sum_{\forall E\in\mathcal{E}}\sum_{\forall D_i\in B_E(D)} c_{poly}(D_i)
\end{equation}

\noindent{\textbf{Example: A Vast Space of Design Opportunities.}}
To get insight into the possible total designs 
we make a few simplifying assumptions.
Assume the same fanout $f$ across all nodes and terminal node size equal to page size $p_{size}$.
Then $N=\lceil\frac{|D|}{p_{size}}\rceil$ is the total number of pages in which we can divide the data
and $h=\lceil log_f(N) \rceil$ is the height of the hierarchy.
We can then approximate the result of Equation \ref{formula:polymorphicRec}
by considering that we have $|E|$ possibilities for the root element, and $f*|E|$ possibilities for its resulting partitions which in turn
have $f*|E|$ possibilities each up to the maximum level of recursion $h = log_f(N)$. This leads to the following result.
\begin{equation}
\label{formula:polymorphic}
c_{poly}(D) \approx |\mathcal{E}|*(f*|\mathcal{E}|)^{\lceil log_f(N) \rceil}
\end{equation}

Most sophisticated data structure designs use only two distinct elements,
each one describing all nodes across groups of levels of the structure,
e.g., B-tree designs use one element for all internal nodes and one for all leaves.
This gives the following design space for most standard designs.
\begin{equation}
\label{formula:standard}
c_{stan}(D) \approx |\mathcal{E}|^2
\end{equation}

Using Equations  \ref{formula:primitives}, \ref{formula:polymorphic} and \ref{formula:standard} we can
get estimations of the possible design space for different kinds of data structure designs.
For example, given the existing library of data layout primitives, and by limiting the domain
of each primitive as shown in Figure \ref{table:layoutprimitivestable} in appendix,
then from  Equation \ref{formula:primitives} we get $|\mathcal{E}|=10^{16}$, meaning we can describe
data structure layouts from a design space of $10^{16}$ possible node elements and their combinations.
\chm{This number includes only valid combinations of layout primitives, i.e., all invalid combinations as defined
by the rules in Figure \ref{table:layoutprimitivestable} are excluded.}
Thus, we have a design space of $10^{32}$ for standard two-element structures (e.g., where B-tree and Trie belong)
and $10^{48}$ for three-element structures
(e.g., where MassTree \cite{Mao2012} and Bounded-Disorder \cite{Litwin1986} belong).
For polymorphic structures, the number of possible designs grows more quickly, and it also depends on the 
size of the training data used
to find a specification, e.g., it is $> 10^{100}$ for $10^{15}$ keys.

The numbers in the above example highlight that data structure design is still a wide-open space with numerous opportunities for innovative designs
as data keeps growing, application workloads keep changing, and hardware keeps evolving. 
\chm{Even with hundreds of new data structures manually designed and published each year, this is a slow pace to test all possible designs and to be able to argue about how the numerous designs compare.
The Data Calculator is a tool that accelerates this process by 1) providing guidance about what is the possible design space, and 2) allowing to quickly test how a given design fits a workload and hardware setting.} 
A more detailed description of the primitives can be found in the appendix.


\section{Data Access Primitives \newline and Cost Synthesis}
\label{sec:access}

We now discuss how the Data Calculator computes the cost (latency) of running a given workload
on a given hardware for a particular data structure specification. 
Traditional cost analysis in systems and data structures happens through experiments and the development of analytical cost models. 
Both options are not scalable when we want to quickly test multiple different parts of the massive design space we define in this paper. They require significant expertise and time, while they are also sensitive to hardware and workload properties. Our intuition is that we can instead synthesize complex operations from their fundamental components as we do for data layouts in the previous section, and then develop a hybrid way (through both benchmarks and models but without significant human effort needed) to assign costs to each component individually;
The main idea is that we learn a small set of cost models for fine-grained data access patterns out of which we can synthesize the cost of complex dictionary operations for arbitrary designs in the possible design space of data structures.

\begin{figure*}[t]
     \includegraphics[width=\textwidth]{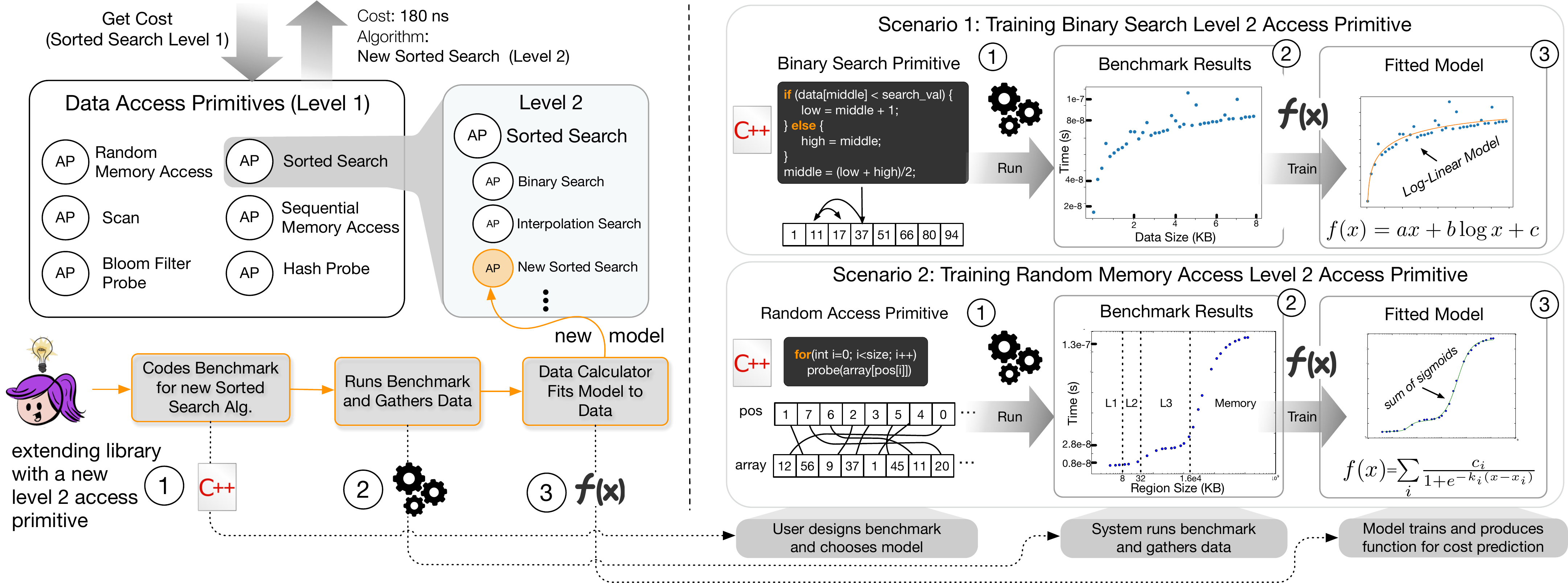}
     \vspace{-8mm}
     \caption {\small \ch{Training and fitting models for Level 2 access primitives and extending the Data Calculator.}}
   \vspace{-3ex}    
  \label{fig:combined}
 \end{figure*}

The middle part of Figure \ref{figure:architecture} depicts the components of the Data Calculator that make cost synthesis possible:
1) the library of data access primitives, 
2) the cost learning module which trains cost models for each access primitive depending on hardware and data properties, 
and 3) the operation and cost synthesis module which synthesizes dictionary operations and their costs from the access primitives and the learned models.
Next, we describe the process and components in detail.

\noindent{\textbf{Cost Synthesis from Data Access Primitives.}}
Each access primitive characterizes one aspect of how data is accessed.
For example, a binary search, a scan, a random read, a sequential read, a random write, 
are access primitives.
The goal is that these primitives should be fundamental enough so that we can 
use them to synthesize operations over arbitrary designs as sequences of such primitives. 
There exist two levels of access primitives.
Level 1 access primitives are marked with white color in Figure \ref{figure:architecture}
and Level 2 access primitives are nested under Level 1 primitives and marked with gray color. 
For example, a scan is a Level 1 access primitive used any time an operation needs to search a block of data where there is no order. 
At the same time, a scan may be designed and implemented in more than one way; this is exactly what Level 2 access primitives represent. 
For example, a scan may use SIMD instructions for parallelization if keys are nicely packed in vectors, 
and predication to minimize branch mispredictions with certain selectivity ranges.
In the same way, a sorted search may use interpolation search if keys are arranged with uniform distribution.   
In this way, each Level 1 primitive is a conceptual access pattern,
while each Level 2 primitive is an actual implementation that signifies a specific set of design choices. 
Every Level 1 access primitive has at least one Level 2 primitive and may be extended with any number of additional ones. 
The complete list of access primitives currently supported by the Data Calculator 
is shown in Table \ref{table:accessPrimitives} in appendix.

\noindent{\textbf{Learned Cost Models.}}
For every Level 2 primitive, the Data Calculator contains one or more models 
that describe its performance (latency) behavior. 
These are not static models; they are trained and fitted for combinations of data and hardware profiles
as both those factors drastically affect performance.
To train a model, each Level 2 primitive includes a minimal implementation that captures the behavior of the primitive, i.e., it isolates the performance effects of performing the specific action.
For example, an implementation for a scan primitive simply scans an array, while an implementation for a random access primitive simply tries to access random locations in memory. These implementations are used to run a sequence of benchmarks to collect data for learning a model for the behavior of each primitive.  Implementations should be in the target language/environment. 

The models are simple parametric models; given the design decision to keep primitives simple (so they can be easily reused), we have domain expertise to expect how their performance behavior will look like. 
For example, for scans, we have a strong intuition they will be linear, for binary searches that they will be logarithmic, and that for random memory accesses that they will be smoothed out step functions (based on the probability of caching). These simple models have many advantages: they are interpretable, they train quickly, and they don't need a lot of data to converge. Through the training process, the Data Calculator learns coefficients of those models that capture hardware properties such as CPU and data movement costs. 

Hardware and data profiles hold descriptive information about data and hardware respectively (e.g., data distribution for data, and CPU, Bandwidth, etc. for hardware). When an access primitive is trained on a data profile, it runs on a sample of such data, and when it is trained for a hardware profile, it runs on this exact hardware. Afterward, though, design questions can get accurate cost estimations on arbitrary access method designs without going over the data or having to have access to the specific machine. Overall, this is an offline process that is done once, and it can be repeated to include new hardware and data profiles or to include new access primitives.

\noindent{\textbf{Example: Binary Search Model.}}
To give more intuition about how models are constructed let us consider the case of a Level 2 primitive of binary searching a sorted array as shown on the upper right part of Figure \ref{fig:combined}. 
The primitive contains a code snippet that implements the bare minimum behavior (Step 1 in Figure \ref{fig:combined}). 
We observe that the benchmark results (Step 2 in Figure \ref{fig:combined}) indicate that performance is related to the size of the array by a logarithmic component. As expected there is also bias as the relationship for small array sizes (such as just 4 or 8 elements) might not fit exactly a logarithmic function. We additionally add a linear term to capture some small linear dependency on the data size. Thus, the cost of binary searching an array of $n$ elements can be approximated as $f(n) = c_{1} n + c_{2} \log n + y_{0}$ where $c_{1}, c_{2},$ and $y_{0}$ are coefficients learned through linear regression. The values of these coefficients help us translate the abstract model, $f(n) = O(\log n)$, into a realized predictive model which has taken into account factors such as CPU speed and the cost of memory accesses across the sorted array for the specific hardware. The resulting fitted model can be seen in Step 3 on the upper right part of Figure \ref{fig:combined}.
The Data Calculator can then use this learned model to query for the performance of binary search within the trained range of data sizes. For example, this would be used when querying a large sorted array as well as a small node of a complex data structure that is sorted. 

Certain critical aspects of the training process can be automated as part of future research. 
For example, the data range for which we should train a primitive depends on the memory hierarchy (e.g., size of caches, memory, etc.) on the target machine and what is the target setting in the application (i.e., memory only, or also disk/flash, etc.). In turn, this also affects the length of the training process. Overall, such parameters can eventually be handled through high-level knobs, letting the system make the lower level tuning choices. 
Furthermore, identification of convergence can also be automated. There exist primitives that require more training than others (e.g., due to more complex code, random access or sensitivity to outliers), and so the number of benchmarks and data points we collect should not be a fixed decision.

\begin{figure*}[t]
\includegraphics[width=\textwidth]{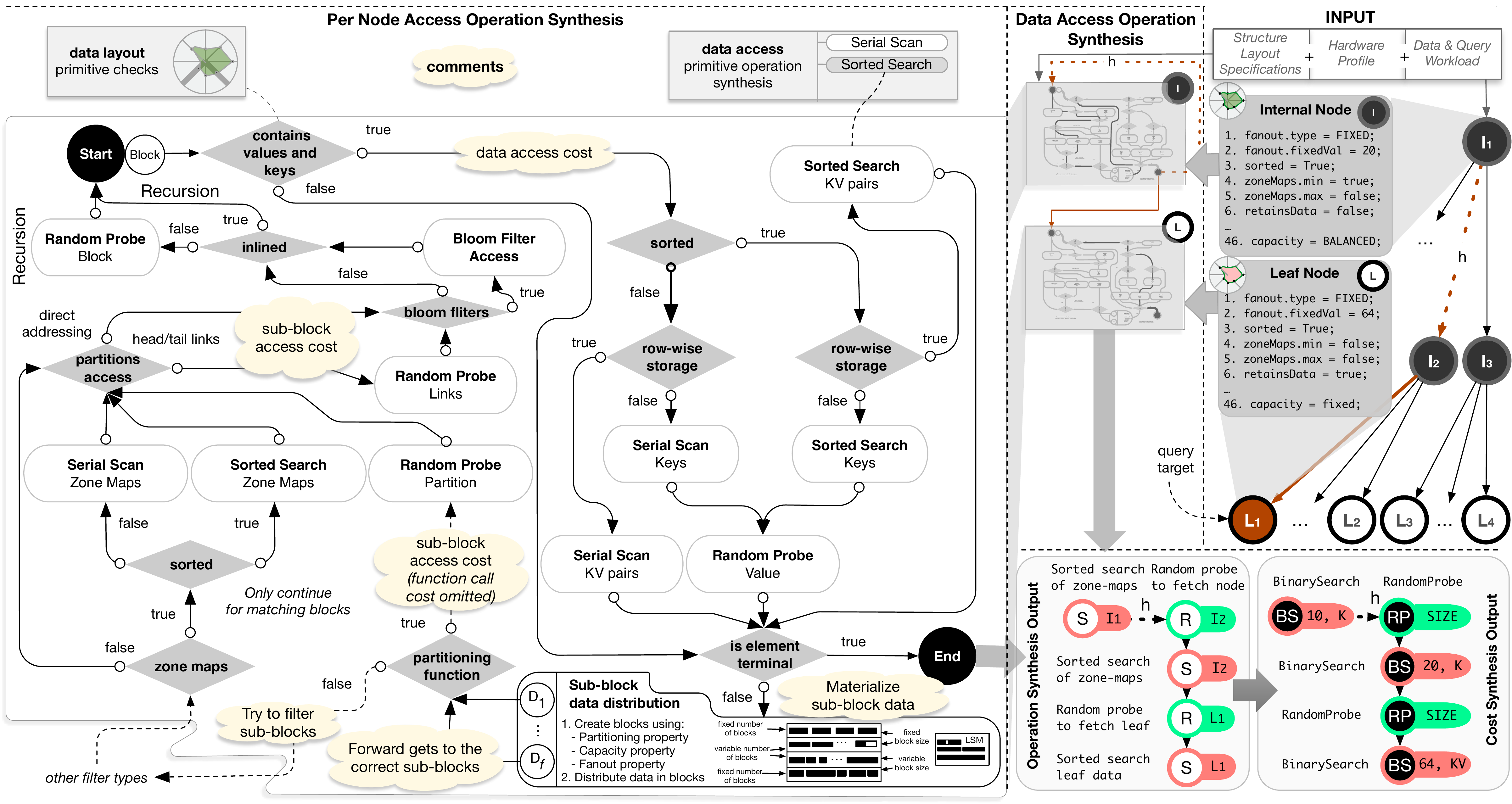}
\vspace{-5ex}    
\caption{\ch{Synthesizing the operation and cost for dictionary operation Get, given a data structure specification.}}
\vspace{-3ex}    
\label{fig:operations}
\end{figure*}

\noindent{\textbf{Synthesizing Latency Costs.}}
\ch{Given a data layout specification and a workload, the Data Calculator uses Level 1 access primitives to synthesize 
operations and subsequently each Level 1 primitive is translated to the appropriate Level 2 primitive
to compute the cost of the overall operation.}  
Figure \ref{fig:operations} depicts this process and an example specifically for the {\tt Get} operation. 
This is an expert system, i.e., a sequence of rules that based on a given data structure specification
defines how to traverse its nodes.\footnote{Figure \ref{fig:operations} is a subset of the expert system. The complete version can be found in the appendix.}
To read Figure \ref{fig:operations} start from the top right corner. 
The input is a data structure specification, a test data set, 
and the operation we need to cost, e.g., {\tt Get key x}.
\ch{The process simulates populating the data structure with the data to figure out how many nodes exist, the height of the structure, etc.
This is because to accurately estimate the cost of an operation, the Data Calculator needs to take into account the 
expected state of the data structure at the particular moment in the workload.
It does this by recursively dividing the data into blocks given the elements used in the specification.}

In the example of Figure \ref{fig:operations} the structure contains two elements, one for internal nodes and one for leaves. 
\ch{For every node, the operation synthesis process takes into account the data layout primitives used.} 
For example, if a node is sorted it uses binary search, but if the node is unsorted, it uses a full scan. 
The rhombuses on the left side of Figure \ref{fig:operations} reflect the data layout primitives that operation {\tt Get} relies on, while
the rounded rectangles reflect data access primitives that may be used.
\ch{For each node the per-node operation synthesis procedure (starting from the left top side of Figure \ref{fig:operations}), first checks if this node is internal or not by checking whether the node contains keys or values; if not, it proceeds to determine which node it should visit next (left side of the figure) and if yes, it continues to process the data and values (right side of the figure).}
A non-terminal element leads to data of this block being split into $f$ new blocks and the process follows the relevant blocks only, i.e., the blocks that this operation needs to visit to resolve.

\ch{
In the end, the Data Calculator generates an  
abstract syntax tree with the access patterns of 
the path it had to go through. This is expressed in terms of Level 1 access primitives
(bottom right part of Figure \ref{fig:operations}).
In turn, this is translated to a more detailed abstract syntax tree where all Level 1 access primitives
are translated to Level 2 access primitives along with the estimated cost for each one
given the particular data size, hardware input, and any primitive specific input. 
The overall cost is then calculated as the sum of all those costs. 
}

\ch{
\noindent{\textbf{Calculating Random Accesses and Caching Effects.}}
A crucial part in calculating the cost of most data structures is capturing random memory access costs, 
e.g., the cost of fetching nodes while traversing a tree, fetching nodes linked in a hash bucket, etc.
If data is expected to be cold, then this is a rather straightforward case, i.e., we may assign
the maximum cost  a random access is expected to  incur on the target machine.
If data may be hot, though, it is a more involved scenario.
For example, in a tree-like structure internal nodes higher in the tree are much more 
likely to be at higher levels of the memory hierarchy during repeated requests. 
We calculate such costs using the random memory access primitive, as shown in the lower right part of Figure \ref{fig:combined}. 
Its input is a ``region size'', which is best thought of as the amount of memory we are randomly accessing into (i.e., we don't know where in this memory region our pointer points to). 
The primitive is trained via benchmarking access to an increasingly bigger contiguous array (Step 1 in Figure \ref{fig:combined}). The results (Step 2 in Figure \ref{fig:combined}) depict a minor jump from L1 to L2 (we can see a small bump just after $10^{4}$ elements). The bump from L2 to L3 is much more noticeable, with the cost of accessing memory going from $0.1 \times 10^{7}$ seconds to $0.3 \times 10^{7}$ seconds as the memory size crosses the 128 KB boundary. Similarly, we see a bump from $0.3 \times 10^{7}$ seconds to $1.3 \times 10^{7}$ seconds when we go from L3 to main memory, at the L3 cache size of 16 MB\footnote{\ch{These numbers are in line with Intel's Vtune.}}. 
We capture this behavior as a sum of sigmoid functions, which are essentially smoothed step functions, using $$c(x) = \sum_{i=1}^{k} f(x) = \sum_{i=1}^{k} \frac{c_{i}}{1 + e^{-k_{i} (\log x - x_{i})}} +  y_{0}.$$ This primitive is used for calculating random access to any physical or logical region (e.g., a sequence of nodes that may be cached together). For example, when traversing a tree, the cost synthesis operation, costs random accesses with respect to the amount of data that may be cached up to this point. That is, for every node we need to access at Level $x$ of a tree, we account for a region size that includes all data in all levels of the tree up to Level $x$. In this way, accessing a node higher in the tree costs less than a node at lower levels. 
The same is true when accessing buckets of a hash table. We give a detailed step by step example below.}

\noindent{\textbf{\ch{Example: Cache-aware Cost Synthesis.}}
Assume a B-tree -like specification as follows: two node types, one for internal nodes and one for leaf nodes.
Internal nodes contain fence pointers, are sorted, balanced, have a fixed fanout of 20 and do not contain any keys or values.
Leaf nodes instead are terminal; they contain both keys and values, are sorted, have a maximum page size of 250 records, 
and follow a full columnar format, where keys and values are stored in independent arrays.
The test dataset consists of $10^5$ records where keys and values are 8 bytes each. 
Overall, this means that we have 400 full data pages, and thus a tree of height 2.
The Data Calculator needs two of its access primitives to calculate the cost of a Get operation.
Every Get query will be routed through two internal nodes and one leaf node: 
within each node, it needs to binary search (through fence pointers for internal nodes and through keys in leaf nodes)
and thus it will make use of the Sorted Search access primitive. 
In addition, as a query traverses the tree it needs to perform 
a random access for every hop.

Now, let us look in more detail how these two primitives are used given the exact specification of this data structure. 
The Sorted Search primitive takes as input the size of the area over which we will binary search and the number of keys.
The Random Access primitive takes as input the size of the path so far which allows us to takes into account caching effects. 
Each query starts by visiting the root node.
The data calculator estimates the size of the path so far to be 312 bytes.
This is because the size of the path so far is in practice equal to the size of the root node which 
containing 20 pointers (because the fanout is 20) and 19 values sums up at $root=internalnode=20*8+19*8=312$ bytes.
In this way, the Data Calculator logs a cost of $RandomAccess(312)$ to access the root node. 
Then, it calculates the cost of binary search across 19 fences, thus logging a cost of $SortedSearch(RowStore,19*8)$.
It uses the ``RowStore'' option as fences and pointers are stored as pairs within each internal node. 
Now, the access to the root node is fully accounted for, and the Data Calculator moves on to cost the access at the next tree level.
Now the size of the path so far is given by accounting for the whole next level in addition to the root node.
This is in total $level2=root+$\emph{fanout}$*internalnode=312+20*312=6552$ bytes (due to fanout being 20 we account for 20 nodes at the next level). 
Thus to access the next node, the Data Calculator logs a cost of $RandomAccess(6552)$ 
and again a search cost of $SortedSearch(RowStore,19*8)$ to search this node.
The last step is to search the leaf level. 
Now the size of the path so far is given by accounting for the whole size of the tree which is 
$level2+400*(250*16)=1606552$ bytes since we have 400 pages at the next level (20x20) and each page has 250 records of key-value pairs (8 bytes each).
In this way, the Data Calculator logs a cost of $RandomAccess(1606552)$ to access the leaf node, 
followed by a sorted search of $SortedSearch(ColumnStore,250*8)$ to search the keys.
It uses the ``ColumnStore'' option as keys and values are stored separately in each leaf in different arrays. 
Finally, a cost of $RandomAccess(2000)$ is incurred to access the target value in the values array (we have $8*250=2000$ in each leaf).

\ch{
\noindent{\textbf{Sets of Operations.}}
The description above considers a single operation. 
The Data Calculator can also compute the latency for a set of operations concurrently in a single pass.}
This is effectively the same process as shown in Figure \ref{fig:operations} only that in every recursion we may follow more than one path and in every step we are computing the latency 
for all queries that would visit a given node. 

\ch{
\noindent{\textbf{Workload Skew and Caching Effects.}}
Another parameter that can influence caching effects is workload skew.
For example, repeatedly accessing the same path of a data structure results in all nodes in this path being cached
with higher probability than others. 
The Data Calculator first generates counts of how many times every node is going to be accessed for a given workload. 
Using these counts and the total number of nodes accessed we get a factor $p=count/total$ that denotes the popularity of a node.
Then to calculate the random access cost to a node for an operation $k$, a weight $w=1/(p*sid)$ is used, where $sid$ is the sequence number of this operation in the workload (refreshed periodically). 
Frequently accessed nodes see smaller access costs and vice versa.
}

\ch{
\noindent{\textbf{Training Primitives.}}
All access primitives are trained on warm caches. This is because they are used to calculate the cost on a node that is already 
fetched. The only special case is the Random Access primitive which is used to calculate the cost of fetching a node. 
This is also trained on warm data, though, since the cost synthesis infrastructure takes care at a higher level
to pass the right region size as discussed; in the case this region is big, this can still result in costing a page fault as large data will not fit in the cache which is reflected in the Random Access primitive model.  
}

\ch{\textbf{Limitations.} For individual queries certain access primitives are hard to estimate precisely without running the actual code on an exact data instance. For example, a scan for a point Get may abort after checking just a few values, or it may need to go all the way to the end of an array. In this way, 
while lower or upper performance bounds can be computed with absolute confidence for both individual queries and sets of queries, actual performance estimation works best for sets. 
}

\noindent{\textbf{More Operations.}} 
The cost of range queries, and bulk loading is synthesized
as shown in Figure \ref{fig:OpRangeBulk} in appendix. 

\begin{figure*}[t]
    \includegraphics[width=2.15\columnwidth]{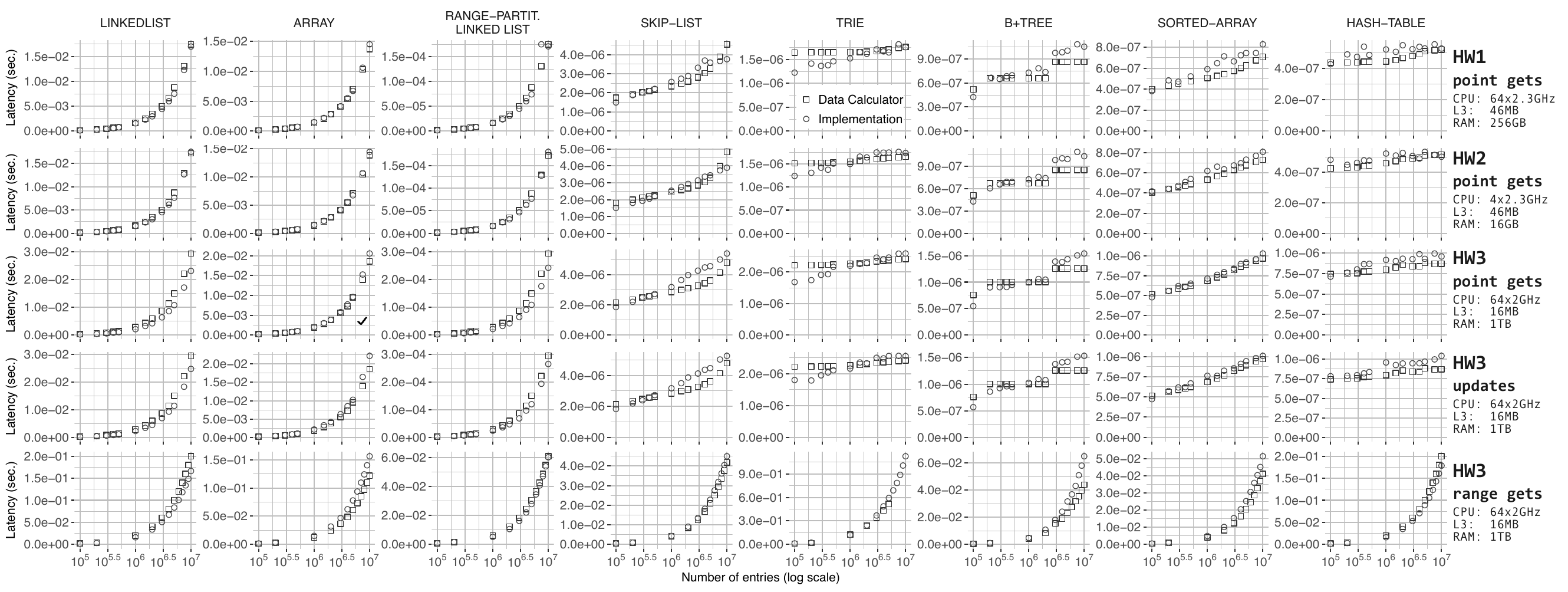}
  \vspace{-6ex}
    \caption{\ch{The Data Calculator can accurately compute the latency of arbitrary data structure designs across a diverse set of hardware and for diverse dictionary operations.}}
     \vspace{-4ex}
    \label{fig:fitting}
\end{figure*}

\noindent{\textbf{Extensibility and Cross-pollination.}}
The rationale of having two Levels of access primitives is threefold. 
First, it brings a level of abstraction allowing higher level cost synthesis algorithms to operate at Level 1 only. 
Second, it brings extensibility, i.e.,  we can add new Level 2 primitives 
without affecting the overall architecture. 
Third, it enhances ``cross-pollination'' of design concepts captured by Level 2 primitives across designs.
Consider the following example. 
An engineer comes up with a new algorithm to perform search over a sorted array, e.g., exploiting new hardware instructions. 
To test if this can improve performance in her B-tree design, where she regularly searches over sorted arrays, she codes up a benchmark for a new sorted search Level 2 primitive and plugs it in the Calculator 
as shown in Figure \ref{fig:combined}. 
Then the original B-tree design can be easily tested
with and without the new sorted search across several workloads and hardware profiles without having to undergo a lengthy implementation phase. 
At the same time, the new primitive can now be considered by any 
data structure design that contains a sorted array such as an LSM-tree with sorted runs, a Hash-table with sorted buckets and so on.
This allows easy transfer of ideas and optimizations across designs, a process that usually requires a full study for each optimization and target design.

\section{What-if Design and Auto-completion}
\label{sec:search}

The ability to synthesize the performance cost of arbitrary designs allows for the development of algorithms 
that search the possible design space. We expect there will be numerous opportunities in this space for techniques that can use this ability to: 
1) improve the productivity of engineers by quickly iterating over designs and scenarios before committing to an implementation (or hardware), 2) 
accelerate research by allowing researchers to easily and quickly test completely new ideas, 3) develop educational tools that allow for rapid testing of concepts, and 4) develop algorithms for offline auto-tuning and online adaptive systems that transition between designs. In this section, we provide two such opportunities for what-if design and auto-completion of partial designs.

\noindent{\textbf{What-if Design.}}
\chm{One can form design questions by varying any one of the input parameters of the Data Calculator:
1) data structure (layout) specification, 2) hardware profile, and 3) workload (data and queries).}
For example, assume one already uses a B-tree-like design for a given workload and hardware scenario. 
The Data Calculator can answer design questions such as 
``What would be the performance impact if I change my B-tree design by adding a bloom filter in each leaf?''
The user simply needs to give as input the high-level specification of the existing design and cost it twice: once 
with the original design and once with the bloom filter variation. In both cases, costing should be done
with the original data, queries, and hardware profile so the results are comparable. 
In other words, users can quickly test variations of data structure designs simply by altering a high level 
specification, without having to implement, debug, and test a new design.
\chm{Similarly, by altering the hardware or workload inputs, a given specification can be tested quickly on alternative environments without having to actually deploy code to this new environment.
For example, in order to test the impact of new hardware the Calculator only needs to train its Level 2 primitives
on this hardware, a process that takes a few minutes. 
Then, one can test the impact this new hardware would have on arbitrary designs by running what-if questions without having implementations of those designs and without accessing the new hardware. 
}

\noindent{\textbf{Auto-completion.}}
The Data Calculator can also complete partial layout specifications 
given a workload, and a hardware profile.
The process is shown in Algorithm \ref{algorithm:search} in the appendix:
The input is a partial layout specification, data, queries, hardware, and the set of the design 
space that should be considered as part of the solution, i.e., a list of candidate elements.
Starting from the last known point of the partial specification, the Data Calculator computes the rest of the missing subtree of the 
hierarchy of elements. 
At each step the algorithm considers a new element as candidate for one of the nodes of the missing subtree
and computes the cost for the different kinds of dictionary operations present in the workload. 
This design is kept 
only if it is better than all previous ones, otherwise it is dropped before the next iteration. 
The algorithm uses a cache to remember specifications and their costs to avoid recomputation.  
This process can also be used to tell if an existing design can be improved by marking
a portion of its specification as ``to be tested''. 
Solving the search problem completely is an open challenge as the design space drawn by the Calculator is massive. 
Here we show a first step which allows dynamic programming algorithms to select from a restricted set of elements but numerous algorithmic options are candidates for future research, e.g., genetic algorithms \cite{IdreosMK17}.


\section{Experimental Analysis}
\label{sec:eval}
We now demonstrate the ability of the Data Calculator to
help with rich design questions by accurately synthesizing performance costs.

\chm{
\noindent{\textbf{Implementation.}}
The core implementation of the Data Calculator is in C++. This includes the expert systems that handle layout primitives and cost synthesis. A separate module implemented in Python is responsible for analyzing benchmark results of Level 2 access primitives and generating the learned models. The benchmarks of Level 2 access primitives are also implemented in C++ such that the learned models can capture performance and hardware characteristics that would affect a full C++ implementation of a data structure. 
The learning process for each Level 2 access primitive is done each time we need to include a new hardware profile; then, the learned coefficients for each model are passed to the C++ back-end to be used for cost synthesis during design questions. 
For learning we use a standard loss function, i.e., least square errors, and the actual process is done via standard optimization libraries, e.g., SciPy's curve fit. For models which have non-convex loss functions such as the {\em sum of sigmoids} model, we algorithmically set up good initial parameters.} 

\noindent{\textbf{Accurate Cost Synthesis.}}
In our first experiment we test the ability to accurately cost arbitrary data structure specifications
across different machines.
To do this we compare the cost generated automatically by the Data Calculator with the cost
observed when testing a full implementation of a data structure.
We set-up the experiment as follows.
To test with the Data Calculator, we manually wrote data structure specifications for eight well known access methods
1) Array, 2) Sorted Array, 3) Linked-list, 4) Partitioned Linked-list,
5) Skip-list, 6) Trie, 7) Hash-table, and 8) B+tree.
The Data Calculator was then responsible for generating the design of operations for each data structure and computing their latency given a workload.
To verify the results against an actual implementation, we implemented all data structures above.
We also implemented algorithms for each of their
basic operations: {\tt Get}, {\tt Range Get},  {\tt Bulk Load and Update}.
The first experiment then starts with a data workload of $10^5$ uniformly distributed integers
and a sequence of $10^2$ {\tt Get} requests, also uniformly distributed.
We incrementally insert more data up to a total of $10^7$ entries and at each step we repeat the query workload.

The top row of Figure~\ref{fig:fitting} depicts results using a machine with 64 cores and 264 GB of RAM.
It shows the average latency per query as data grows as computed by the Data Calculator and as observed
when running the actual implementation on this machine.
For ease of presentation results are ranked horizontally from slower to faster (left to right).
The Data Calculator gives an accurate estimation of the cost
across the whole range of data sizes and regardless of the complexity of the designs both in terms of the data structure. It can accurately compute the latency of both simple traversals in a plain array
and the latency of more complex access patterns such as descending a tree and performing random hops in memory.

\noindent{\textbf{Diverse Machines and Operations.}}
The rest of the rows in Figure~\ref{fig:fitting} repeat the same experiment as above
using different hardware in terms of both CPU and memory properties (Rows 2 and 3) and different operations (Rows 4 and 5).
The details of the hardware are shown on the right side of each row in Figure~\ref{fig:fitting}.
Regardless of the machine or operation, the Data Calculator can accurately cost any design.
By training its Level 2 primitives on individual machines and maintaining a profile for each one of them,
it can quickly test arbitrary designs over arbitrary hardware and operations.
Updates here are simple updates that change the value of a key-value pair and so they are effectively the same as a point query with an additional write access. More complex updates that involve restructures are left for future work
both in terms of the design space and cost synthesis. 
Finally, Figure~\ref{fig:bulk}a) depicts that the Data Calculator can accurately synthesize the bulk loading costs for all data structures.

\noindent{\textbf{Training Access Primitives.}}
Figure~\ref{fig:bulk}b) depicts the time needed to train all Level 2 primitives
on a diverse set of machines.
Overall, this is an inexpensive process. It takes merely a few minutes to train
multiple different combinations of data and hardware profiles.

\begin{figure}
    \includegraphics[width=0.65\columnwidth]{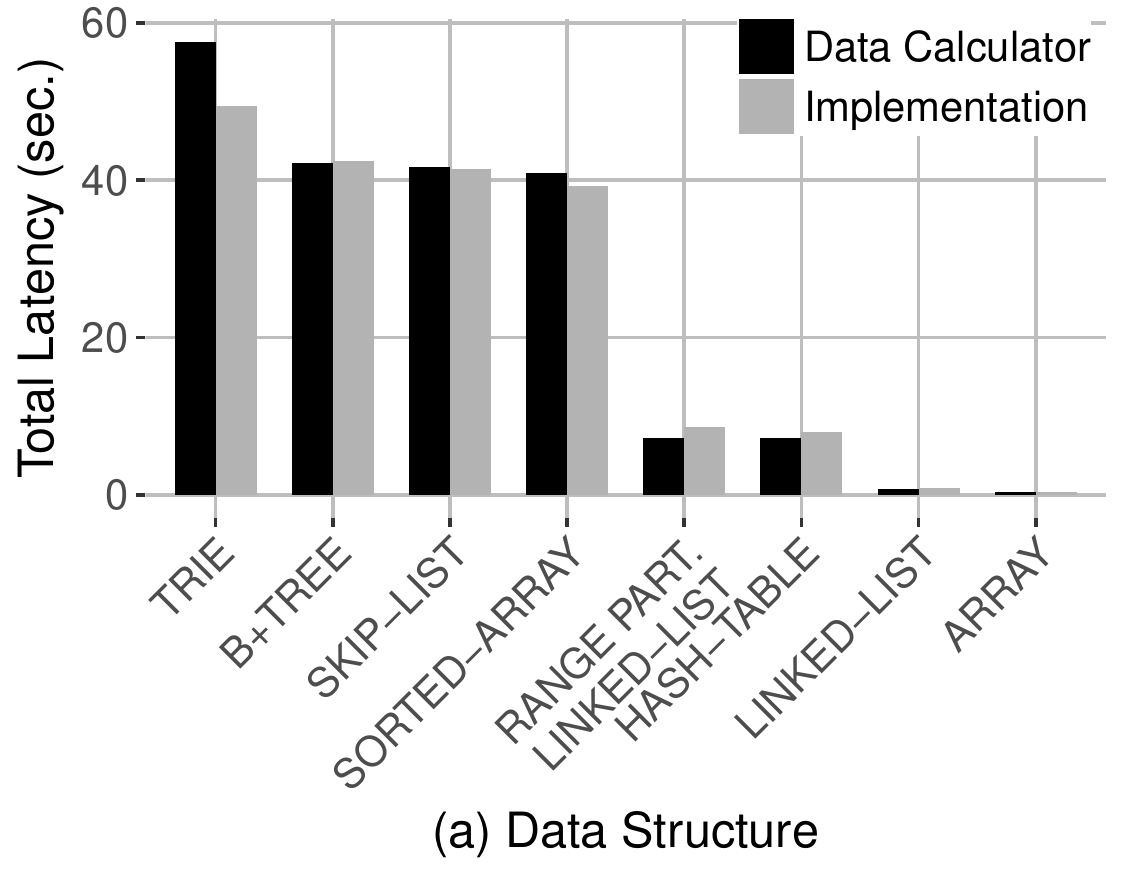}
        \includegraphics[width=0.33\columnwidth]{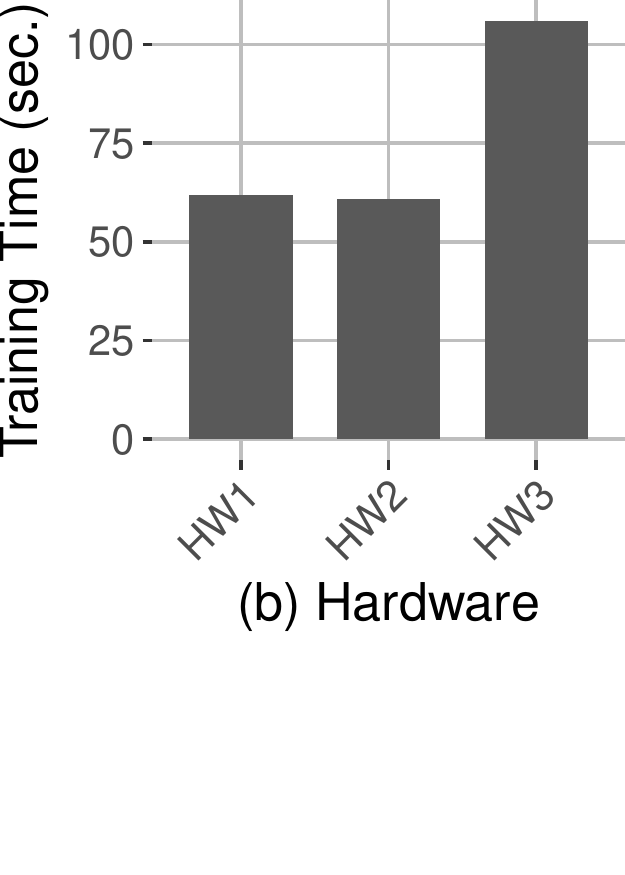}
        \vspace{-0.4cm}
   \caption{Computing Bulk-loading cost (left) and Training cost across diverse hardware (right).}
           \vspace{-0.4cm}
    \label{fig:bulk}
\end{figure}
\begin{figure}
  \includegraphics[height=8.5em]{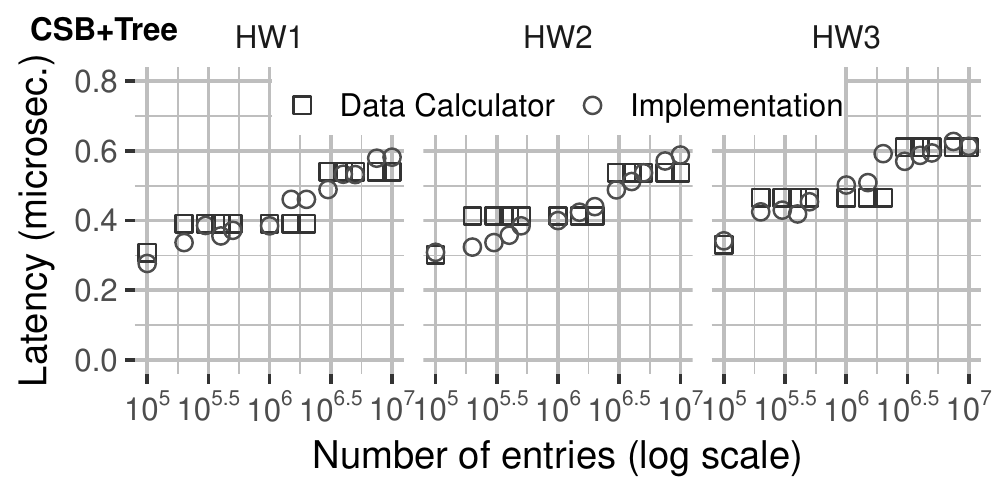}
  \includegraphics[height=8.5em]{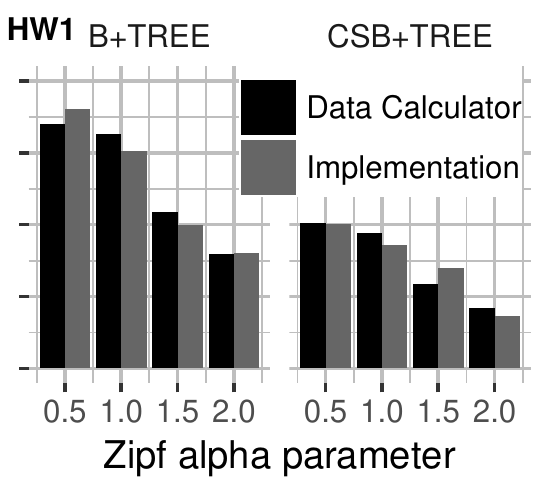}
          \vspace{-0.8cm}
   \caption{\ch{Accurately computing the latency of cache conscious designs in diverse hardware and workloads.}}

    \label{fig:csbfit}
\end{figure}

\ch{
\noindent{\textbf{Cache Conscious Designs and Skew.}}
In addition, Figure \ref{fig:csbfit} repeats our base fitting experiment
using a cache-conscious design, Cache Conscious B+tree (CSB). 
Figure \ref{fig:csbfit}a) depicts that the Data Calculator accurately predicts
the performance behavior across a diverse set of machines, capturing caching effects
of growing data sizes and design patterns where the relative position of nodes affects
tree traversal costs. 
We use the ``Full'' design from Cache Conscious B+tree \cite{Rao2000}.
Furthermore, Figure \ref{fig:csbfit}b) tests the fitting when the workload exhibits skew. 
For this experiment Get queries were generated with a 
Zipfian distribution: $\alpha = \{0.5, 1.0, 1.5, 2.0\}$. 
Figure \ref{fig:csbfit}b) shows that for the implementation results, workload skew improves performance and in fact it improves more for the standard B+tree.
This is because the same paths are more likely to be taken by queries resulting in these nodes being cached more often.
Cache Conscious B+tree improves but at a much slower rate as it is already optimized for the
cache hierarchy. The Data Calculator is able to synthesize these costs accurately, capturing skew and the related caching effects. 
}

\noindent{\textbf{Rich Design Questions.}}
In our next, experiment we provide insights about the kinds of design questions possible and how long 
they may take, working over a B-tree design and a
workload of uniform data and queries: 1 million inserts and 100 point Gets.
The hardware profile used is HW1 (defined in Figure \ref{fig:fitting}).
The user asks "What if we change our hardware to HW3?".
It takes the Data Calculator only 20 seconds (all runs are done on HW3) to compute that the performance would drop.
The user then asks: "Is there a better design for this new hardware and workload
if we restrict search on a specific set of five possible elements?" (from the pool of element on right side of Figure \ref{figure:nodePrimitives}).
It takes only 47 seconds for the Data Calculator to compute the best choice.
The user then asks ``Would it be beneficial to add a bloom filter in all B-tree leaves?''
The Data Calculator computes in merely 20 seconds that such a design change would be beneficial for the current workload and hardware.
The next design question is:
"What if the query workload changes to have skew targeting just 0.01\% of the key space?"
The Data Calculator computes in 24 seconds that
this new workload would hurt the original design and it computes
a better design in another 47 seconds.

\begin{figure}[t]
    \includegraphics[width=\columnwidth]{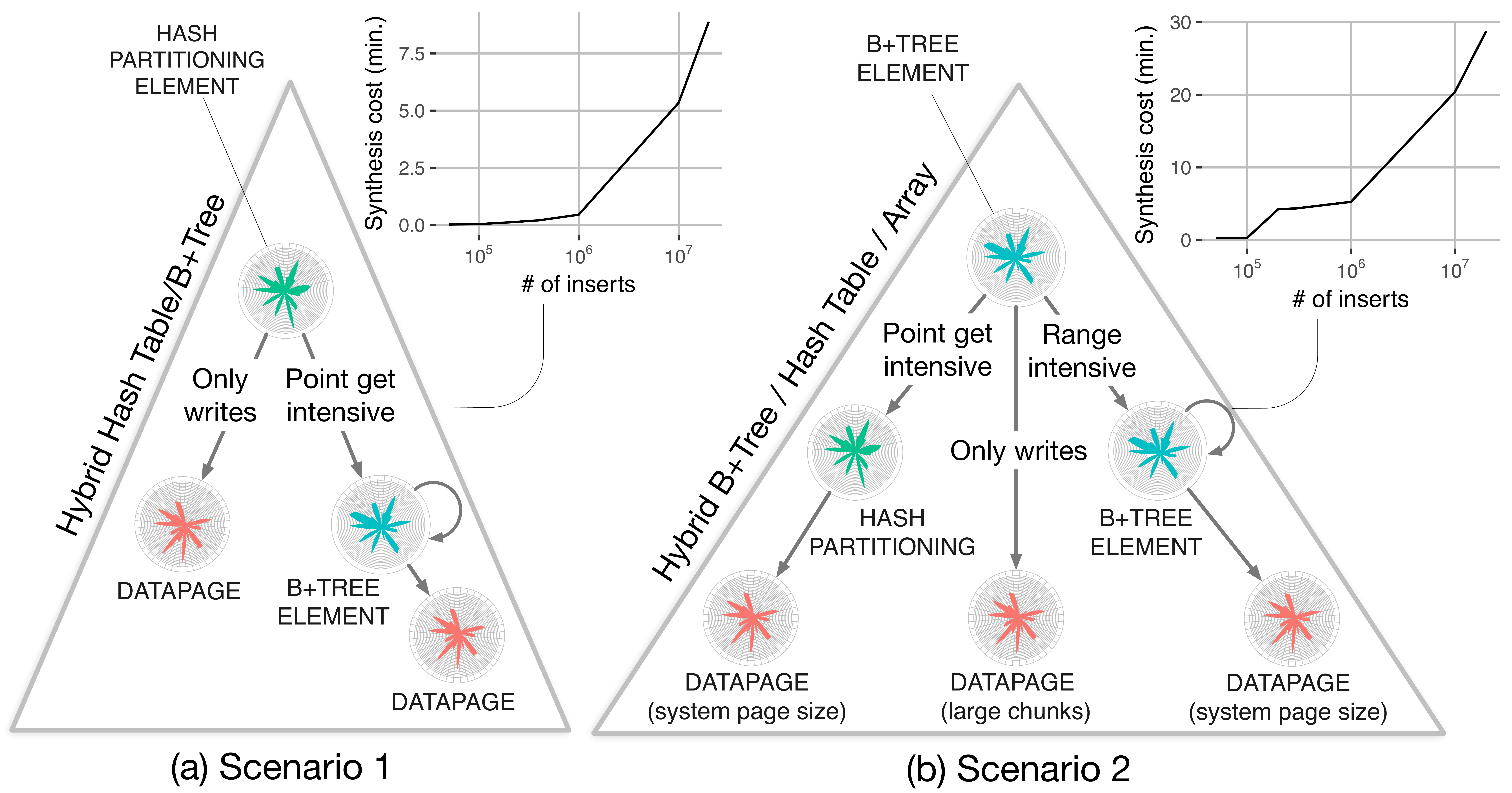}
         \vspace{-5ex}
    \caption{\ch{The Data Calculator designs new hybrids of known data structures to match a given workload.}}
         \vspace{-1ex}

    \label{fig:creativity}
\end{figure}

In two of the design phases the user asked ``give me a better design if possible''.
We now provide more intuition for this kind of design questions regarding the cost and scalability of computing such designs
as well as the kinds of designs the Data Calculator may produce to fit a workload.
We test two scenarios for a workload of mixed reads and writes (uniformly distributed inserts and point reads)
and hardware profile HW3.
In the first scenario, all reads are point queries in 20\% of the domain.
In the second scenario, 50\% of the reads are point reads and touch 10\% of the domain, while the other half are range queries
and touch a different (non intersecting with the point reads) 10\% of the domain.
We do not provide the Data Calculator with an initial specification.
Given the composition of the workload our intuition is that a mix of hashing, B-tree like indexing (e.g., with quantile nodes and sorted pages), and a simple log (unsorted pages) might lead to a good design, and so we instruct the Data Calculator to use those four elements to construct a design (this is done using Algorithm \ref{algorithm:search} but starting with an empty specification.
Figure \ref{fig:creativity} depicts the specifications of the resulting data structures 
For the first scenario (left side of Figure \ref{fig:creativity}) the Data Calculator
computed a design where a hashing element at the upper levels of the hierarchy allows to quickly access data
but then data is split between the write and read intensive parts of the domain to simple unsorted pages (like a log)
and B+tree -style indexing for the read intensive part.
For the second scenario (right side of Figure \ref{fig:creativity}), the Data Calculator produces a design which
similarly to the previous one takes care of read and writes separately but this time also distinguishes between range and point gets by allowing
the part of the domain that receives point queries to be accessed with hashing and the rest via B+tree style indexing.
The time needed for each design question was in the order of a few seconds up to 30 minutes depending on the size of the sample workload (the synthesis costs are embedded in Figure \ref{fig:creativity} for both scenarios). Thus, the Data Calculator quickly answers complex questions that would normally take humans days or even weeks to test fully.

\section{Related Work}
\label{sec:related}
To the best of our knowledge this is the first work to discuss the problem of interactive
data structure design and to compute the impact on performance. However, there
are numerous areas from where we draw inspiration and with which we share concepts. 

\noindent{\textbf{Interactive Design.}}
Conceptually, the work on Magic for layout on integrated circuits \cite{Ousterhout1984} comes closest to our work. 
Magic uses a set of design rules to quickly verify transistor designs so they can be simulated by designers.
In other words, a designer may propose a transistor design and Magic will determine if this is correct or not. Naturally, this is a huge step especially for hardware design where actual implementation is extremely costly. 
The Data Calculator pushes interactive design one step further to incorporate cost estimation as part of the design phase
by being able to estimate the cost of adding or removing individual design options 
which in turn also allows us to build design algorithms for automatic discovery of good and bad designs
instead of having to build and test the complete design manually.

\noindent{\textbf{Generalized Indexes.}}
One of the stronger connections is the work on Generalized Search Tree Indexes (GiST) \cite{Hellerstein1995, Aoki1998, Aoki1999, Kornacker1997, Kornacker1999,Kornacker1998,Kornacker2003}. 
GiST aims to make it easy to extend data structures and tailor them to specific problems and data with minimal effort. It is a template, an abstract index definition that allows designers and developers to implement a large class of indexes. The original proposal focused on record retrieval only but later work added support for concurrency \cite{Kornacker1997}, a more general API \cite{Aoki1998}, improved performance \cite{Kornacker1999}, selectivity estimation on generated indexes \cite{Aoki1999} and even visual tools that help with debugging  \cite{Kornacker1998,Kornacker2003}. While the Data Calculator and GiST share motivation, they are fundamentally different: GiST is a template to implement tailored indexes while the Data Calculator is an engine that computes the performance of a design enabling rich design questions that compute the impact of design choices
before we start coding, making these two lines of work complementary.

\noindent{\textbf{Modular/Extensible Systems and System Synthesizers.}}
A key part of the Data Calculator is its design library, breaking down a design space in components and then being able to use any set of those components as a solution. As such the Data Calculator shares concepts with the stream of work on modular systems, an idea that has been explored in many areas of computer science: 
in databases for easily adding data types \cite{Goldhirsch1987,Graefe1994,McPherson1987,Orborn1987,Stonebraker1987} with minimal implementation effort,
or plug and play features and whole system components with clean interfaces \cite{Levandoski2013,Levandoski2013a,Chaudhuri2000,Batory1988,Carey1987, Klonatos2014},
as well as in software engineering \cite{Parnas1979}, computer architecture \cite{Ousterhout1984},
and networks \cite{Kohler2000}. 
Since for every area the output and the components are different, there are particular challenges that have to do with defining the proper components, interfaces and algorithms. 
The concept of modularity is similar in the context of the Data Calculator. The goal and application of the concept is different though.

\noindent{\textbf{Additional Topics.}}
Appendix B discusses additional related topics such as auto-tuning systems 
and data representation synthesis in programming languages.

\section{Summary and Next Steps}
\label{sec:summary}
Through a new paradigm of first principles of data layouts and learned cost models, the Data Calculator allows
researchers and engineers to interactively and semi-automatically 
navigate complex design decisions when designing or re-designing data structures, considering new workloads, and hardware.
The design space we presented here includes basic layout primitives 
and primitives that enable cache conscious designs by dictating the relative 
positioning of nodes, focusing on read only queries.
The quest for the first principles of data structures needs to continue to find the primitives for additional significant classes of designs including updates, compression, concurrency, adaptivity, graphs, spatial data, version control management, and replication. 
Such steps will also require innovations for cost synthesis. For every design class added (or even for every single primitive added), the knowledge gained in terms of the possible data structures designs grows exponentially. 
Additional opportunities include full DSLs for data structures, compilers for code generation and eventually certified code \cite{Qiu2017,Wang2017}, new classes of adaptive systems that can change their core design on-the-fly, and machine learning algorithms that can search the whole design space.

\section{Acknowledgments}
We thank the reviewers for valuable feedback and direction. 
Mark Callaghan provided the quotes on the importance of data structure design. 
Harvard DASlab members Yiyou Sun, Mali Akmanalp and Mo Sun helped with parts of the implementation and the graphics.
This work is partially funded by the USA National Science Foundation project IIS-1452595.



\appendix

\section{Appendix Introduction}
This appendix provides more detailed specifications for internal components and the design space of the Data Calculator. Section \ref{sec:morerelated} discusses additional related work. Sections~\ref{sec:dl} and ~\ref{sec:da} provide detailed specifications of the layout and access primitives respectively.
In Section~\ref{sec:cs}, we describe the complete expert system which is used for cost synthesis for the Get operation.
In Section~\ref{sec:inex}, we provide the complete specification of all data structures used 
as input to the Data Calculator for the experiments and the corresponding output cost (Section~\ref{sec:outex}).
Finally, we provide examples of the exact output which includes a dot and a JSON file that describe the design and an instance of a resulting data structure.

\begin{figure}[t]
  \centering
 \includegraphics[width=\columnwidth]{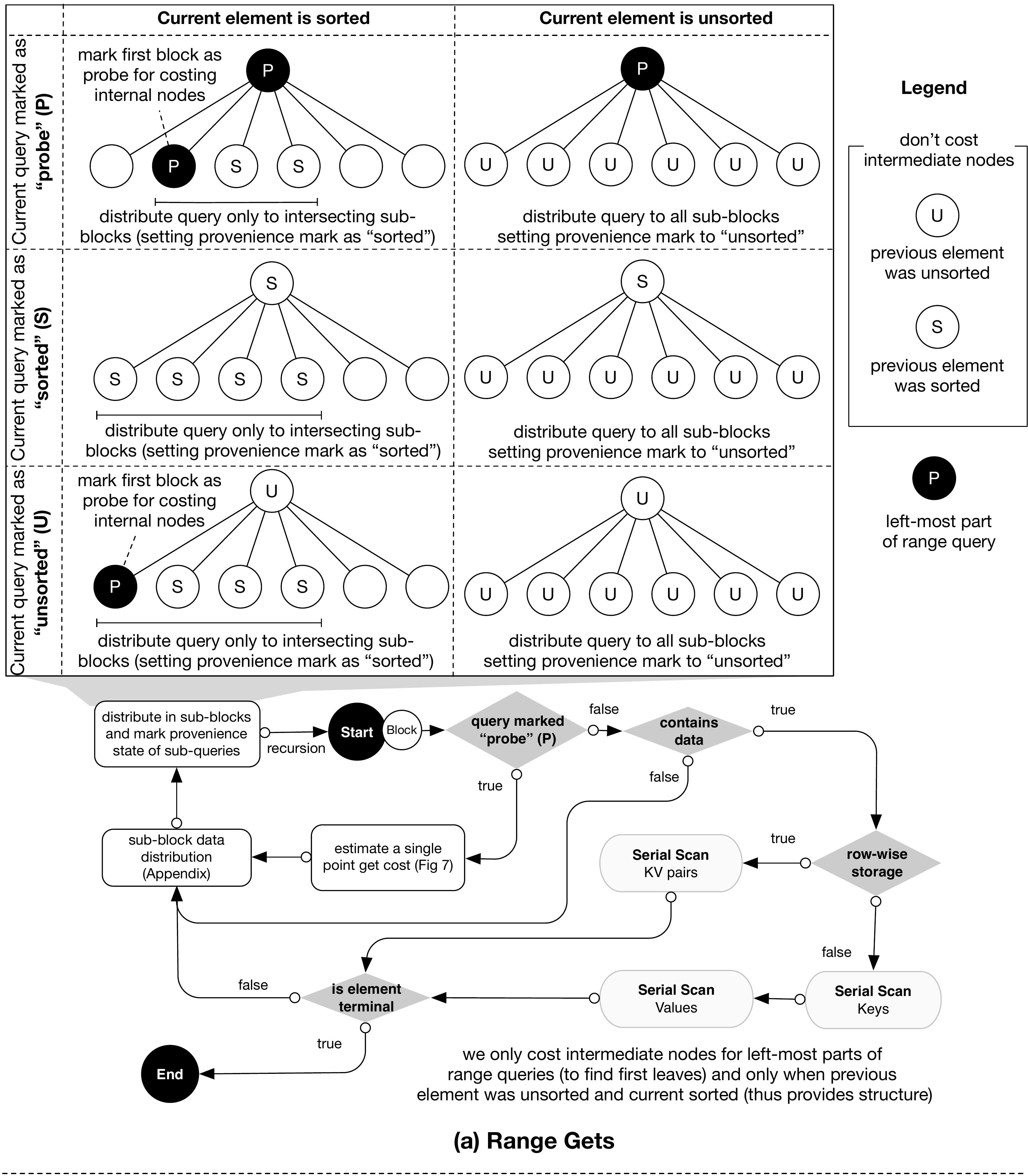}
 \includegraphics[width=0.95\columnwidth]{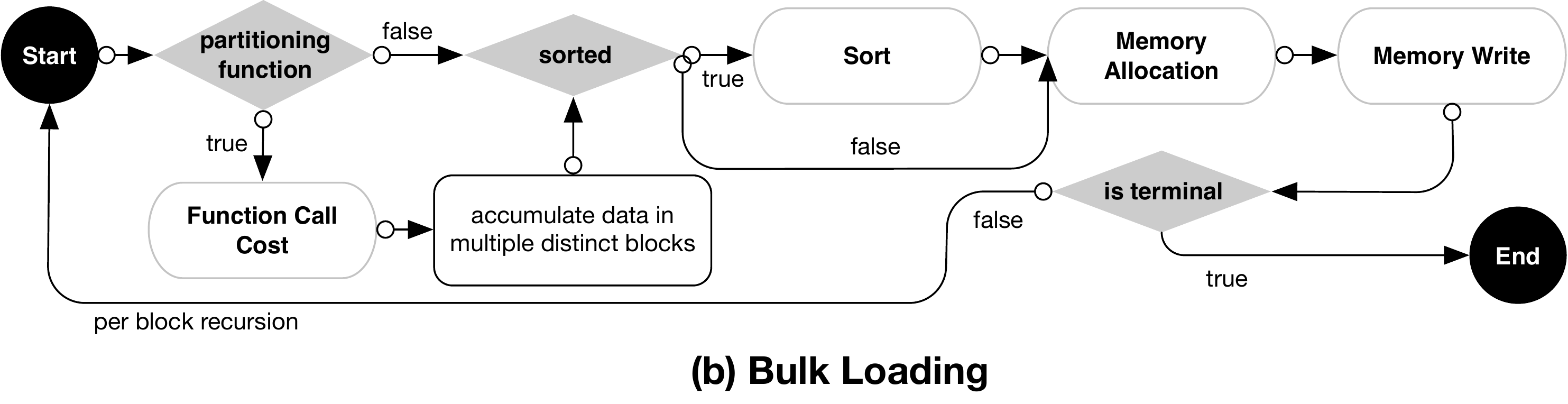}
 \vspace{-1em}
 \caption{Cost synthesis Range Gets and Bulk Loading.}
 \label{fig:OpRangeBulk}
\end{figure}

\begin{algorithm}[t]
\footnotesize
 \Fn{CompleteDesign ($Q, \mathcal{E}, l, currentPath=[], H$)}{
      \If{blockReachedMinimumSize$(H.page\_size)$}{
        return END\_SEARCH\;
      }
      \If{!meaningfulPath($currentPath, Q, l$)} {
        return END\_SEARCH\;
      }
      \If{(cacheHit = cachedSolution($Q, l, H$)) != null}{
        return cacheHit\;
      }

      bestSolution = initializeSolution(cost=$\infty$)\;
      \For{$E \in \mathcal{E}$}{

        tmpSolution = initializeSolution()\;

          tmpSolution.cost = synthesizeGroupCost($E, Q$)\;
          updateCost($E, Q$, tmpSolution.cost);

        \If{createsSubBlocks(E)} {
              $Q'= createQueryBlocks(Q)$\;
              currentPath.append($E$)\;
              subSolution = CompleteDesign($Q', \mathcal{E}, l+1, currentPath$)\;
              \If{subSolution.cost != END\_SEARCH} {
                tmpSolution.append(subSolution)\;
              }
            }

        \If{tmpSolution.cost $\leq$ bestSolution.cost} {
            bestSolution = tmpSolution \;
        }
     }
     cacheSolution($Q, l, bestSolution$)\;
     return bestSolution\;
}
\caption{Complete a partial data structure layout specification.}
\label{algorithm:search}
\end{algorithm}

\section{Additional Related Areas}
\label{sec:morerelated}
\noindent{\textbf{Auto-tuning and Adaptive Systems.}}
Work on tuning \cite{Ioannidis1987, Chaudhuri1997} and adaptive systems is also relevant as conceptually any adaptive technique tunes along a part of the design space. For example, work on hybrid data layouts and adaptive indexing automates selection of the right layout \cite{Arulraj2016,Alagiannis2014, Hankins2003,Idreos2007, Idreos2010a, Idreos2007a, Dittrich2011,Schuhknecht2013,Alvarez2014, Liu2016, Dayan2017, Dayan2018, Pirk2014, Graefe2012a,Petraki2015,Zoumpatianos2014,Idreos2009, Idreos2011, Idreos2011a, Karras2016, Graefe2014, Graefe2010b, Kennedy2015,Sleator1985, Borovica-Gajic2015}. Typically, in these lines of work the layout adapts to incoming requests. Similarly works on tuning via experiments \cite{Babu2009}, learning \cite{Anderson2013}, and tuning via machine learning \cite{Aken2017,Heimel2015} can adapt parts of a design using feedback from tests.
While there are shared concepts with these lines of work, they are all restricted to much smaller design spaces, typically to solve a very specific systems bottleneck, e.g., incrementally building a specific index or smoothly transitioning among specific layouts.
The Data Calculator, on the other hand, provides a generic framework to argue about the whole design space
of data layouts.
Its capability to quickly test the potential performance of a design can potentially lead to new adaptive techniques that will also leverage experience in existing adaptive systems literature to adapt across the massive space drawn by the Data Calculator.

\noindent{\textbf{Data Representation Synthesis.}}
Data representation synthesis aims for programming languages that automate data structure selection. SETL \cite{Schonberg1979,Schonberg1981} was the first language to generate structures in the 70s
as combinations of existing data structures: array, and linked hash table.
A series of works kept providing further functionality, and expanding on the components used \cite{Smaragdakis1997, Cohen1993, Shacham2009, Hawkins2011,Hawkins2012, Loncaric2016}.
Cozy \cite{Loncaric2016} is the latest system; it supports complex
 retrieval operations such as disjunctions, negations, and inequalities
 and by uses a library of five data structures: array (sorted and plain), linked list, binary tree, and hash map.
These works compose data structure designs out of a small set of existing data structures. This is parallel to the tuning and access path selection problem in databases. The Data Calculator introduces a new vision for what-if design and focuses on two new dimensions: 1) design out of fine-grained primitives, and 2) calculation of the performance cost given a hardware profile and a workload. The focus on fine-grained primitives enables exploration of a massive design space. For example, using the equations of Section 2 for homomorphic two-node designs, a fixed design space of 5 possible elements can generate 25 designs, while the Data Calculator can generate $10^{32}$ designs. The gap grows for polymorphic designs, i.e, $2*10^9$ for a 5 element library, while the Data Calculator can generate up to $1.6*10^{55}$ valid designs (for a 10M dataset and 4K pages). In addition, the focus on cost synthesis through learned models of fine-grained access primitives means that we can capture hardware and data properties for arbitrary designs.
Array Mapped Tries \cite{Steindorfer2016}
use fine-grained primitives, but the focus is only on trie-based collections and without cost synthesis.


\section{Data Layout Primitives}
\label{sec:dl}

In this section we list and describe in detail all 21 primitives in the design space currently supported by the Data Calculator. A summary of the primitives is also shown in Figure \ref{table:layoutprimitivestable}.

\begin{enumerate}

    \item \textbf{Key retention}~\\
    \textbf{Domain:} yes | no | func~\\
    Elements can hold data within them apart from partitioning them in sub-blocks.
    This data ``retention'' can either be full (by setting the value to yes), like in simple arrays, where the keys are stored, or partial (defined by a function), like tries where only a part of the key is stored. Alternatively, when ``no'' is set, they merely act as partitioners without retaining any data, like in hash-maps.
    \textbf{Examples:}~\\
    - Data pages and arrays fully store keys ~\\
    - A hash map contains no data ~\\
    - Tries retain only a part of the key at every level.

    \item \textbf{Value retention}~\\
    \textbf{Domain:} yes | no | func~\\
    Elements can hold data within them apart from partitioning them in sub-blocks.
    This data ``retention'' can either be full (by setting the value to yes), like in simple arrays, where the values are stored, or partial (defined by a function).
    Alternatively, when ``no'' is set, they merely act as partitioners without retaining any data, like in hash-maps.

    \textbf{Examples:}~\\
    - Data pages and arrays fully store values ~\\
    - A hash map contains no data

    \item \textbf{Key-value layout}~\\
    \textbf{Domain:} columnar | row-wise | col-row-groups~\\
    Keys and values can either be stored in a columnar, a row-wise fashion, or a row groups fashion.
    In columnar storage all keys are stored contiguously and all values the same.
    In row-wise storage they are stored as key value pairs.
    In col-row-groups storage data are grouped in chunks by row, but within each chunk they are stored row-wise.

    \begin{center}\includegraphics[width=\linewidth]{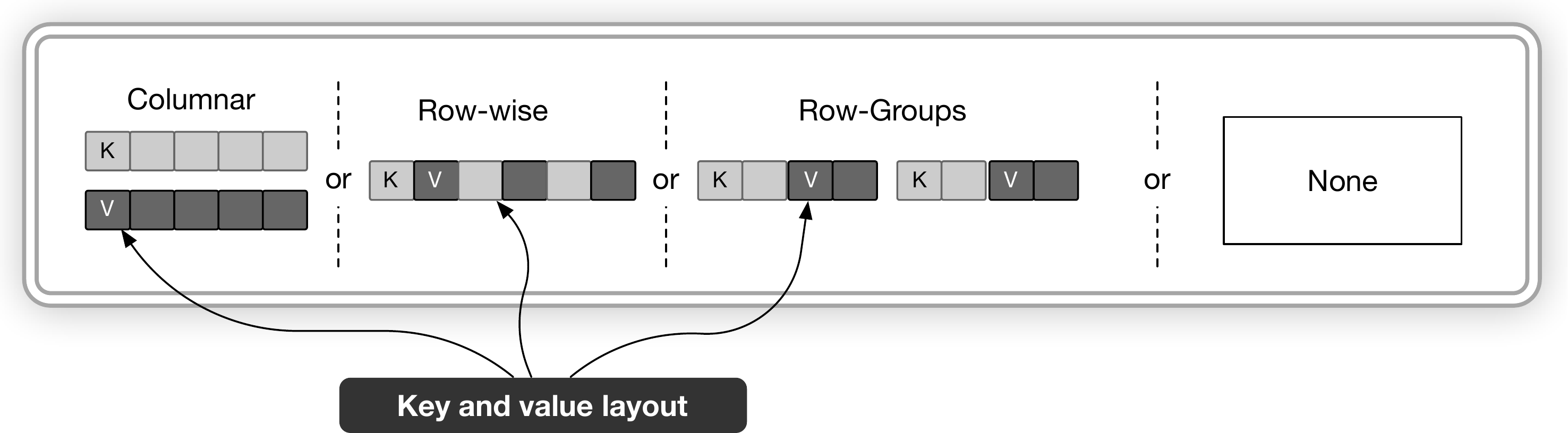}\end{center}

    \item \textbf{Intra-node access}~\\
    \textbf{Domain:} direct | head-link | tail-link | func ~\\
    \textbf{Description:} Determines how we can access sub-blocks. Direct access means that we can directly access each distinct sub-block. Head-link means that we can find the first sub-block through a link. Tail-link means that we can find the last sub-block through a link. Function means that we can get a link to one or more of the sub-blocks using a function.~\\
    \textbf{Examples:}
    - A linked list has head or tail links.
    - A B+Tree internal node allows us to directly address each sub-tree.

    \begin{center}\includegraphics[width=\linewidth]{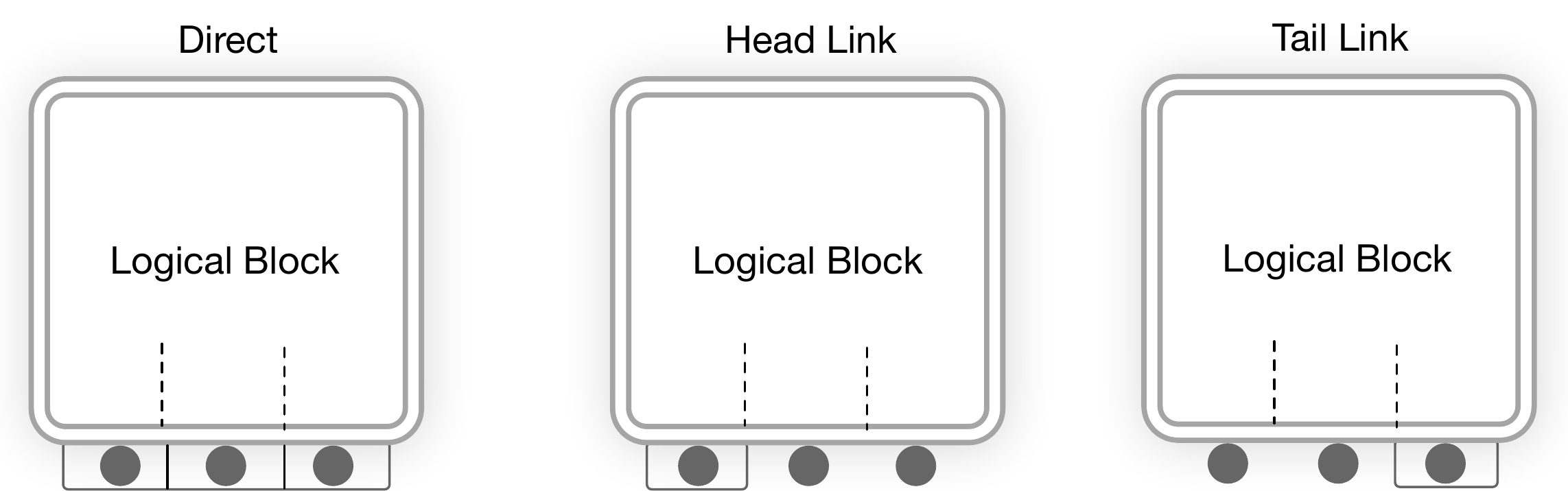}\end{center}

    \item \textbf{Utilization}~\\
    \textbf{Domain:} none | $>=$ | func~\\
    Utilization constraints, how much empty space is allowed.
    It can be specified as an arbitrary function or as a ``less or equal to capacity'' check.

    \textbf{Examples:}~\\
    - B+Trees can have a constraint of half-empty leaves.

    \begin{center}\includegraphics[width=0.66\linewidth]{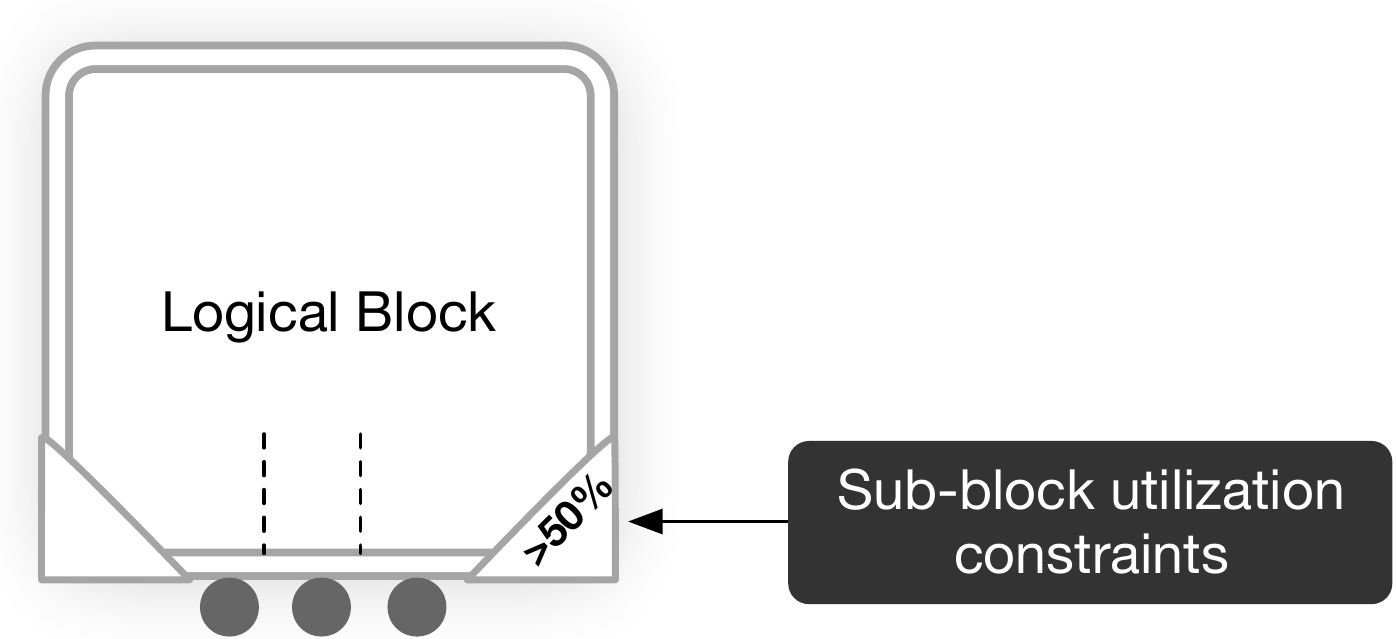}\end{center}

     \item \textbf{Bloom filters}~\\
    \textbf{Domain:} off | on(num-hashes, num-bits)~\\
    If this is set to on, bloom filters per sub-block are used to filter reads.
    The parameters specified as arguments to on are used for the bloom filters.
    \textbf{Examples:}~\\
    - LSM trees use bloom filters per sub-block (run or level) to skip large chunks of data.

    \item \textbf{Zone map filters}~\\
    \textbf{Domain:} min | max | both | exact | off~\\
    If this is set to min, the minimum key under a sub-block is kept for filtering reads (as a fence pointer).
    If this is set to max, the maximum key under a sub-block is kept for filtering reads (as a fence pointer).
    If this is set to both, both minimum and maximum keys are kept.
    If this is set to exact, the exact key of only each sub-block is kept.
    \textbf{Examples:}~\\
    - B+Trees trees use zone maps in internal nodes to guide search.

    \begin{center}\includegraphics[width=0.66\linewidth]{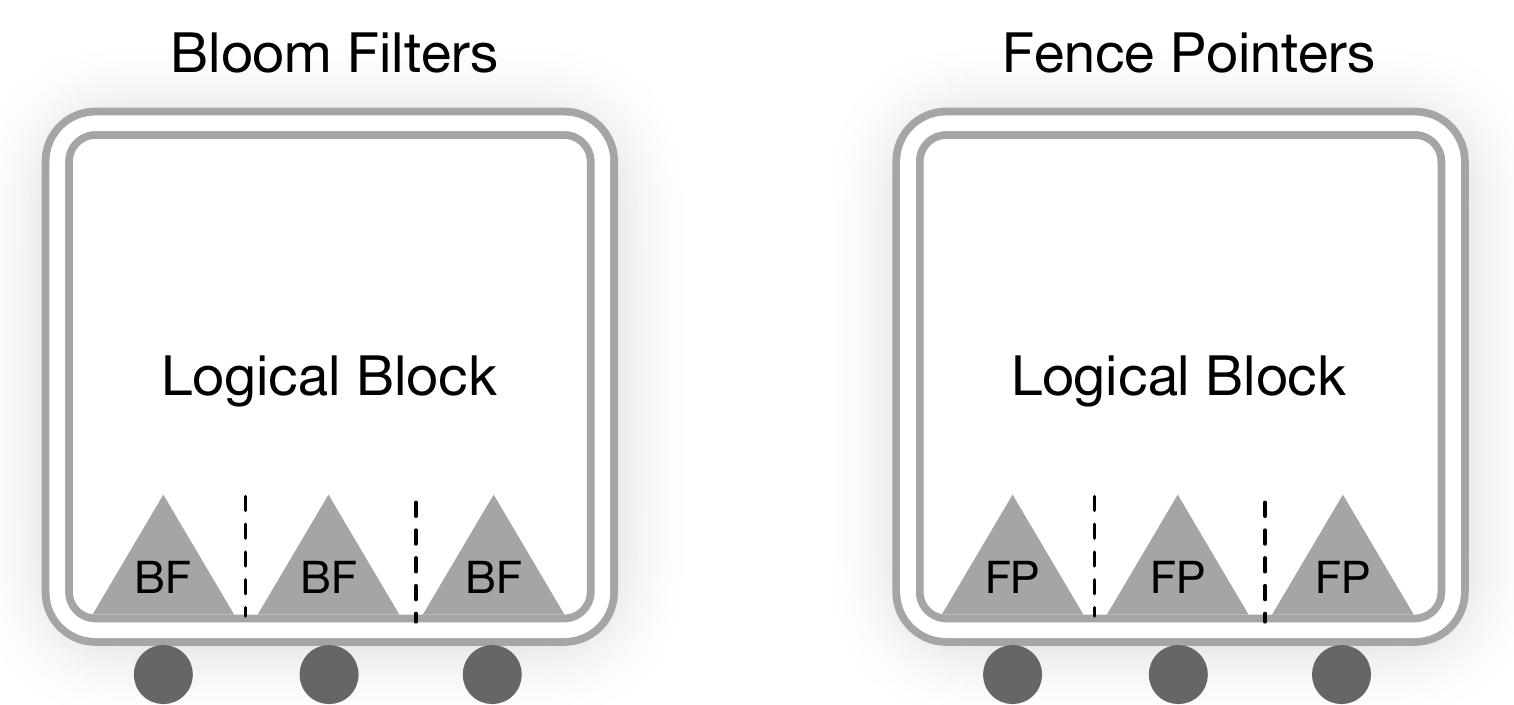}\end{center}

    \item \textbf{Filters memory layout}~\\
    \textbf{Domain:} consolidate | scatter~\\
    Filters like zone maps or bloom filters can either be stored contiguously for an element or scattered and stored per sub-block.
    For example all minimum values could be consolidated in a single array thus allowing binary search.
    Alternatively, filters can be scattered per sub-block and stored with it.
    \textbf{Examples:}~\\
    - B+Trees consolidate zone maps ~\\
    - Linked list of very large pages could include per-page headers that contain scattered zone maps.

    \begin{center}\includegraphics[width=0.66\linewidth]{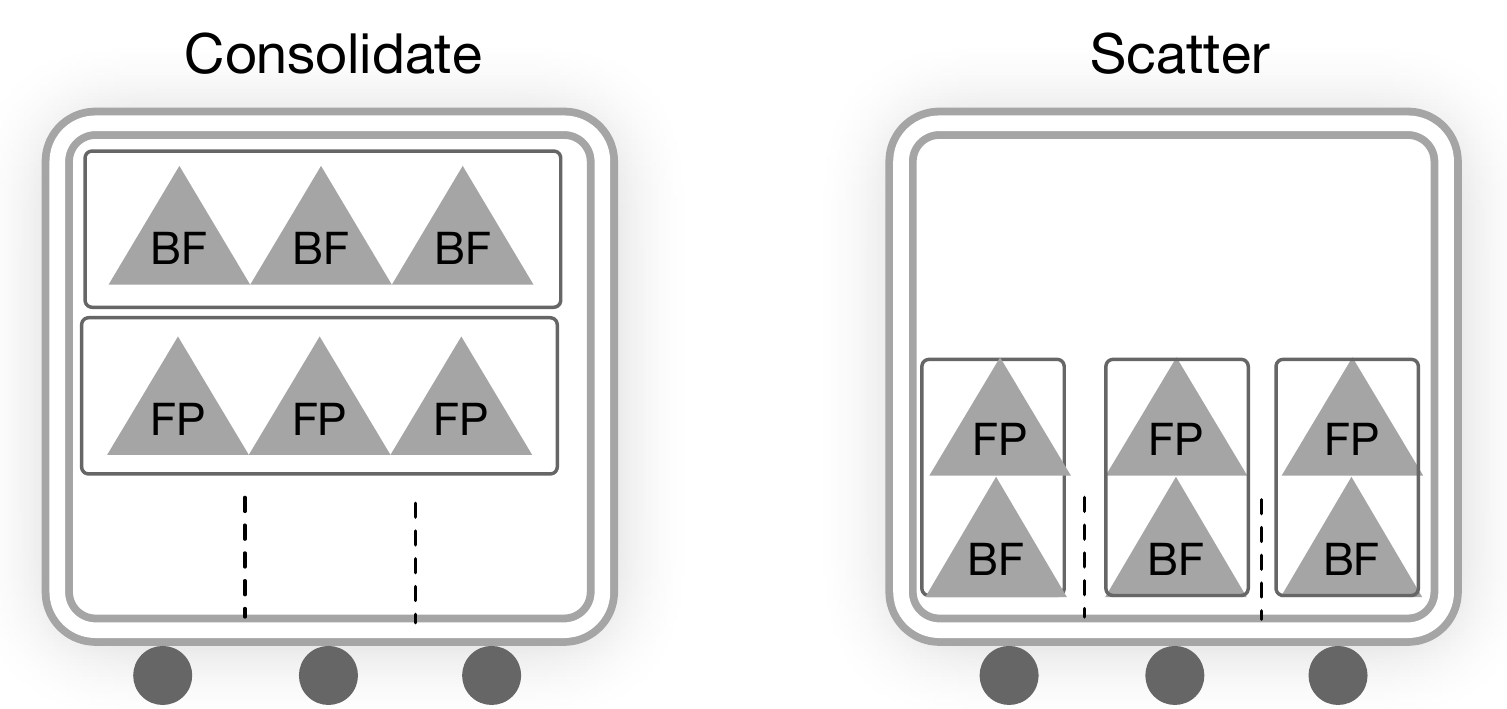}\end{center}

    \item \textbf{Fanout/Radix}~\\
    \textbf{Domain:} fixed(size) | func | unlimited | terminal(capacity) ~\\
    \textbf{Description:} This primitive allows us to specify the type and
    number of sub-blocks for this element. Unlimited means any number of
    sub-blocks is allowed. For fixed, the number of sub-blocks is static. If it
    is set to function, then the number of sub-blocks is determined by the
    result of a function, while terminal means that there are no sub-blocks and this element absorbs all data up to a certain capacity.
    ~\\
    \textbf{Examples:}~\\
    - Linked-lists have an infinite number of sub-blocks, as new sub-blocks can
      always be appended to the structure.~\\
    - Traditional B+Tree nodes have a fixed number of sub-blocks, as the fanout
      of each node is fixed a priori. ~\\
    - Partitioned arrays with fixed partition size can have a functional fanout
      determined by the maximum number of data to be inserted.

     \begin{center}\includegraphics[width=0.7\linewidth]{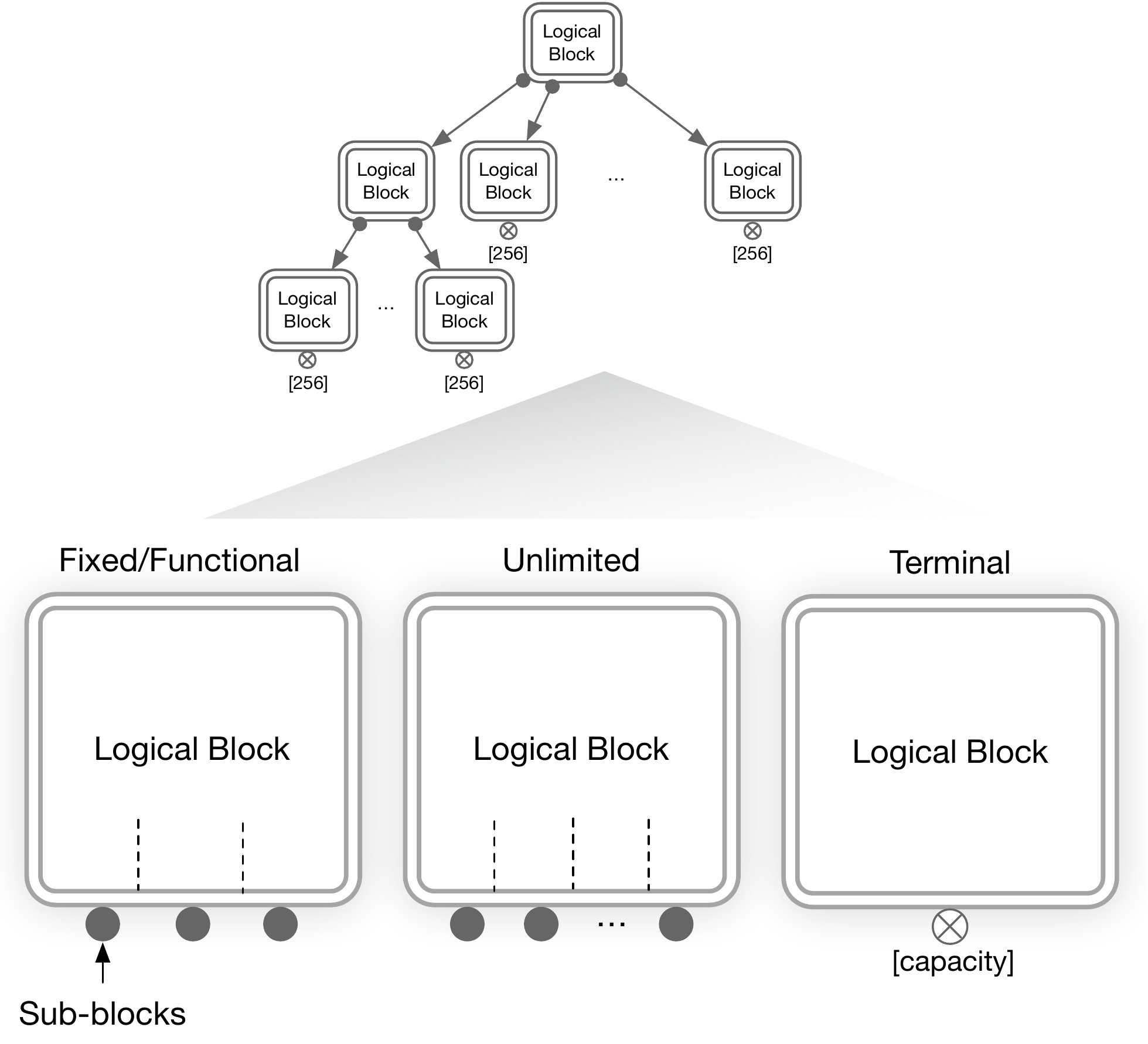}\end{center}

    \item \textbf{Key/fence partitioning}~\\
    \textbf{Domain:} append(fw|bw) | data-dependent(sorted | k-ary | func) | data-independent(range | radix | func) | temporal(size-ratio, policy) ~\\
    \textbf{Description:} This is used to define data are placed in order.
    This defines both the order of elements with sub-blocks and the order of data within a terminal node.
    Such predefined ways decide the location of a new insert or the order of accesses during reads.
    For example a data independent functional partitioning will insert data in the right sub-block after the key of the new tuple is passed through the partitioning function.
    Such arbitrary functions return a block id.
    Append means that inserts and reads will have to be placed in forward or backward order.
    Finally data dependent partitioning specifies that some kind of order has to be maintained. This is either in the form of sorting the data or some other kind of functional or k-ary order.
    K-ary specifies that the sub-blocks are ordered and that the fence keys are stored in a k-ary tree.
    Each internal node of a k-ary tree has k-1 fence keys and k-children; leaf nodes have k keys and k values.
    At each internal node, label its children $\ell = (1,1, \dots, k)$. Then for a node at position $m$, its k children are located at position $(k * m) + (k -1) * \ell$.
    This assumes that the root is at position 0.

    For the classic example of a binary tree, this gives that a nodes left child is at $2m + 1$ and its right child is at $2m + 2$. This is shown below.
    An in-order layout specifies that the sub-blocks are ordered and that the fence keys are laid out via an in-order traversal of a binary tree.
    In-order layout does not make sense for a tree with $k$ larger than 2, as a node has children which are between its smallest and largest fence keys.
    
    Finally, temporal partitioning partitions blocks by time of insertion in runs and levels in an LSM style way.
    During inserts the most recent run is used, during reads we process one run at a time and one level at a time (top to bottom) until a hit is found.
    ~\\
    \textbf{Examples:}~\\
    - A hash-map uses a data-independent partitioning function to identify the correct bucket~\\
    - A linked list has no predefined partitioning and data are appended in the end of the list in the order of arrival. ~\\
    - An LSM tree has temporal/log structured partitioning.
   
   \begin{center}\includegraphics[width=\linewidth]{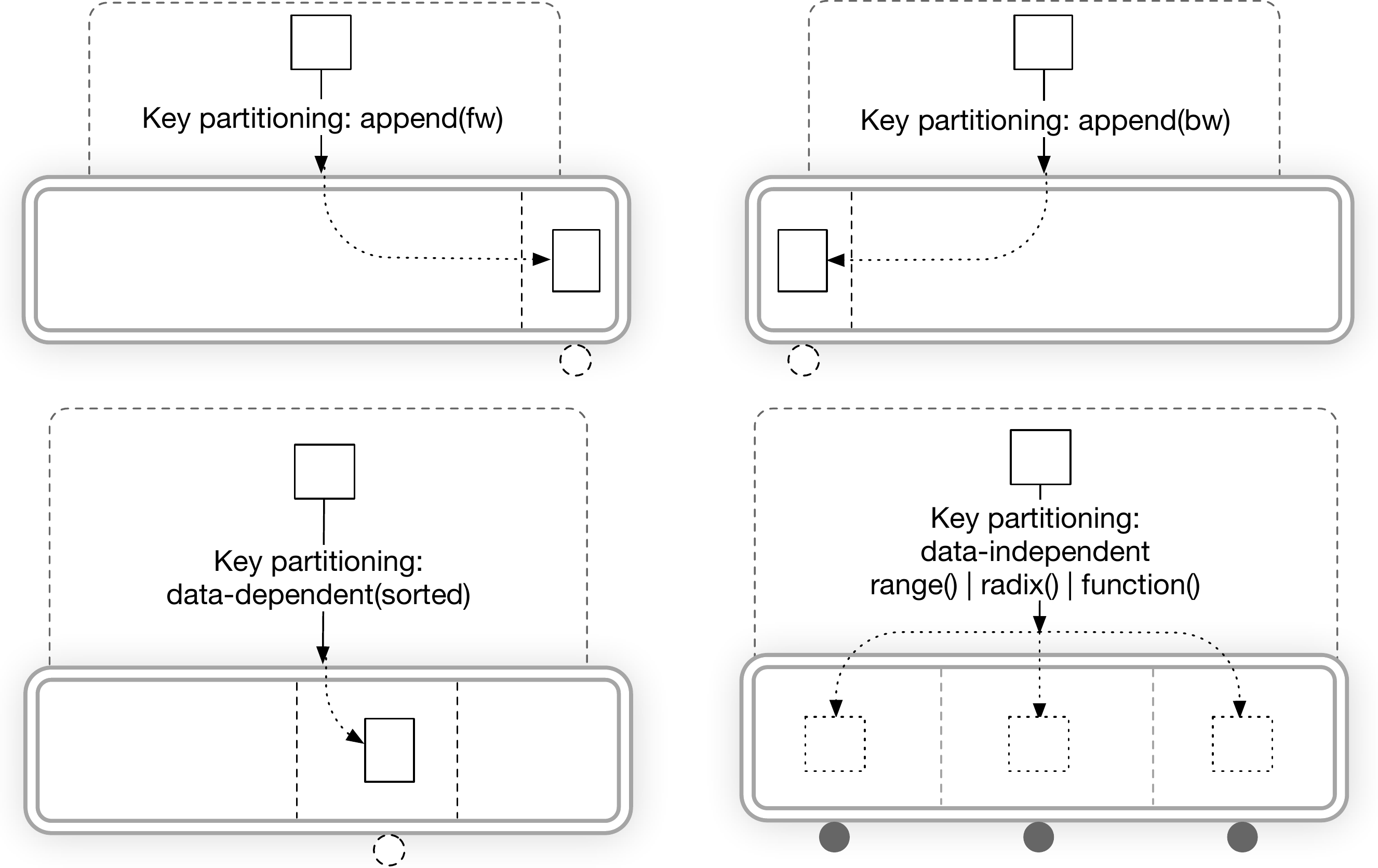}\end{center}

    \item \textbf{Sub-block capacity}~\\
    \textbf{Domain:}  fixed | balanced | unrestricted | func ~\\
    \textbf{Description:}
     This is used to define the capacity of each sub-block.
     This is the amount of tuples that can (at maximum) end up under each one of an element's sub-blocks.
     One can have functional capacity, fixed capacity, unrestricted capacity (where capacity can arbitrarily grow in each sub-block) and balanced capacity where the total number of records is equally divided across all sub-blocks.
        ~\\
    \textbf{Examples:}~\\
    - A hash-map uses unrestricted capacity as the amount of elements in each bucket is not known a-priori and buckets should arbitrarily grow~\\
    - A linked list has a fixed capacity, as each sub-block can hold 1 or more (in the case of linked lists of pages) elements.
    When one sub-block overflows a new sub-block should be created. ~\\
    - A B+Tree node has balanced capacity as under the capacity of each sub-tree of an internal is equal to its siblings.
 
    \begin{center}\includegraphics[width=0.5\linewidth]{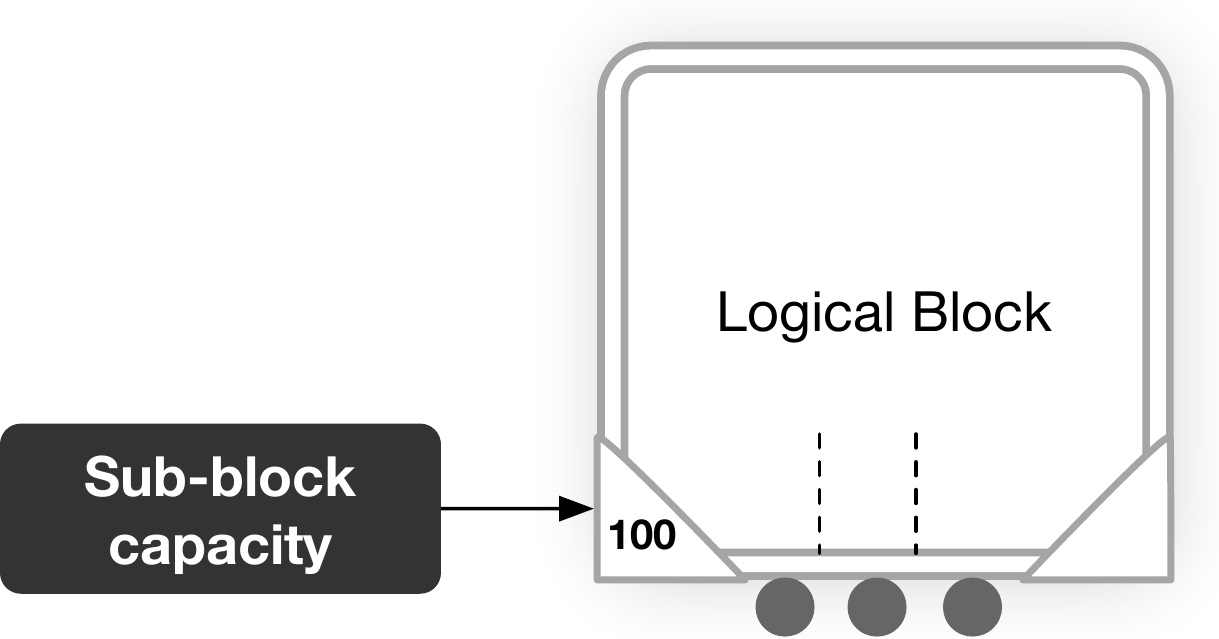}\end{center}

    \item \textbf{Immediate node links}~\\
    \textbf{Domain:} next | previous | both | none ~\\
    \textbf{Description:} This indicates that each sub-block includes a
    pointer to its immediate sibling in the ``forward direction'' (defined by the structure), in the ``backward direction'', or both. ~\\
    \textbf{Examples:}
    - A linked list contains sub-blocks that are linked to each other.

    \item \textbf{Skip node links}~\\
    \textbf{Domain:} perfect | randomized | func | none ~\\
    This is set to none when no skip links are used.
    Skip links can navigate from one sub-block to a sub-block further away from it.
    When set to perfect, all links that allow for binary-search style navigation are materialized.
    When set to randomized, only a probabilistic set of skip links is included, enough to provide some performance guarantees.
    When set to functional, an arbitrary function defines the links that are going to be materialized.
    \textbf{Examples:}~\\
    - Classic perfect skip-lists contain perfect skip-links.~\\
    - Randomized skip-lists contain randomized skip-lins.

    \item \textbf{Area links}~\\
    \textbf{Domain:} forward | backward | both | none
    Each sub-tree can be connected with another sub-tree at the leaf level through area links. This can happen in a forward or backward way or both.~\\
    \textbf{Examples:}~\\
    - The linked leaves of a B+Tree are connected to each other for efficient range queries.

    \begin{center}\includegraphics[width=0.8\linewidth]{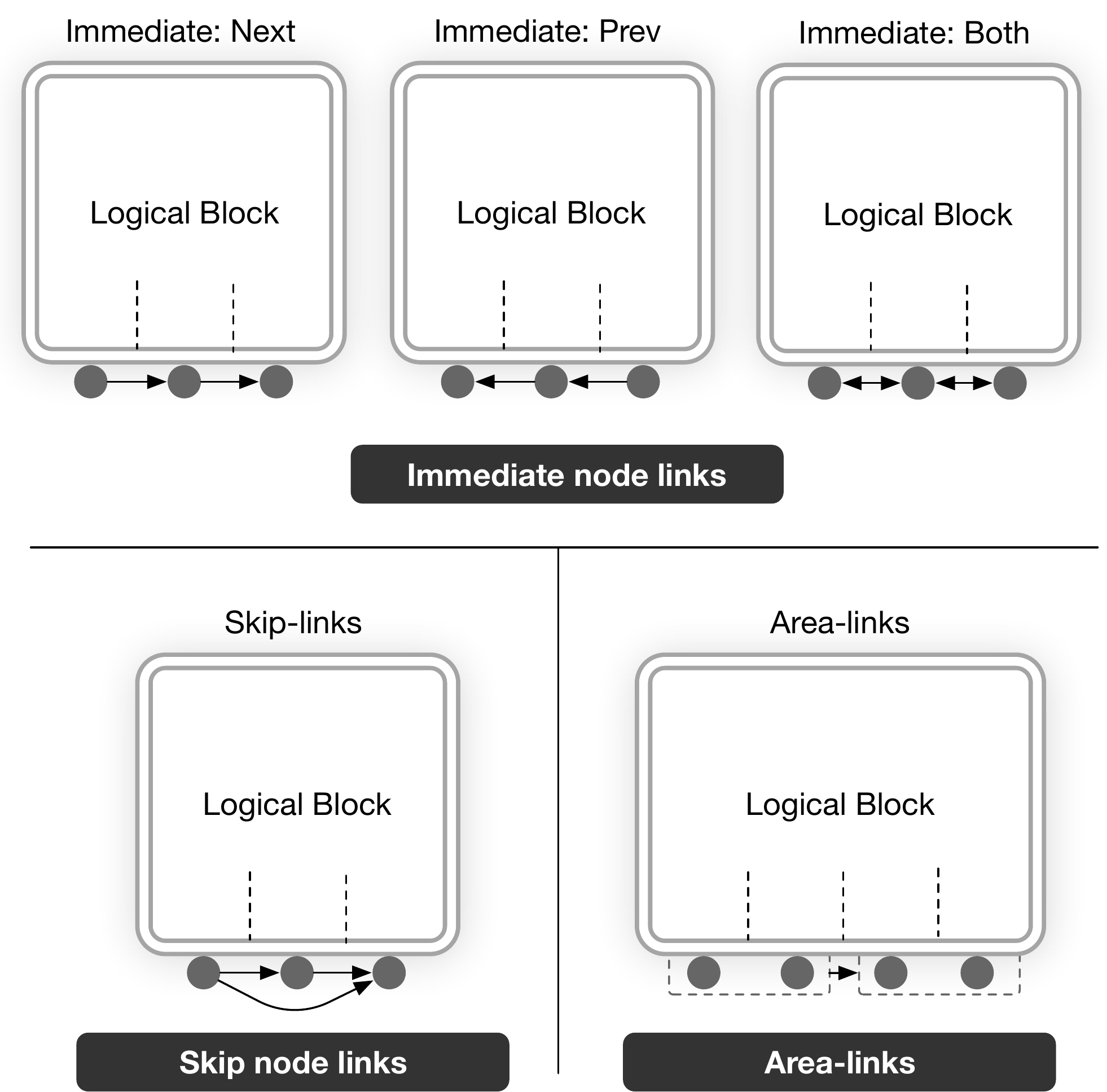}\end{center}

    \item \textbf{Sub-block physical location}~\\
    \textbf{Domain:} inline | pointed | double pointed | none ~\\
    \textbf{Description:}
     This is used to define the physical location of sub-blocks, are they inlined (i.e., physically contained within the parent element) or
     pointed to using pointers to arbitrary locations in memory. 
     Double pointed means that the sub-block points back to the parent as well.
     None means that sub-blocks are either nonexistent (i.e., the element is terminal) or not stored.     ~\\
    \textbf{Examples:}~\\
    - A classic B+Tree uses pointed sub-blocks where sub-trees are pointed to by their parents.
    - A hash-map followed by per bucket linked-lists can contain the linked list head per bucket inlined.

    \begin{center}\includegraphics[width=\linewidth]{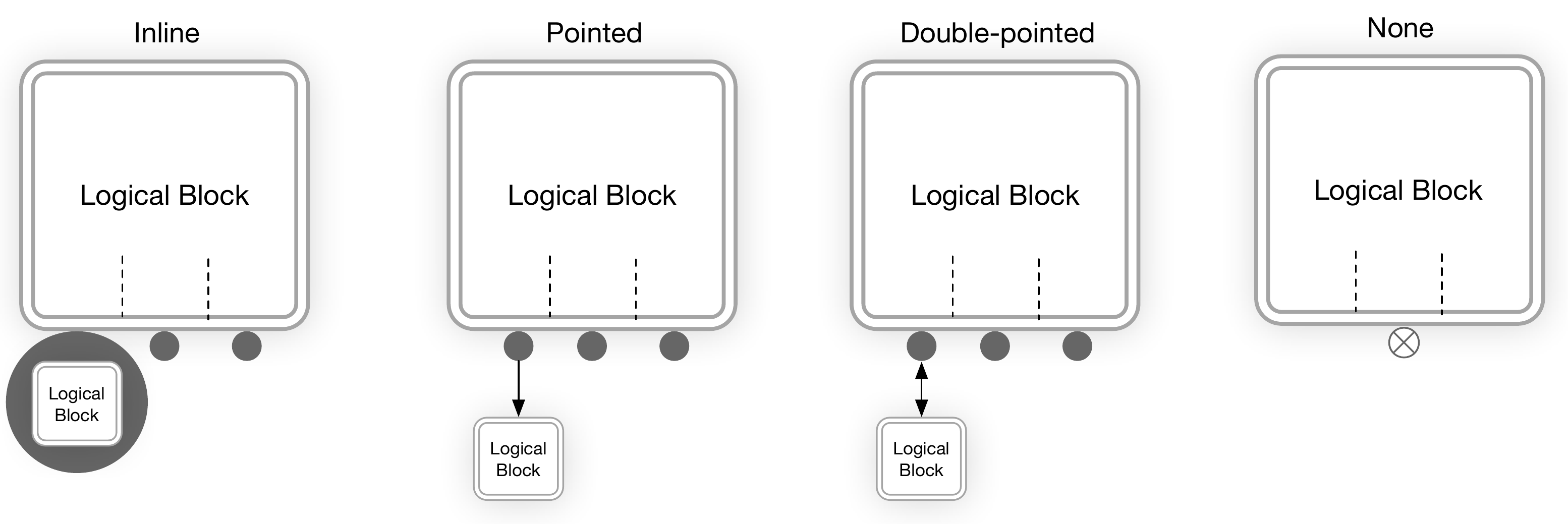}\end{center}

    \item \textbf{Sub-block physical layout}~\\
    \textbf{Domain:} BFS | BFS layer grouping | Scatter ~\\
    \textbf{Description:}
    This represents the physical layout of sub-blocks. Scatter: random placement in memory. BFS: laid out in a breadth-first
    layout. BFS layer grouping: hierarchical level nesting of BFS layouts.      ~\\
    \textbf{Examples:}~\\
    - A cache conscious B+Tree uses BFS to layout nodes.
    - A linked list can have blocks that are randomly scattered in main memory.

    \begin{center}\includegraphics[width=\linewidth]{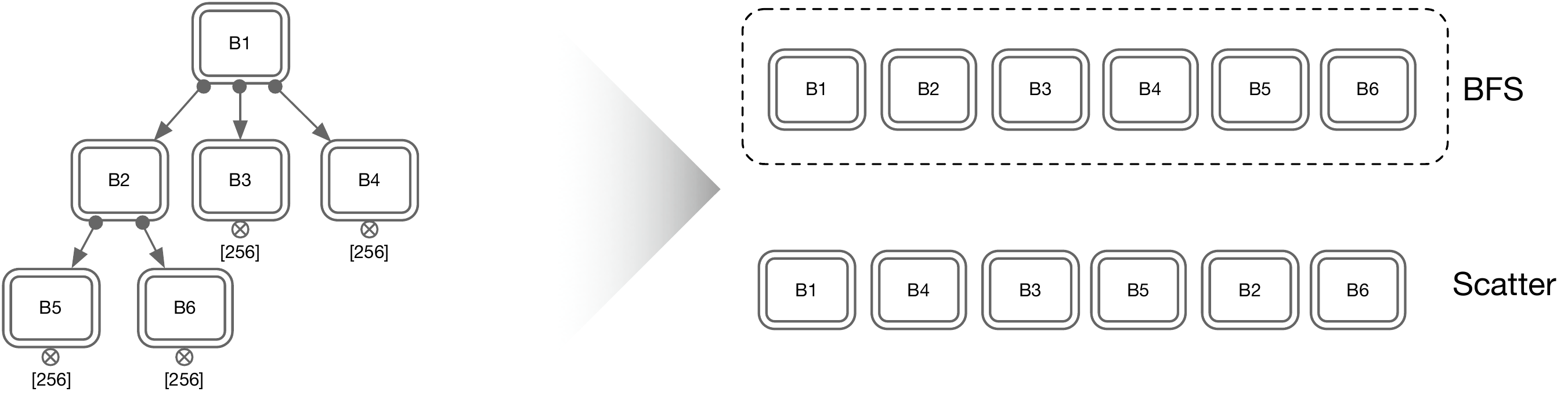}\end{center}

    \item \textbf{Sub-blocks homogeneous}~\\
    \textbf{Domain:} boolean ~\\
    This is set to true in the case where the sub-blocks have the same definition.
    \textbf{Examples:}~\\
    - Data structures like bounded disorder and ART are not homogeneous, as sibling nodes can be of different types.~\\
    - A classic B+Tree is homogeneous as all children of a node are of the same type.

    \begin{center}\includegraphics[width=0.6\linewidth]{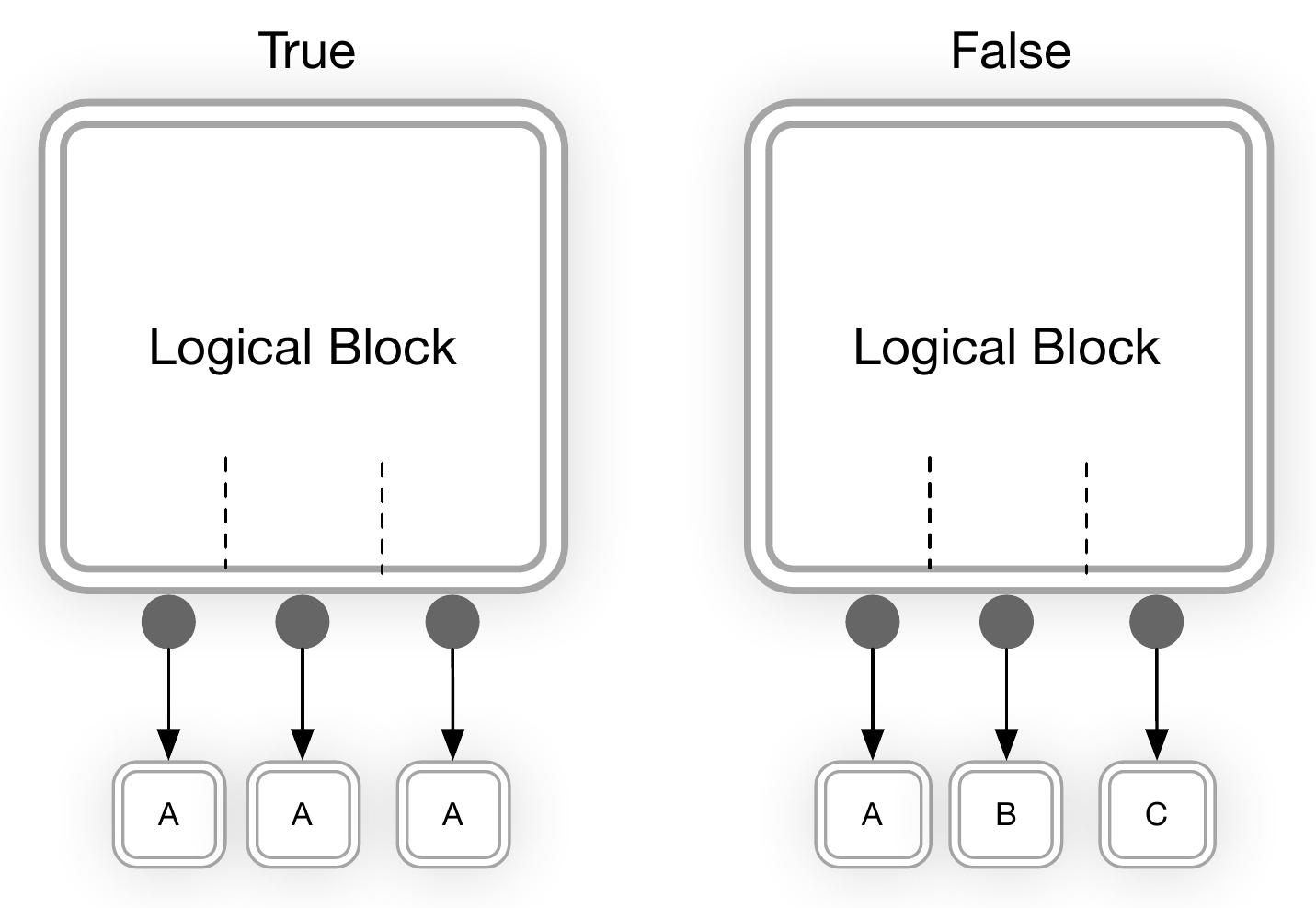}\end{center}

    \item \textbf{Sub-block consolidation}~\\
    \textbf{Domain:} boolean~\\
    If the node has only one sub-block, this node is suppressed.

    \begin{center}\includegraphics[width=\linewidth]{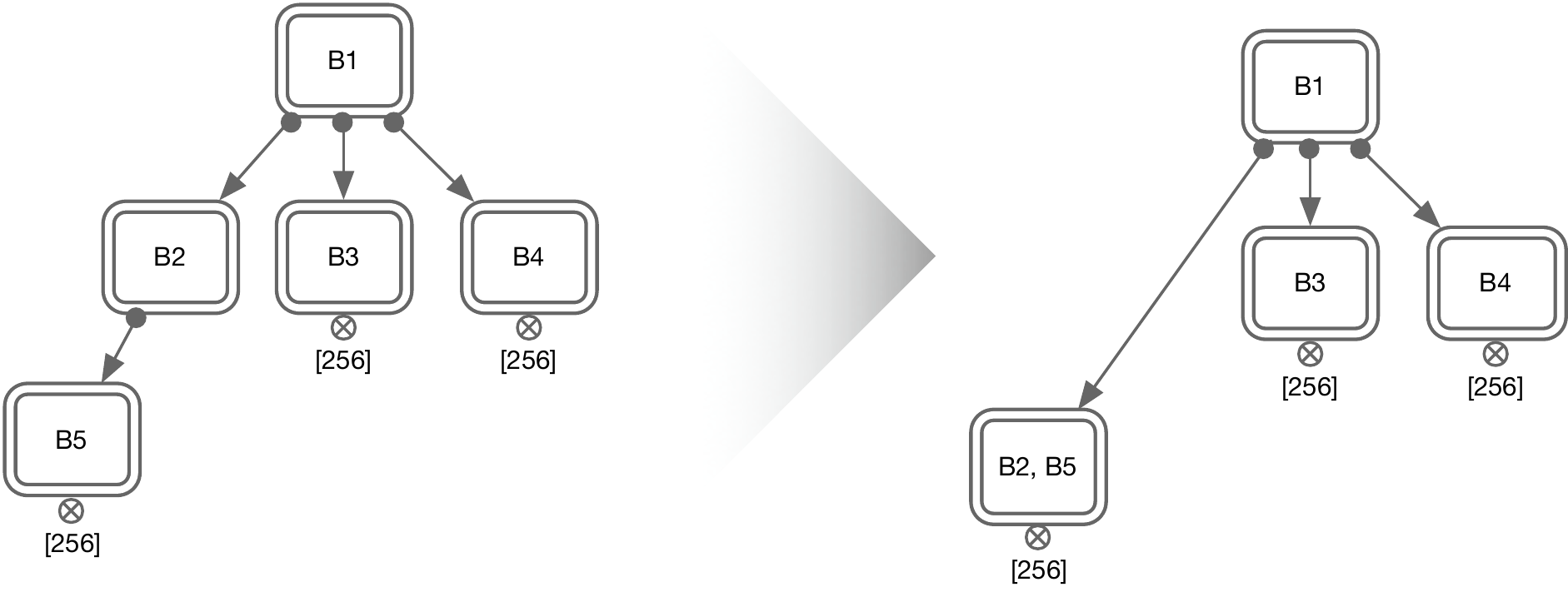}\end{center}

    \item \textbf{Sub-block instantiation}~\\
    \textbf{Domain:} lazy | eager~\\
    If a sub-block is empty, it is represented with a null pointer when set to lazy, or instantiated in any case when set to eager.

    \begin{center}\includegraphics[width=\linewidth]{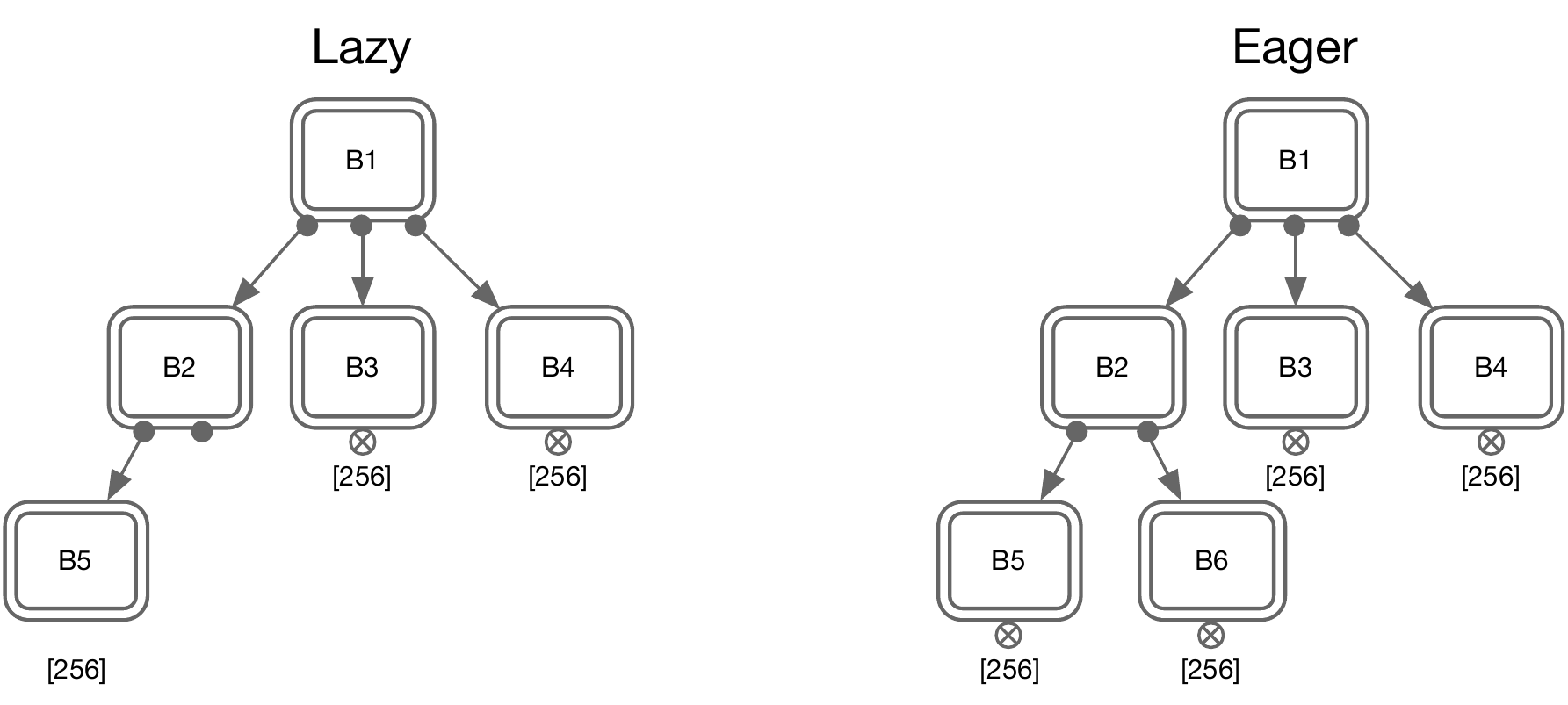}\end{center}

    \item \textbf{Links location}~\\
    \textbf{Domain:} consolidate | scatter ~\\
    In case links are used this tells us how they are stored.
    Consolidate means that they are stored as a contiguous array, scatter means that they are stored per partition.
    \textbf{Examples:}~\\
    - Linked lists have scattered links, where each sub-block contains the links. ~\\
    - Unix inodes consolidate within the current inode links to blocks or indirect blocks.
    
    \begin{center}\includegraphics[width=0.6\linewidth]{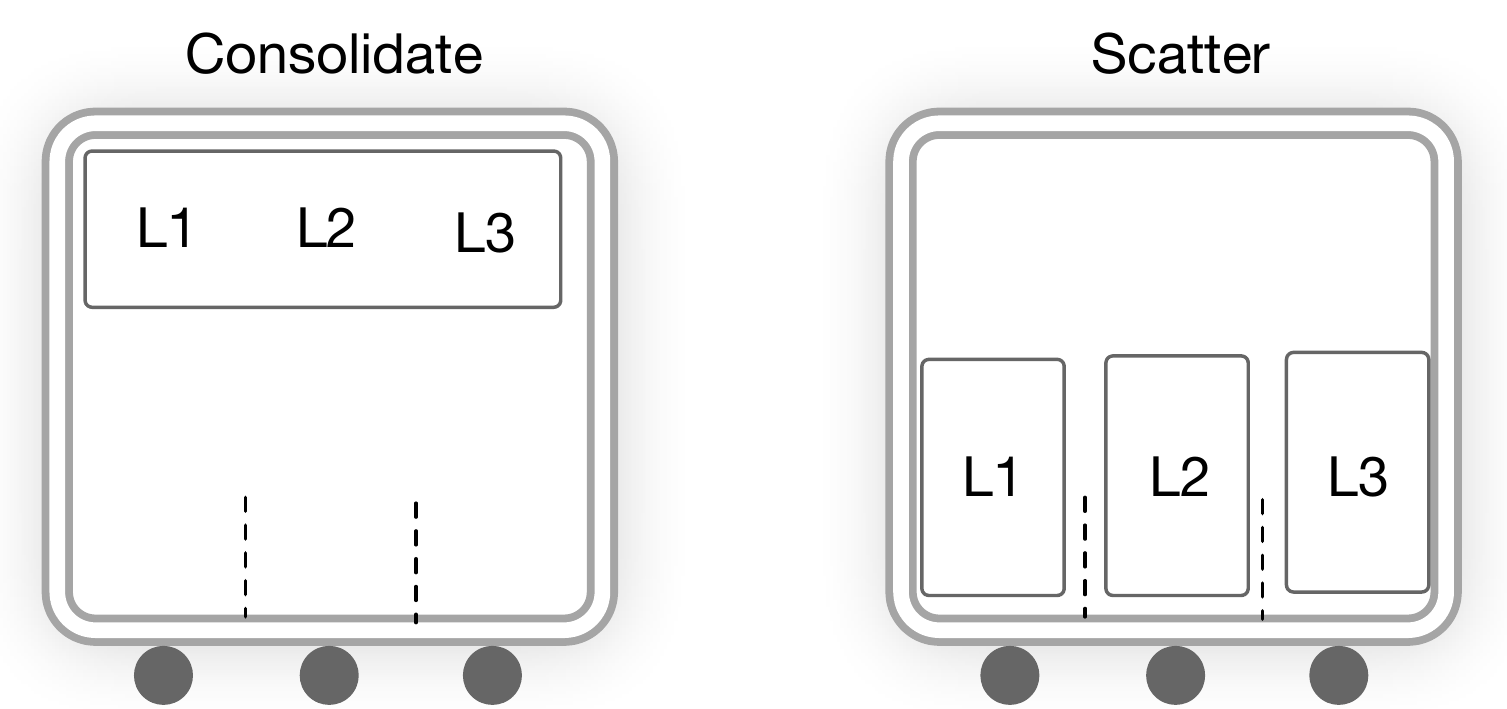}\end{center}

    \item \textbf{Recursion}~\\
    \textbf{Domain:} yes(func) | no ~\\
   If yes, sub-blocks will be subsequently inserted into a node of the same type until a maximum depth (expressed as a function) is reached. Then the terminal node type of this data structure will be used.

    \begin{center}\includegraphics[width=0.4\linewidth]{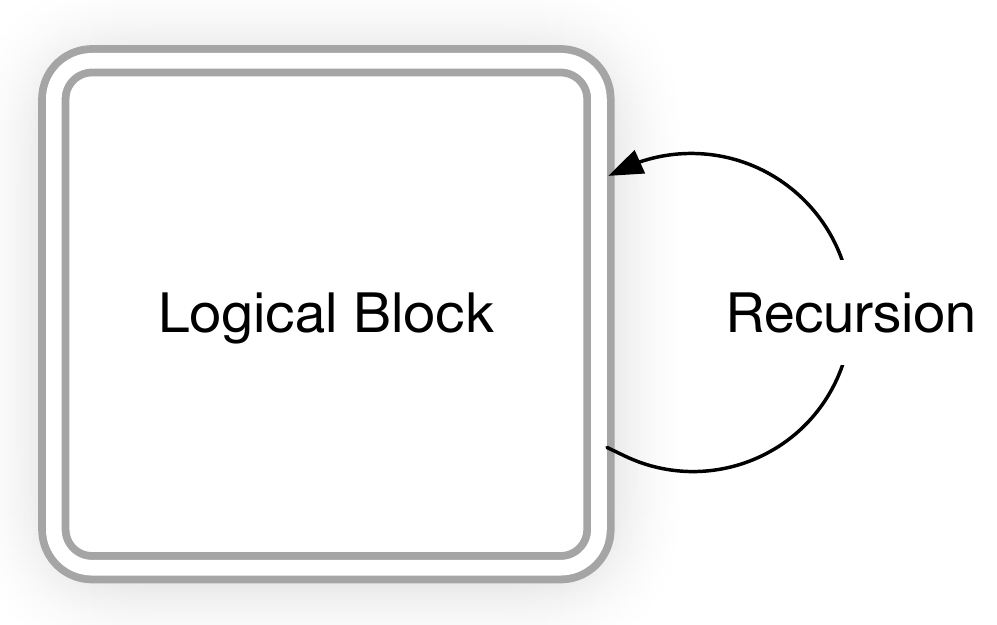}\end{center}

\end{enumerate}

\begin{figure*}[h]
    \includegraphics[width=2.02\columnwidth]{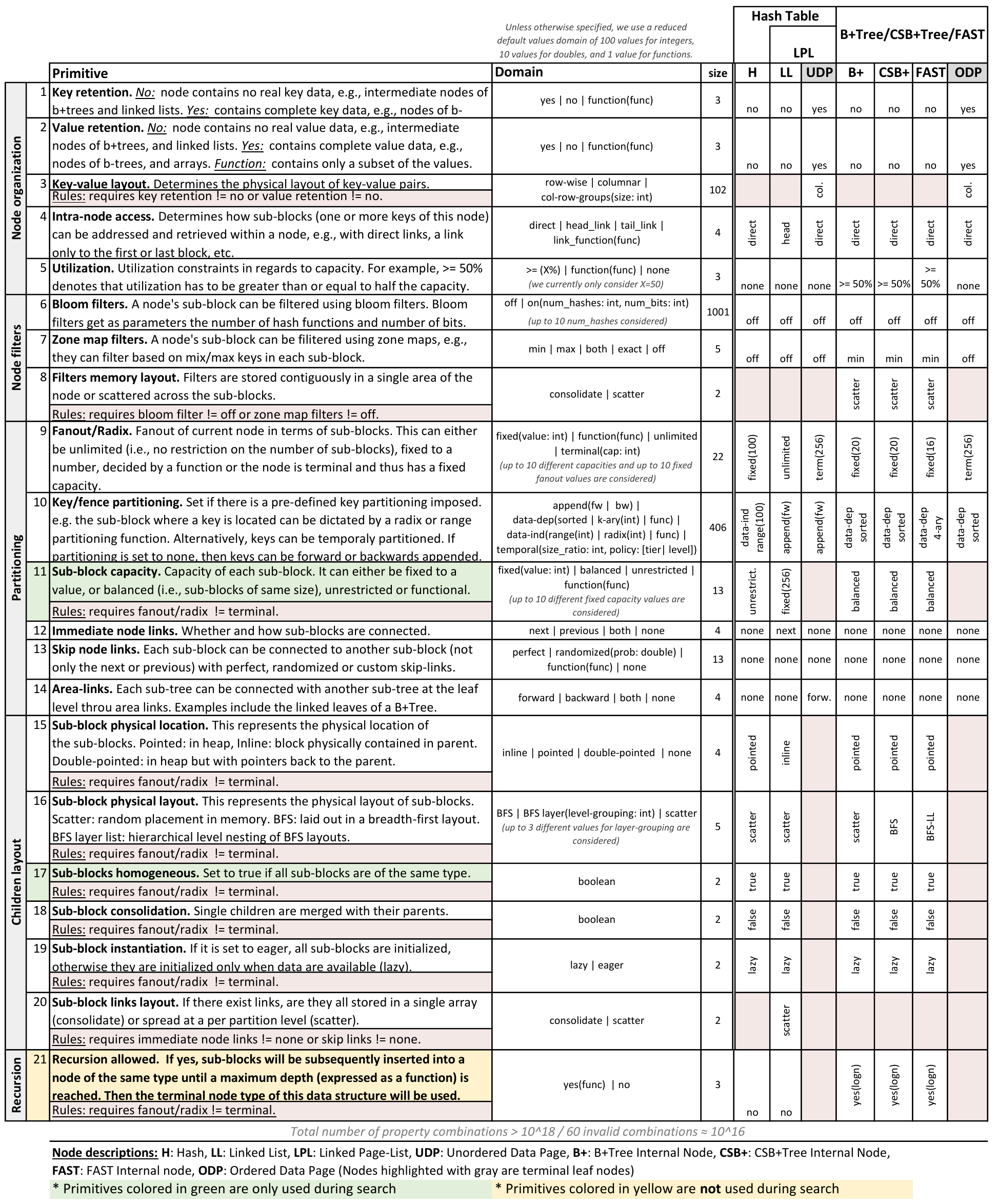}
        \vspace{-3ex}
 \caption{\ch{Data layout primitives and synthesis examples of data structures.}}
 \label{table:layoutprimitivestable}
\end{figure*}


\section{Data Access Primitives \\ and Learned Cost Models}
\label{sec:da}

We now present in detail the Level 2 access primitives and discuss how to generate learned cost models to perform cost synthesis. Each data access primitive has three stages. \\

\noindent\textbf{Stage 1:} In the first stage, C++ code implementing the Level
2 access primitive is run. This implementation usually consists of doing $n$
iterations of the required action in a tight loop; as an example, we might
perform $n=10000$ binary searches an array for some set of 10000 chosen values
sampled with replacement from the values within the array. After doing the
required action $n$ times, the benchmark latency is observed and normalized
(i.e. latency = total latency / $n$).

This implementation benchmark is run multiple times with controlled values for
various input parameters such as data size. The benchmark latency is observed
and recorded for each set of input parameters. Afterwards, these values are
 collected into 1) a matrix of the parameter values $X$ and 2) a
corresponding array of the performance (latency) $Y$ observed in each benchmark. In the joined matrix $(X | Y)$,
each row contains a single configuration of parameters and a single benchmark
latency. A column in $X$ contains all the configuration values of a parameter,
and the $Y$ column is the benchmark latencies. Below are two examples of
benchmarks and their corresponding data matrixes: \\

Example 1 (Binary Search): Matrix of parameter values is $(d \times 1)$ vector,
with $d = $ the number of input sizes. A concrete example might be sizes =
\{128,256,512,1024\}. Additionally, a $d \times 1$ vector of benchmark
latencies is passed. \\

Example 2 (Bloom Filter): The matrix of parameter values in this case is $d
\times 2$ (recall that variables such as hash function are chosen via the
Level 1 Benchmark), with the two variables being data size and number of hash
functions. A concrete example might be $$\begin{pmatrix} 128 & 1 \\ 128 & 2 \\
256 & 1 \\ 256 & 2 \end{pmatrix}$$. The set of benchmark latencies is still a
$d \times 1$ vector, with the recorded latencies passed in the same order as
the parameter of matrix values (i.e., the first benchmark latency corresponds
to the first row of parameters in the matrix of parameter values). \\

\noindent\textbf{Stage 2:} A model-specific Python function is then called to
train the model. The matrix of parameter values and corresponding array of
latencies is passed to the function. Depending on the model, some combination
of NumPy, SciPy, and PyTorch is used to train the models. For simple models
this can be as simple as a call to the least squares regression function; for
others, this requires some pre-processing of data or training via gradient
descent. \\

Example 1 (Interpolation Search): As input, Python is given the required input
of data sizes and benchmark latencies. It performs two basic actions, which
essentially provide new features. These two actions are to compute $\log$ data
size and $\log (\log \text{data size})$. Thus each row in $X$ now contains
three input features, data size, $\log$ data size, and $\log (\log \text{data
size})$. At this point we run regular linear regression via NumPy. We record
the optimal coefficients and return them as the return values of our train
function back to the main cost synthesis routine of the Data Calculator in C++. \\

Example 2 (Random Memory Access): The benchmark for a random memory access is fit by a sum of sigmoids
model: $f(x) = \sum_{i=1}^{k} \frac{c_{i}}{1 + e^{-k_{i} (\log x - x_{i})}} +
y_{0}$ with all coefficients learned as usual. This model is no longer convex
in its L2 loss and cannot be learned via off the shelf techniques such as least
squares regression. Instead, we first pre-process the data by calculating rates
of change for intervals of some size $z$. We find the $k$ highest local maxima
for this graph of rates of change and set these as our initial guesses for the
$x_{i}$ values. For the values of the $k_{i}, c_{i}$ we initialize them to
random positive numbers between $0$ and $1$. Our initial guess for $y_{0}$ is
just the first point. At this point, we use SciPy's curve fit package to train
the parameters and observe the results.

The learning of the coefficients for the sum of sigmoid model requires good
initial conditions for the guesses of the $x_{i}$. It is not sensitive to
initial guess for the parameters of $k_{i}$ or $c_{i}$. We found that for each
of $z = 0.1,0.5$, and $1$, the initial guesses for the $x_{i}$ were good enough
so that the fitted model produces accurate results.\\

\noindent\textbf{Stage 3:} These coefficients are then stored in a model
specific class such as ``SimpleLinearModel''. The class then has a prediction
function which uses the coefficients to calculate costs given input parameters.
Thus, even though the training is in Python, the prediction is made in C++
classes. \\

Example 1 (Interpolation Search): In the case of interpolation search, the
corresponding model is LogLogLinear model. The model is defined by $f(x) =
c_{1}x + c_{2} \log x + c_{3} \log \log x + c_{0}$. It's class constructor
takes in 4 coefficients, and stores them. Then for
prediction, it takes as input $x$, and computes $f(x)$ using the given
coefficients. In the case of Interpolation Search, the input variable $x$ is
data size and so the model is producing a prediction for search cost given the
data size. \\

Example 2 (Weighted Nearest Neighbors): Nearest neighbors is our only
non-parametric model at the moment, although we expect to add more in the
future. Unlike the previous example, which is a parametric model (defined by
the coefficients $c_{1},c_{2},c_{3},c_{0})$, the non parametric model needs to
keep track of its data. Thus in weighted nearest neighbors, its constructor
takes in the matrix $X$ and $Y$ and predictions are made using these values
explicitly via the function: $f(x) = \frac{1}{\sum_{i=1}^{k}
\frac{1}{d(x,x_{i})}} \sum_{i=1}^{k} \frac{1}{d(x,x_{k})} y_{k}$.

\subsection{Notes for Presentation}

For the example graphs presented next, each fitted model is from the same
machine and used 8 byte key and value sizes. The graphs are given with Data
Size as the dependent variable for various graphs, although the input into the
Level 2 primitives can be either the number of elements or data size (for fixed
size elements). For fixed size elements, the graphs are more easily read as
number of elements. Finally, in the following models, there is some small
amount of terminology that makes naming of the benchmarks easier. We provide
the terms below.

RowStore vs. ColumnStore: This refers to how the data is laid out. In general
RowStore means that we have arrays of key-value pairs, where as a column store
means we have separate arrays, one for keys and one for values. This is the
input of the Data Layout option that comes into higher level benchmarks. (i.e.
to cost a scan over data that is stored as a ``row store'', you pass Data
Layout = ``row store'' into the higher level benchmark).

Hash Family: For the purpose of benchmarking and cost modeling, we take the
general form of the hash function and randomly generate the coefficient values
during execution. That is, no particular value of the coefficients is assumed
and we do not attempt (currently) to learn optimal values of these
coefficients. As an example, for the multiply-shift scheme, the form is assumed
to be $h(x) = (a * x ) >> (w - M)$, where $b = 2^M$ is the number of bins and
$w$ is the word size (both fixed). During benchmarking, we conduct multiple
runs and generate the value of $a$ randomly during each run.
\begin{figure}
\begin{center}\includegraphics[width=0.9\columnwidth]{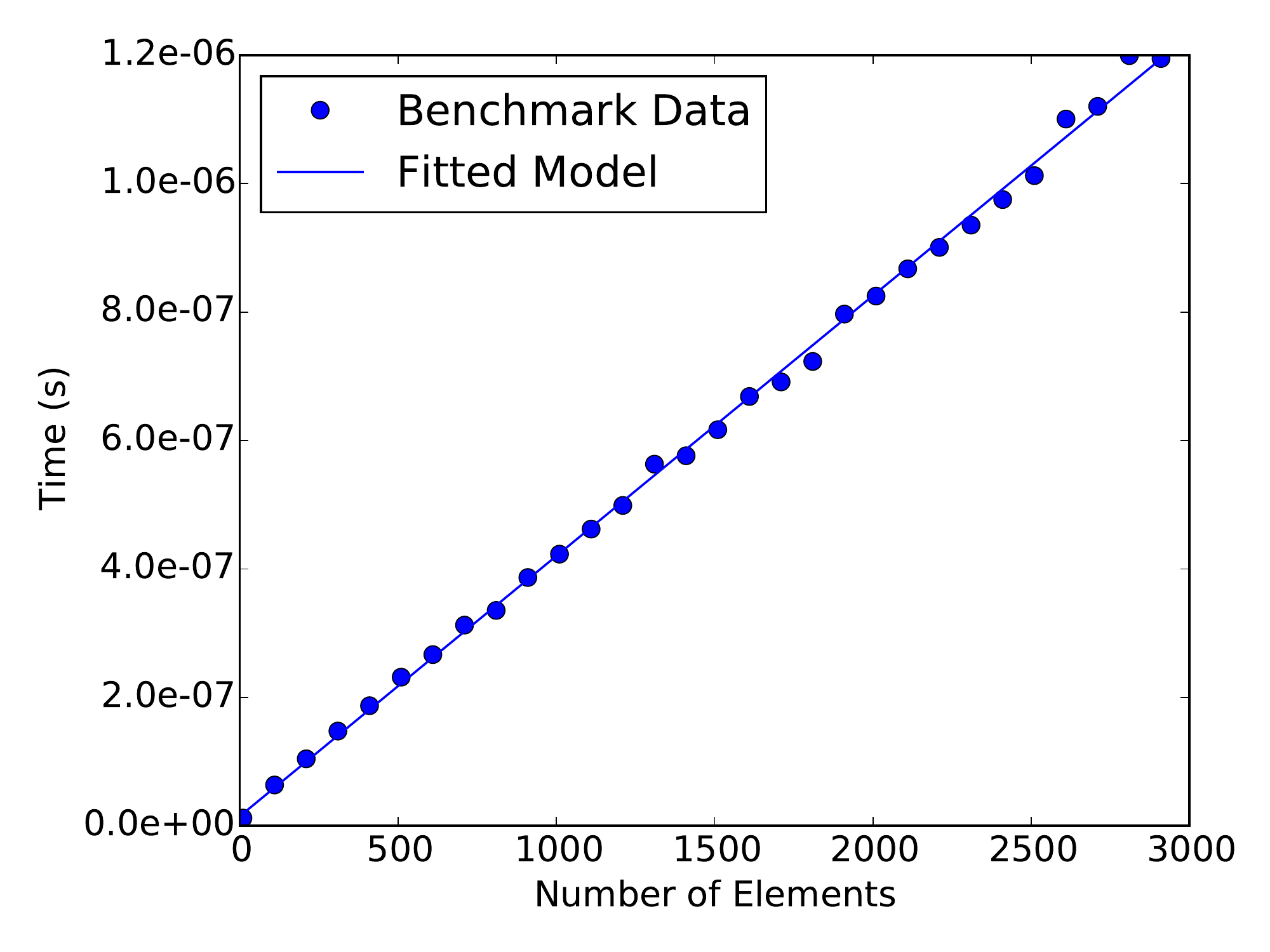}\end{center}
\caption{Example of Benchmark (1)}\label{fig:bench1}
\end{figure}

\subsection{Learned Cost Models}

\begin{enumerate}

    \item \textbf{Scalar Scan} (RowStore, Equal)\\

    \textbf{Description:} This benchmark performs a \textsc{for} loop
    over an array of key-value pairs. At each array index, we check whether the
    key at that position matches the desired key.\\

    \textbf{Model:} The benchmark is fit by a linear model: $f(x) = ax + b$
    with a and b being learned coefficients. Both coefficients are constrained
    to be non-negative.\\

\vspace{-2ex}
{\center Benchmarking PseudoCode}\\
\doublerule

\begin{algorithmic}[1]
\REQUIRE array of $n$ key-value pairs, search val $x$
\FOR{$i$ in $1, \dots n$}
\IF {array[i].key == x}
\RETURN array[i].val
\ENDIF
\ENDFOR
\end{algorithmic}

\hrulefill \\

\textbf{Example of Learned Cost Model}: Figure \ref{fig:bench1} \\

    \item \textbf{Scalar Scan} (ColumnStore, Equal)\\

    \textbf{Description:} This benchmark performs a \textsc{for} loop
    over an array of keys. At each array index, we check whether the key at
    that position matches the desired key. If yes, we
    return the value at the same index in the values array. \\

    \textbf{Model:} The benchmark is fit by a linear model: $f(x) = ax +
    b$ with a and b being learned coefficients. Both coefficients are
    constrained to be non-negative.\\

{\center Benchmarking PseudoCode}\\
\doublerule

\begin{algorithmic}[1]
\REQUIRE array of $n$ keys, array of $n$ values, search val $x$
\FOR{$i$ in $1, \dots n$}
\IF {keys[i] == x}
\RETURN values[i]
\ENDIF
\ENDFOR
\end{algorithmic}

\hrulefill

\begin{figure}
\begin{center}\includegraphics[width=0.9\columnwidth]{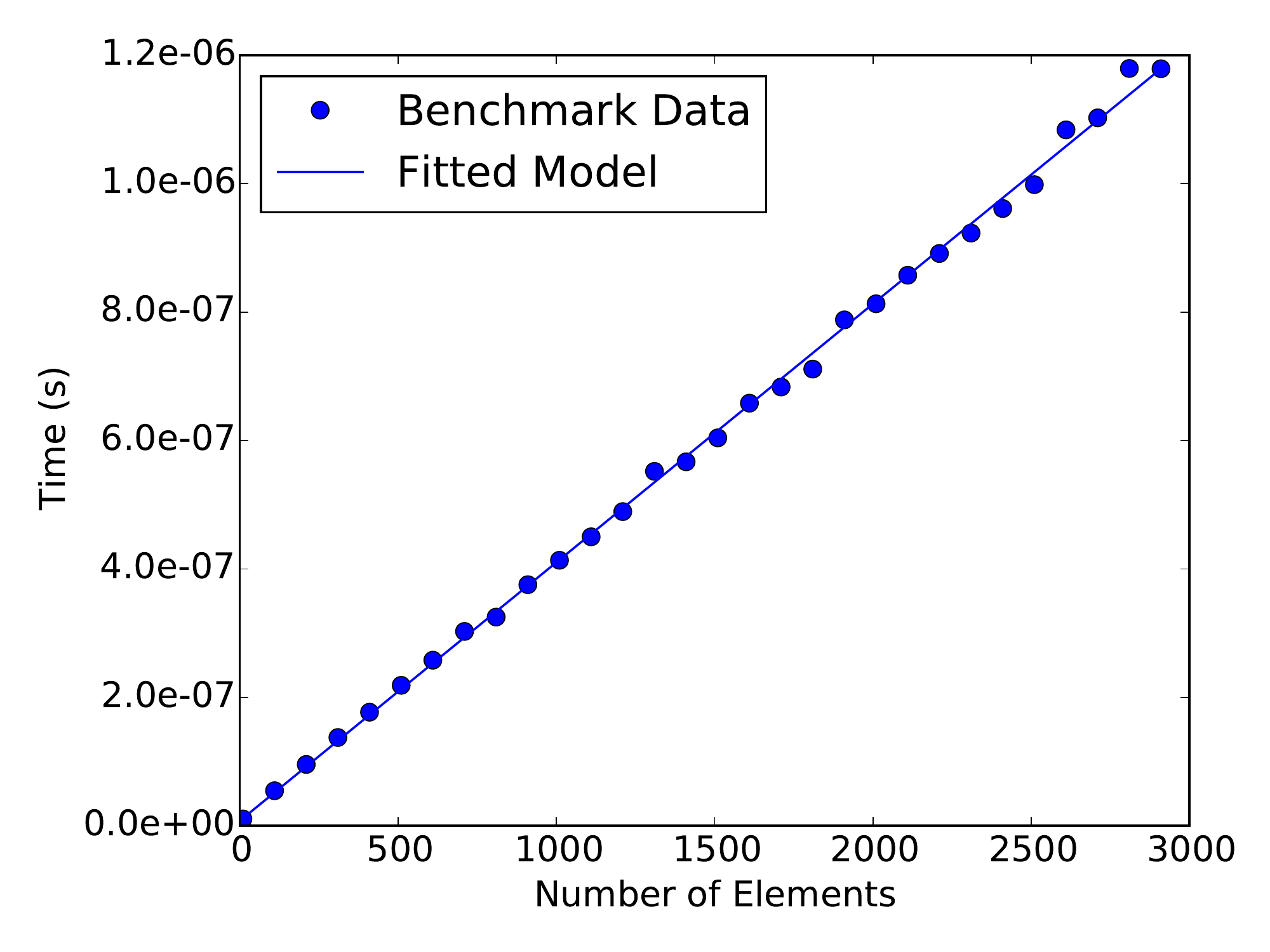}\end{center}
\caption{Example of Benchmark (2)}\label{fig:bench2}
\end{figure}

     The graphs for the scans tend to look very similar, even down to
     what initially appears to be noise. Each graph is in fact different. For
     instance, you can see that the upper right two points for this graph are
     different than the upper two right points in the prior graph. As well, it
     makes sense that the two graphs are very similar, they have almost
     identical code and the process is computationally bound. \\ 
     
     \textbf{Example of Learned Cost Model}: Figure \ref{fig:bench2} \\

    \item \textbf{Scalar Scan} ( RowStore, Range)\\

    \textbf{Description:} This benchmark performs a for loop over an
    array of key-value pairs. At each array index, we check whether the key at
    that position is less than the desired key. If the key satisfies the
     predicate, the corresponding value is added to a set of values to be
    returned. After looping through the whole array, we return the list of
    values that had keys less than the desired predicate.  Similar benchmarks
    exist for between predicates. \\

    \textbf{Model:} The benchmark is fit by a linear model: $f(x) = ax +
    b$ with a and b being learned coefficients. Both coefficients are
    constrained to be non-negative.\\

{\center Benchmarking PseudoCode}\\
\doublerule

\begin{algorithmic}[1]
\REQUIRE array of $n$ key-value pairs, search val $x$
\ENSURE output array of values
\STATE count = 0
\FOR{$i$ in $1, \dots n$}
\IF {array[i].key $<$ x}
\STATE output[count] = array[i].val
\STATE count++
\ENDIF
\ENDFOR
\RETURN output
\end{algorithmic}

\hrulefill \\

\textbf{Example of Learned Cost Model}: Figure \ref{fig:bench3} \\

\begin{figure}
\begin{center}\includegraphics[width=0.9\columnwidth]{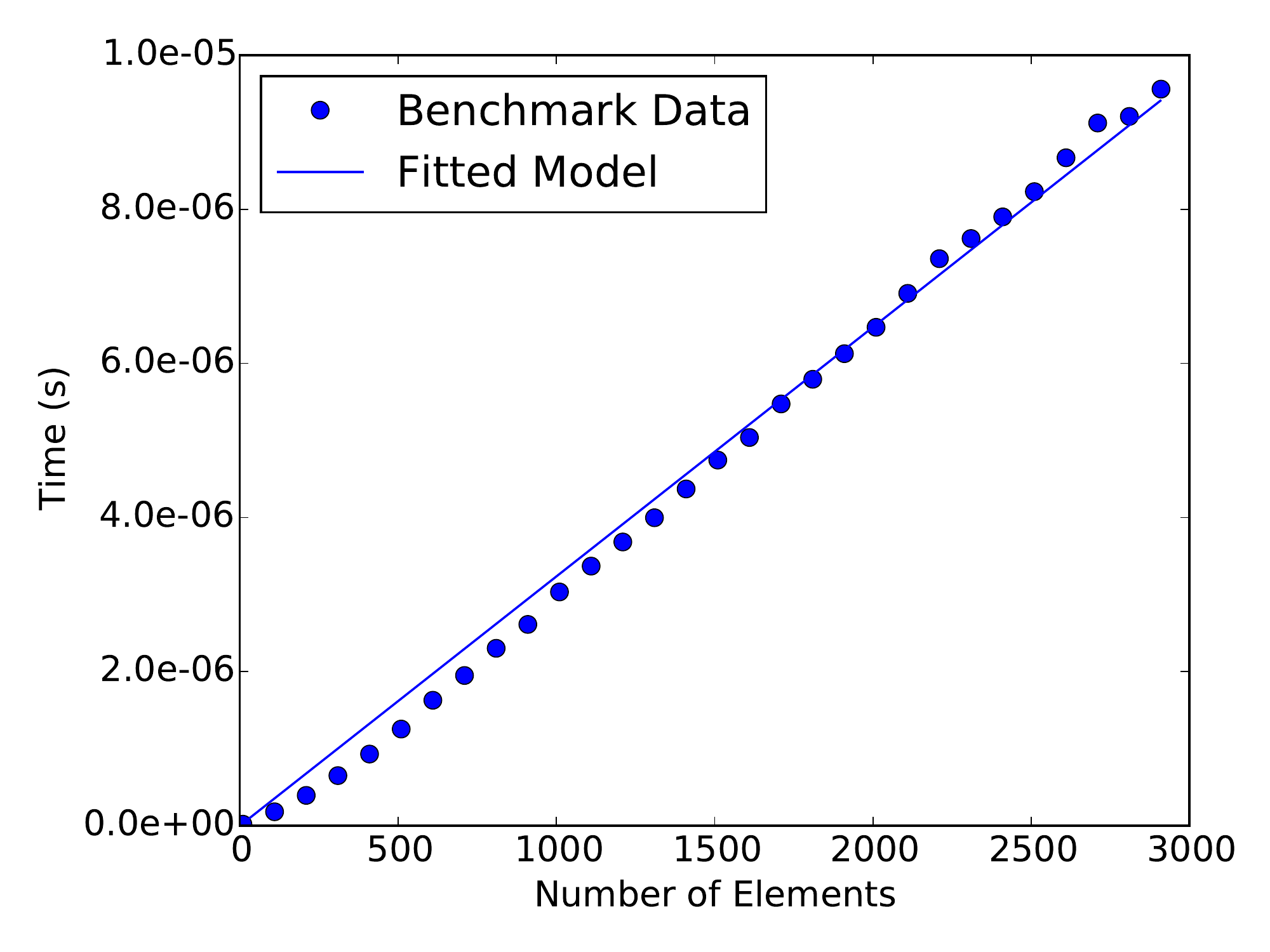}\end{center}
\caption{Example of Benchmark (3)}\label{fig:bench3}
\end{figure}

    \item \textbf{Scalar Scan} (ColumnStore, Range)\\

    \textbf{Description:} This benchmark performs a for loop over an
    array of keys. At each array index, we check whether the key at that
    position is less than the desired key. If the key is less than the desired
    key, we add the corresponding value in the values array to a set of values
    to be returned. After looping through all the values, the set of values is
    returned. Similar benchmarks exist for between predicates. \\

    \textbf{Model:} The benchmark is fit by a simple linear model: $f(x) = ax +
    b$ with a and b being learned coefficients. Both coefficients are
    constrained to be non-negative.\\

{\center Benchmarking PseudoCode}\\
\doublerule

\begin{algorithmic}[1]
\REQUIRE array of $n$ keys, array of $n$ values, search val $x$
\ENSURE output array of values
\STATE count = 0
\FOR{$i$ in $1, \dots n$}
\IF {keys[i] $<$ x}
\STATE output[count] = values[i]
\STATE count++
\ENDIF
\ENDFOR
\RETURN output
\end{algorithmic}

\hrulefill \\

\textbf{Example of Learned Cost Model}: Figure \ref{fig:bench4} \\

\begin{figure}
\begin{center}\includegraphics[width=0.9\columnwidth]{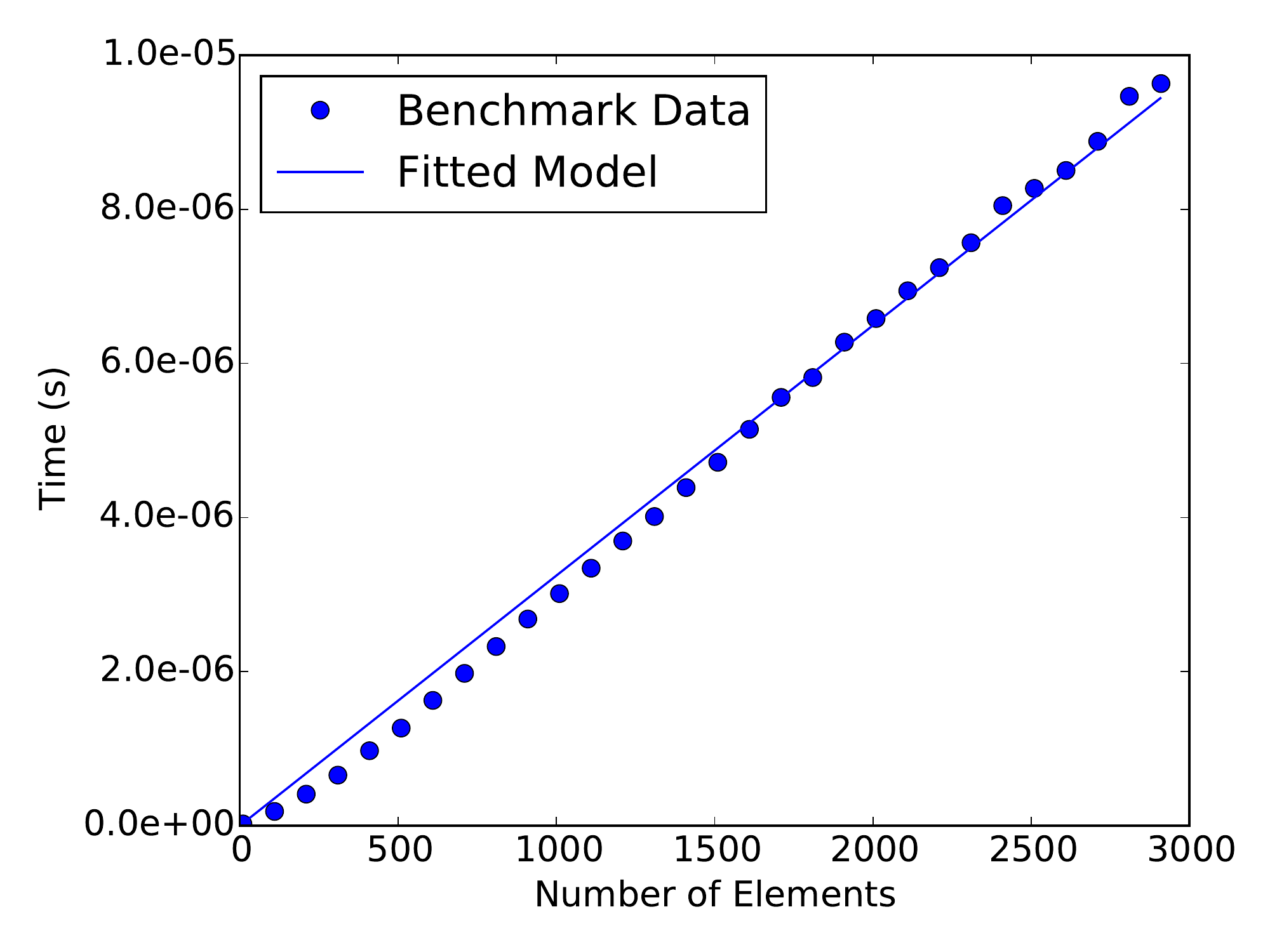}\end{center}
\caption{Example of Benchmark (4)}\label{fig:bench4}
\end{figure}

    \item \textbf{Simd-AVX Scan} (ColumnStore, Equal)\\

    \textbf{Description:} This benchmark performs a SIMDified loop over an
    array of keys. At each point, we compare a SIMD bank of keys to the desired
    value. If a matching key is found, we break the loop and return the desired
    value at the corresponding position.\\

    \textbf{Model:} The benchmark is fit by a linear model: $f(x) = ax +
    b$ with a and b being learned coefficients. Both coefficients are
    constrained to be non-negative. \\

{\center Benchmarking PseudoCode}\\
\doublerule

\begin{algorithmic}[1]
\REQUIRE array of $n$ keys, array of $n$ values, search val $x$.
\STATE \COMMENT {SIMD bank is of width b = 128 / element size.}
\FOR{each SIMD bank}
\STATE load memory into SIMD register
\STATE comp\_bank = SIMD compare equal
\IF {test\_not\_all\_zeros(comp\_bank)}
\STATE get mask from comp\_bank and use this to pull value from values array
\RETURN pulled value
\ENDIF
\ENDFOR
\end{algorithmic}

\hrulefill \\

\textbf{Example of Learned Cost Model}: Figure \ref{fig:bench5} \\

\begin{figure}
\begin{center}\includegraphics[width=0.9\columnwidth]{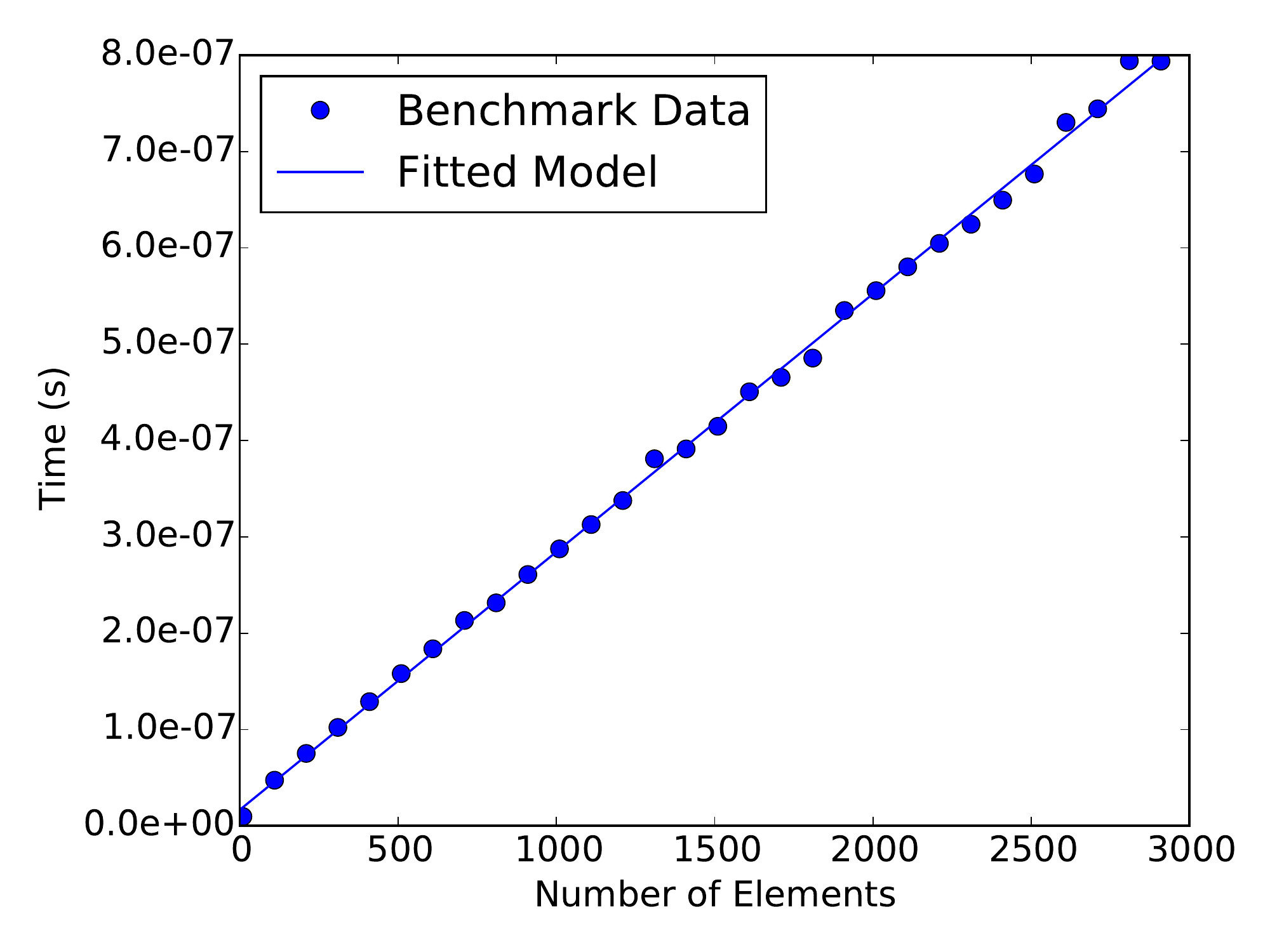}\end{center}
\caption{Example of Benchmark (5)}\label{fig:bench5}
\end{figure}

    \item \textbf{Simd-AVX Scan} (ColumnStore, Range)\\

    \textbf{Description:} This benchmark performs a SIMDified loop over an
    array of keys. We loop through the array, performing actions one bank of
    SIMD values at a time. For each bank of SIMD values, we perform a less than
    predicate on the bank of keys, comparing it with the predicate value. After
    finding which keys satisfy the predicate, we perform a swizzle instruction
    to put the corresponding values to the matching keys at the beginning of a
    separate SIMD bank. We then store the matching values. Similar benchmarks
    exist for between predicates. More than the other benchmarks, this
    benchmark depends heavily on the input key size as the code changes with
    each key size. In particular, the way in which we perform the swizzle
    instructions depends on the width of the key.\\

    \textbf{Model:} The benchmark is fit by a linear model: $f(x) = ax +
    b$ with a and b being learned coefficients. Both coefficients are
    constrained to be non-negative. \\

{\center Benchmarking PseudoCode}\\
\doublerule

\begin{algorithmic}[1]
\REQUIRE array of $n$ keys, array of $n$ values, search val $x$.
\ENSURE output array of values
\STATE \COMMENT {SIMD bank is of width b = 128 / element size.}
\STATE count = 0
\FOR{each SIMD bank}
\STATE load memory into SIMD register
\STATE comp\_bank = SIMD comp. less than
\STATE pull mask from comp\_bank
\STATE use mask to swizzle around values
\STATE store values into output[count]
\STATE count += popcnt mask
\ENDFOR
\RETURN output
\end{algorithmic}

\hrulefill \\

\textbf{Example of Learned Cost Model}: Figure \ref{fig:bench6} \\

\begin{figure}
\begin{center}\includegraphics[width=0.9\columnwidth]{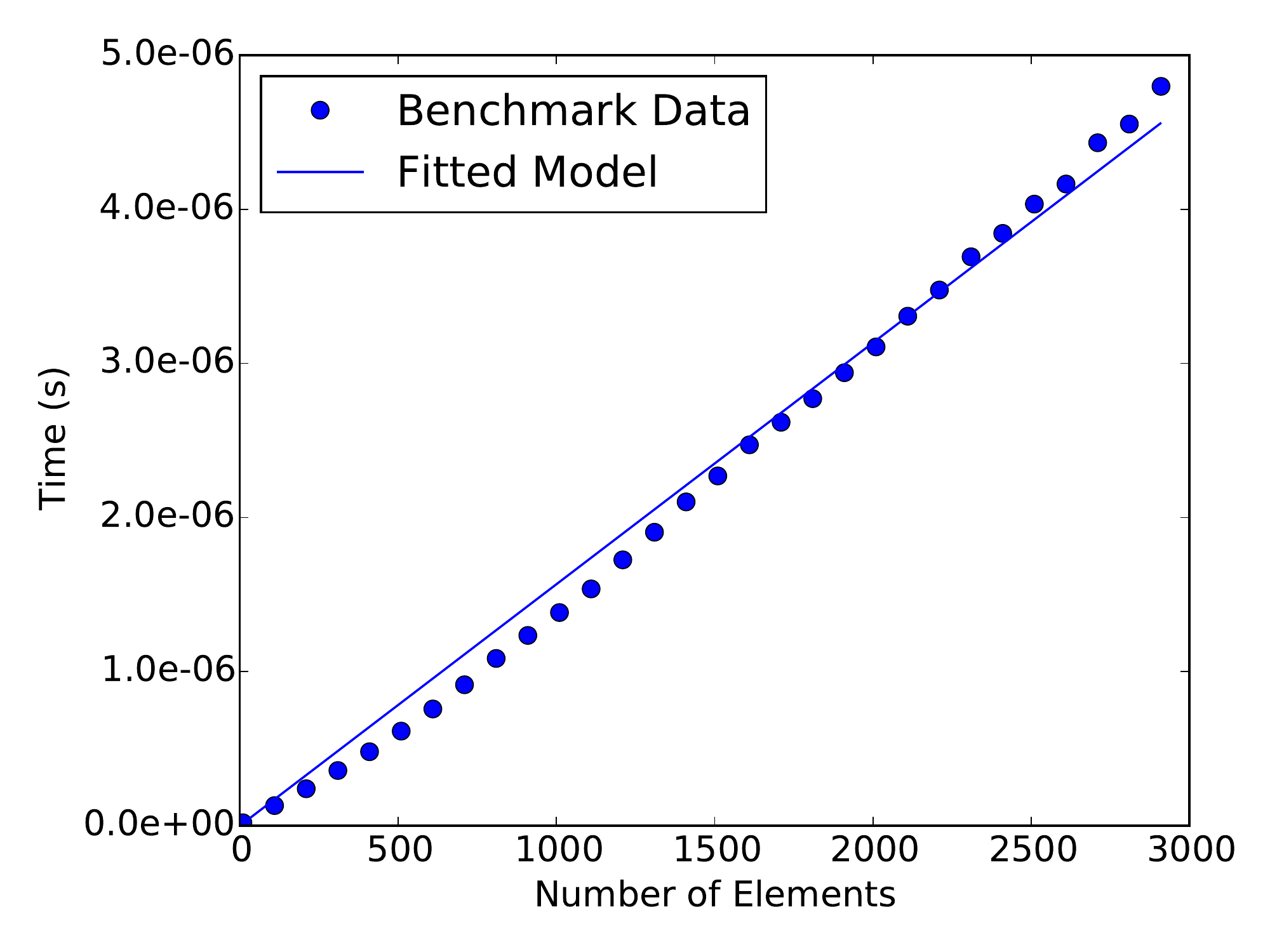}\end{center}
\caption{Example of Benchmark (6)}\label{fig:bench6}
\end{figure}

    \item \textbf{Binary Search} (RowStore)\\

    \textbf{Description:} This benchmark performs standard binary search over
    an array of key-value pairs.  \\

    \textbf{Model:} The benchmark is fit by a log-linear model: $f(x) = ax + b
    \log x + c$ with a,b, and c being learned coefficients. a,b,c are all
    constrained to non-negative. \\

{\center Benchmarking PseudoCode}\\
\doublerule

\begin{algorithmic}[1]
\REQUIRE  array of $n$ key-value pairs, search val $x$
\STATE low = 0, high = array.size - 1
\STATE middle = (low + high) / 2
\WHILE{low $<$ high}
\IF {array[middle].key $<$ x}
\STATE low = middle + 1
\ELSE
\STATE high = middle
\ENDIF
\STATE middle = (low + high) / 2
\ENDWHILE
\IF {array[middle].key == x}
\RETURN array[middle].val
\ELSE
\RETURN NOT\_FOUND
\ENDIF
\end{algorithmic}

\hrulefill \\

\textbf{Example of Learned Cost Model}: Figure \ref{fig:bench7} \\

\begin{figure}
\begin{center}\includegraphics[width=0.9\columnwidth]{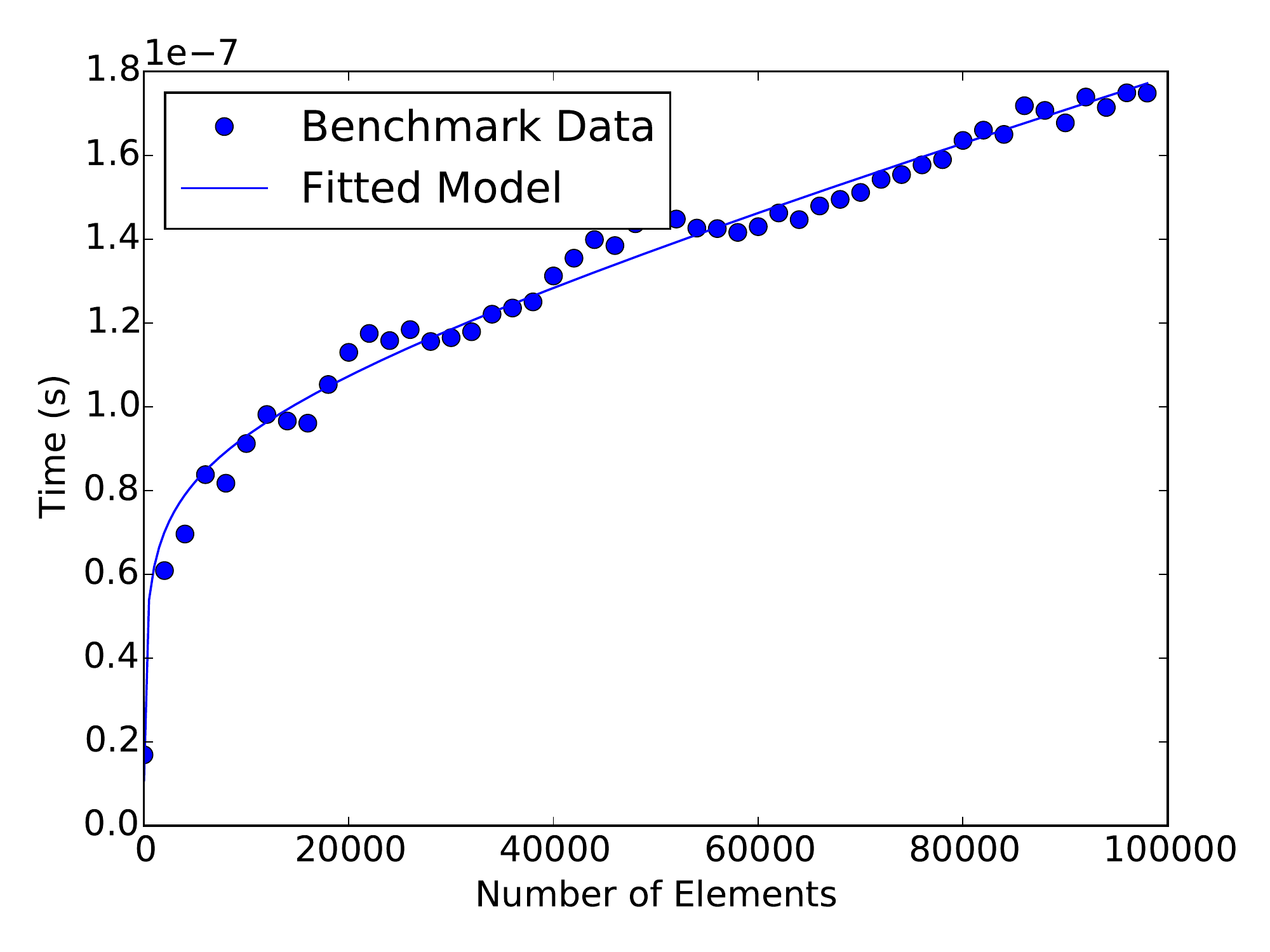}\end{center}
\caption{Example of Benchmark (7)}\label{fig:bench7}
\end{figure}

    \item \textbf{Binary Search} (ColumnStore)\\

    \textbf{Description:} This benchmark performs standard binary search over
    an array of keys. After, it returns the corresponding value.  \\

    \textbf{Model:} The benchmark is fit by a log-linear model: $f(x) = ax + b
    \log x + c$ with a,b, and c being learned coefficients. a,b,c are all
    constrained to non-negative.\\

{\center Benchmarking PseudoCode}\\
\doublerule

\begin{algorithmic}[1]
\REQUIRE array of $n$ keys, array of $n$ values, search val $x$.
\STATE low = 0, high = keys.size - 1
\STATE middle = (low + high) / 2
\WHILE{low $<$ high}
\IF {keys[middle] $<$ x}
\STATE low = middle + 1
\ELSE
\STATE high = middle
\ENDIF
\STATE middle = (low + high) / 2
\ENDWHILE
\IF {keys[middle] == x}
\RETURN values[middle]
\ELSE
\RETURN NOT\_FOUND
\ENDIF
\end{algorithmic}

\hrulefill \\

\textbf{Example of Learned Cost Model}: Figure \ref{fig:bench8} \\

\begin{figure}
\begin{center}\includegraphics[width=0.9\columnwidth]{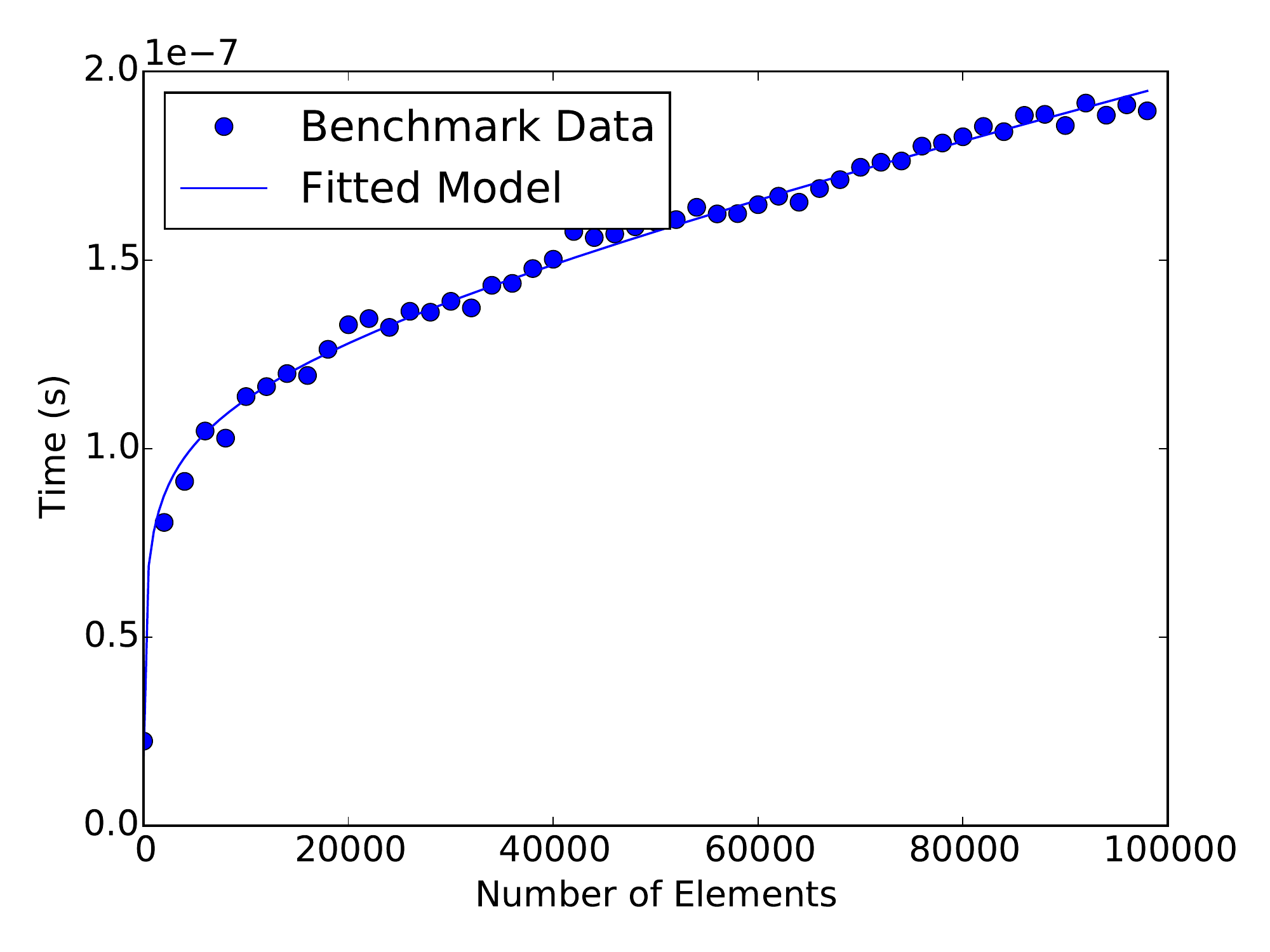}\end{center}
\caption{Example of Benchmark (8)}\label{fig:bench8}
\end{figure}

    \item \textbf{Interpolation Search} (RowStore)\\

    \textbf{Description:} This benchmark performs interpolation search over an array of key-value pairs, searching by key.  \\

    \textbf{Model:} The benchmark is fit by a loglog-linear model: $f(x) = ax
    + b \log x + c \log \log x + d$ with a,b,c and d being learned
    coefficients. \\

{\center Benchmarking PseudoCode}\\
\doublerule

\begin{algorithmic}[1]
\REQUIRE  array of $n$ key-value pairs, search val $x$
\STATE low = 0, high = array.size - 1
\STATE key\_low = array[low].key, key\_high = array[high].key
\STATE diff = key\_high - key\_low
\WHILE{low $<$ high}
\STATE si = (high - low) * ((x - val\_low) / diff)
\IF {array[si].key $<$ x}
\STATE low = si + 1
\ELSIF {array[si].key == x}
\RETURN array[si].val
\ELSE
\STATE high = si
\ENDIF
\ENDWHILE
\RETURN NOT\_FOUND
\end{algorithmic}

\hrulefill \\

\textbf{Example of Learned Cost Model}: Figure \ref{fig:bench9} \\

\begin{figure}
\begin{center}\includegraphics[width=0.9\columnwidth]{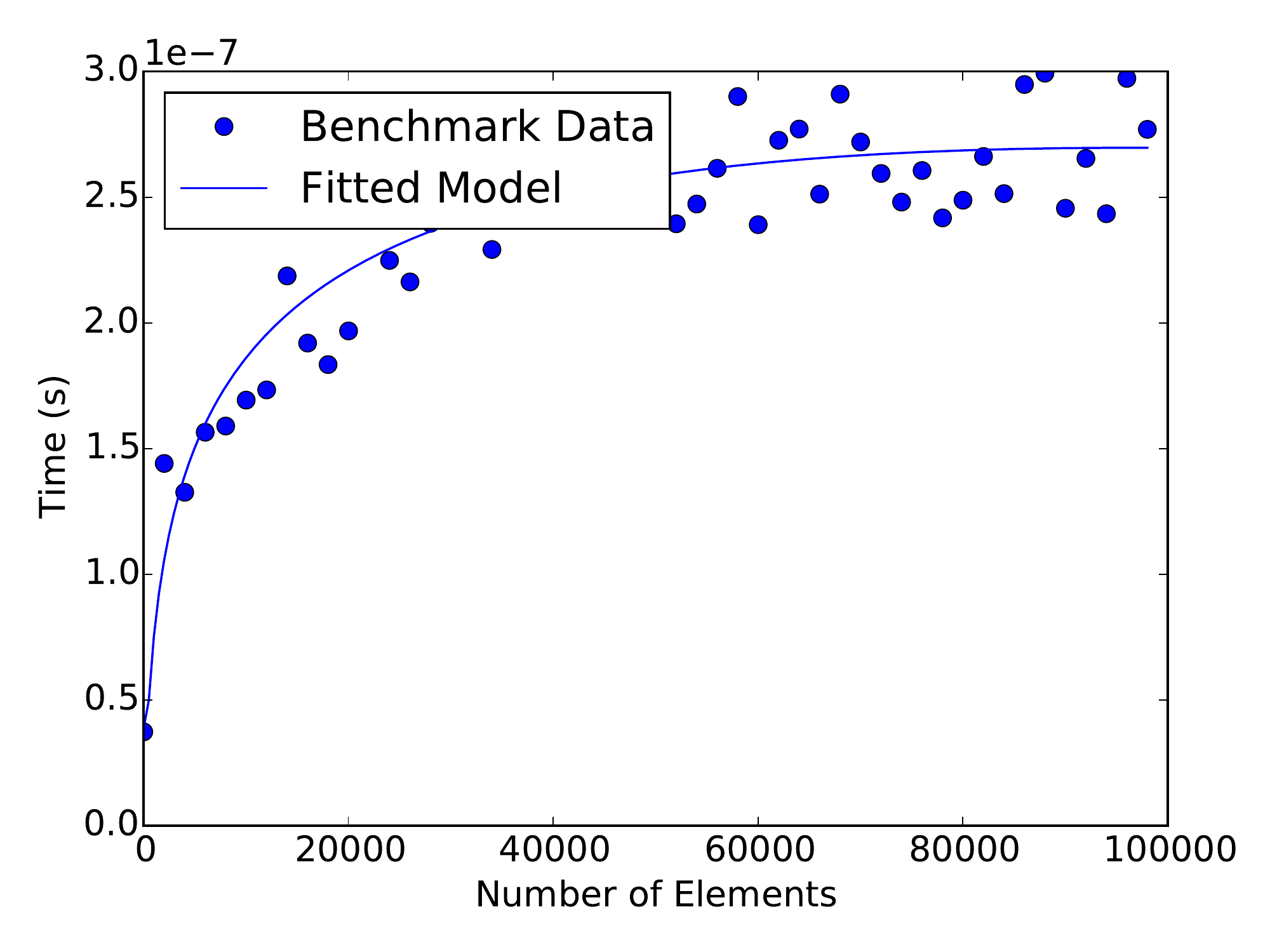}\end{center}
\caption{Example of Benchmark (9)}\label{fig:bench9}
\end{figure}

    \item \textbf{Interpolation Search} (ColumnStore)\\

    \textbf{Description:} This benchmark performs interpolation search over an
    array of keys. After, it returns the corresponding value.  \\

    \textbf{Model:} The benchmark is fit by a loglog-linear model: $f(x) = ax
    + b \log x + c \log \log x + d$ with a,b,c and d being learned
    coefficients. \\

{\center Benchmarking PseudoCode}\\
\doublerule

\begin{algorithmic}[1]
\REQUIRE array of $n$ keys, array of $n$ values, search val $x$.
\STATE low = 0, high = keys.size - 1
\STATE key\_low = keys[low], key\_high = keys[high]
\STATE diff = key\_high - key\_low
\WHILE{low $<$ high}
\STATE si = (high - low) * ((x - val\_low) / diff)
\IF {keys[si] $<$ x}
\STATE low = si + 1
\ELSIF {keys[si] == x}
\RETURN vals[si]
\ELSE
\STATE high = si
\ENDIF
\ENDWHILE
\RETURN NOT\_FOUND
\end{algorithmic}

\hrulefill \\

\textbf{Example of Learned Cost Model}: Figure \ref{fig:bench10} \\

\begin{figure}
\begin{center}\includegraphics[width=0.9\columnwidth]{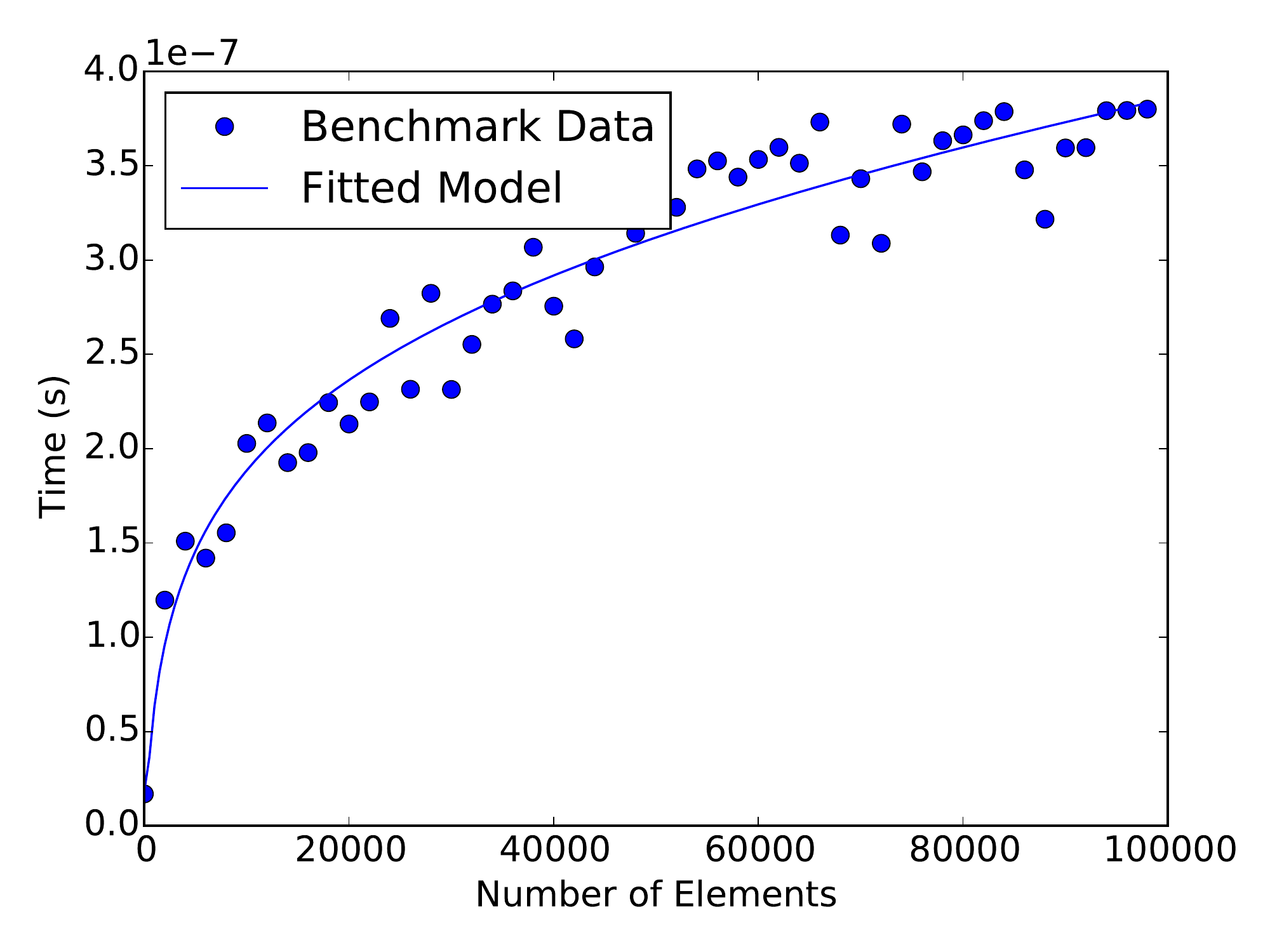}\end{center}
\caption{Example of Benchmark (10)}\label{fig:bench10}
\end{figure}

\item \textbf{Hash Probe (Multiply Shift)}\\

\textbf{Description:} This benchmark computes a hash value and probes the bucket of that hash value. This is used, for example, in hash partitioning. We note that in this model there are no collisions, which is why the benchmark is a hash probe and not a model of a hash table. 

We generate an array of length $k$, which is the number of elements needed to make a structure size of input parameter Structure Size. We call this array the probe array. Each element in the array stores a random number. A second array of length $n$, which is the number of trials we will perform, is created as well, with each element in that array also random. This array is called the ``scan array''. A full justification of why two arrays are needed can be seen in the description of the Random Memory Access Benchmark. 

To perform the benchmark we compute the necessary hash value for each trial. The hash value is dependent on 1) the value stored at the previously probed hash bucket and 2) the value in the scan array for the current trial. 1) is required so that the request for memory at trial n requires the value for trial n-1, so that the cpu can't issue multiple memory requests in parallel. 2) is required so that cycles don't occur in the access patterns for hash probing (which would throw off the expected cache latencies for large arrays of hash buckets). 

The previous paragraphs are true for all hash probe benchmarks. The following is true only for the multiply shift hash benchmark. 1) The input size must create a number of buckets which is a power of 2. 2) If $S$ is the size of the created hash array, let $s = \log_{2} S$. The equation for the multiply shift hash function is $$h(x) = a * x >> (64 - s)$$ where it is assumed that the processor uses 64 bit words. In the equation, $a$ is a randomly drawn odd integer. \\

\textbf{Model:} The model for the hash probe can be a sum of sigmoids or the weighted nearest neighbor model. See the random memory access benchmark for more details on training these models. 

{\center Benchmarking PseudoCode}\\
\doublerule

\begin{algorithmic}[1]
\REQUIRE pa: array of $k$ longs, sa: array of $k$ longs, i: numAccesses
\STATE fill each slot of pa with random x=Uniform(0,k-20)
\STATE fill each slot of sa with random x=Uniform(0,20)
\STATE x = 0, s = $log_{2} k$
\FOR{$i = 0, \dots n-1$}
\STATE x= [a * (pa[x] + sa[i])] >> (64 - s)
\ENDFOR
\STATE print x
\end{algorithmic}

\hrulefill \\

\textbf{Example of Learned Cost Model}: Figure \ref{fig:bench11} \\

\begin{figure}
\begin{center}\includegraphics[width=0.85\columnwidth]{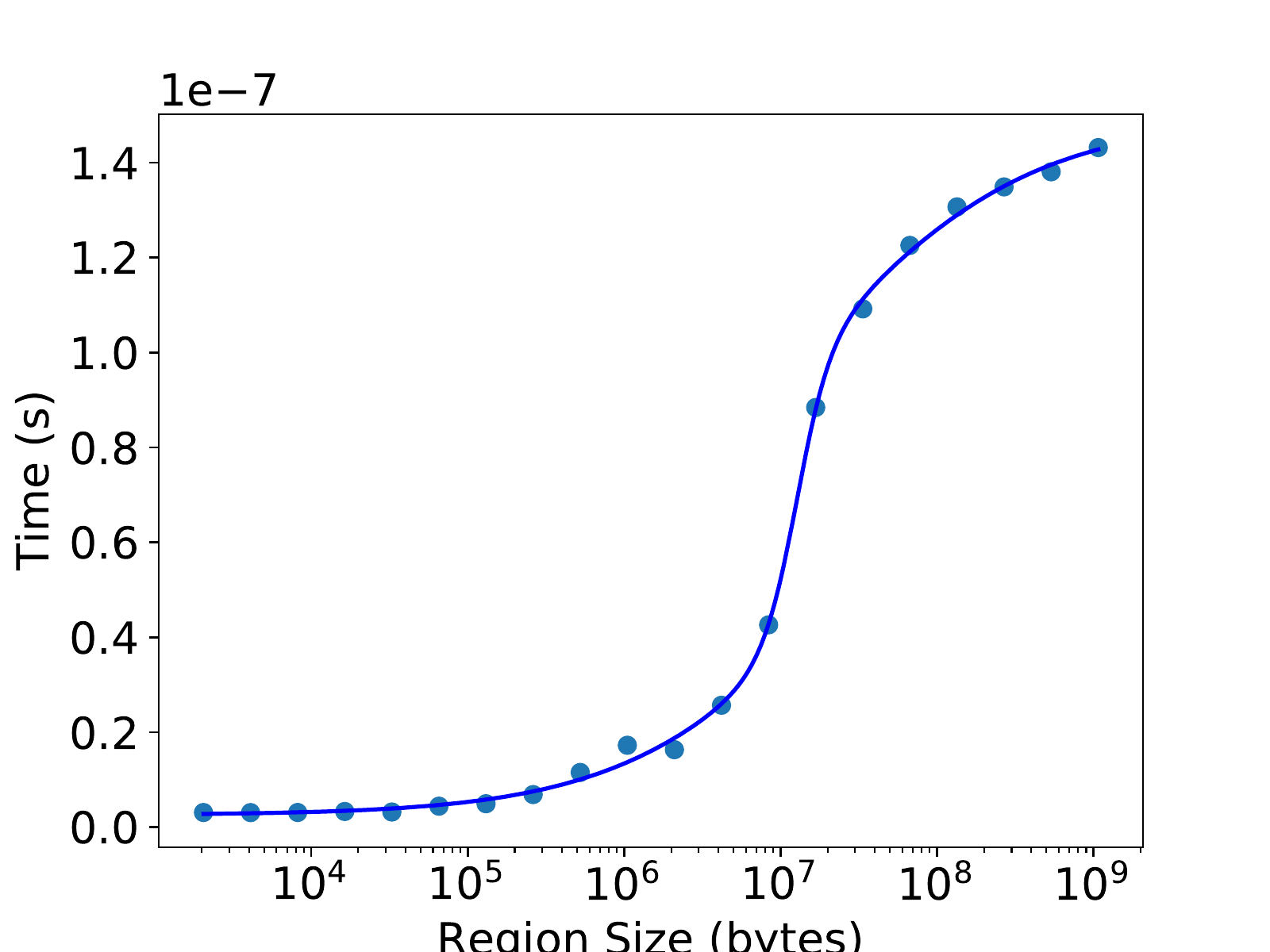}\end{center}
\caption{Example of Benchmark (11)}\label{fig:bench11}
\end{figure}

\item \textbf{Hash Probe (k-wise independent)}\\

\textbf{Description:} For a description of the common setup of all hash benchmarks, see the benchmark above. 

The following is true only for the 2-wise independent hash probe benchmark. 1) The input size must create a number of buckets which is a prime number. We will denote this number by $p$. 2) The equation for the multiply shift hash function is $$h(x) = (a * x + b) \% p$$ where it is assumed that the processor uses 64 bit words. In the equation, $a$ and $b$ are randomly drawn integers between 0 and $p$. \\

\textbf{Model:} The model for the hash probe can be a sum of sigmoids or the weighted nearest neighbor model. See the random memory access benchmark for more details on training these models. 

{\center Benchmarking PseudoCode}\\
\doublerule

\begin{algorithmic}[1]
\REQUIRE pa: array of $k$ longs, sa: array of $k$ longs, i: numAccesses
\STATE fill each slot of pa with random x=Uniform(0,k-20)
\STATE fill each slot of sa with random x=Uniform(0,20)
\STATE x = 0
\FOR{$i = 0, \dots n-1$}
\STATE x = [a * (pa[p] + sa[i]) + b] \% p
\ENDFOR
\STATE print x
\end{algorithmic}

\hrulefill \\

\textbf{Example of Learned Cost Model}: Figure \ref{fig:bench12} \\

\begin{figure}
\begin{center}\includegraphics[width=0.85\columnwidth]{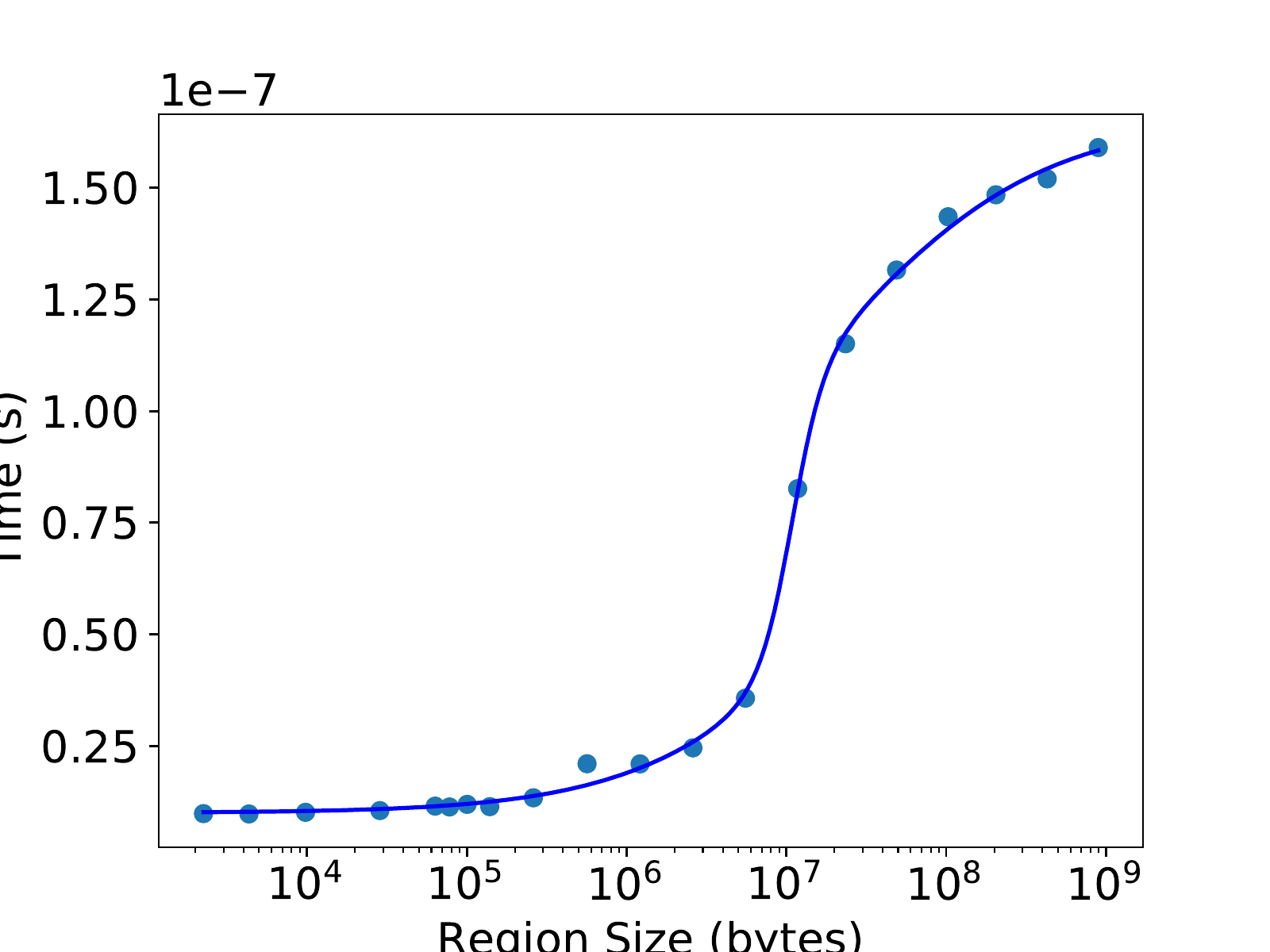}\end{center}
\caption{Example of Benchmark (12)}\label{fig:bench12}
\end{figure}

\item \textbf{Bloom Filter Probe (Muliply Shift)}\\

\textbf{Description:} This benchmark first builds a bloom filter by computing $k$ hash values for each entry and setting the bits at the computed $k$ hash values to 1.  It then probes the bloom filter using the same $k$ hash functions. In this benchmark, each of the $k$ hash functions is from the multiply shift family. 

There are two inputs the bloom filter benchmark, the size of the filter and the number of hash functions. For each of the $n$ trials, we randomly generated a value to probe for. 

The following describes the multiply shift hash function. 1) The bloom filter size must be a power of 2. 2) If $S$ is the size of the created hash array, let $s = \log_{2} S$. The equation for the multiply shift hash function is $$h(x) = a * x >> (64 - s)$$ where it is assumed that the processor uses 64 bit words. In the equation, $a$ is a randomly drawn odd integer. \\

\textbf{Model:} The model for the bloom filter is the sum of sum of sigmoids model or the weighted nearest neighbor model. The model is trained similarly to the sum of sigmoids model. More details on the model training can be found in the description for the random memory access benchmark. 

{\center Benchmarking PseudoCode}\\
\doublerule

\begin{algorithmic}[1]
\REQUIRE pa: array of $k$ longs, n: numAccesses, p: size
\STATE build bloom filter BF of size p, with p prime (build is similar to probe shown below)
\STATE fill each slot of pa with random x. 
\STATE count = 0
\FOR{$i = 0, \dots n-1$}
\STATE possiblyInSet = true
\FOR{$j = 0 \dots, k-1$}
\STATE hashBit = [a[j] * (pa[x])] >> (64 - s)
\STATE bitOffset = hashBit \& 7
\STATE notInSet = !((1 << bitOffset) \& BF[hashBit>>3])
\IF{notInSet}
\STATE count += 1
\ENDIF
\ENDFOR
\ENDFOR
\STATE print count
\end{algorithmic}

\hrulefill \\

\textbf{Example of Learned Cost Model}: The Bloom Filter Benchmark needs a two dimensional model. Figure \ref{fig:bench13} shows slices where the size is kept constant and where the number of hash functions is kept constant. \\

\begin{figure}
\begin{center}\includegraphics[width=0.45\columnwidth]{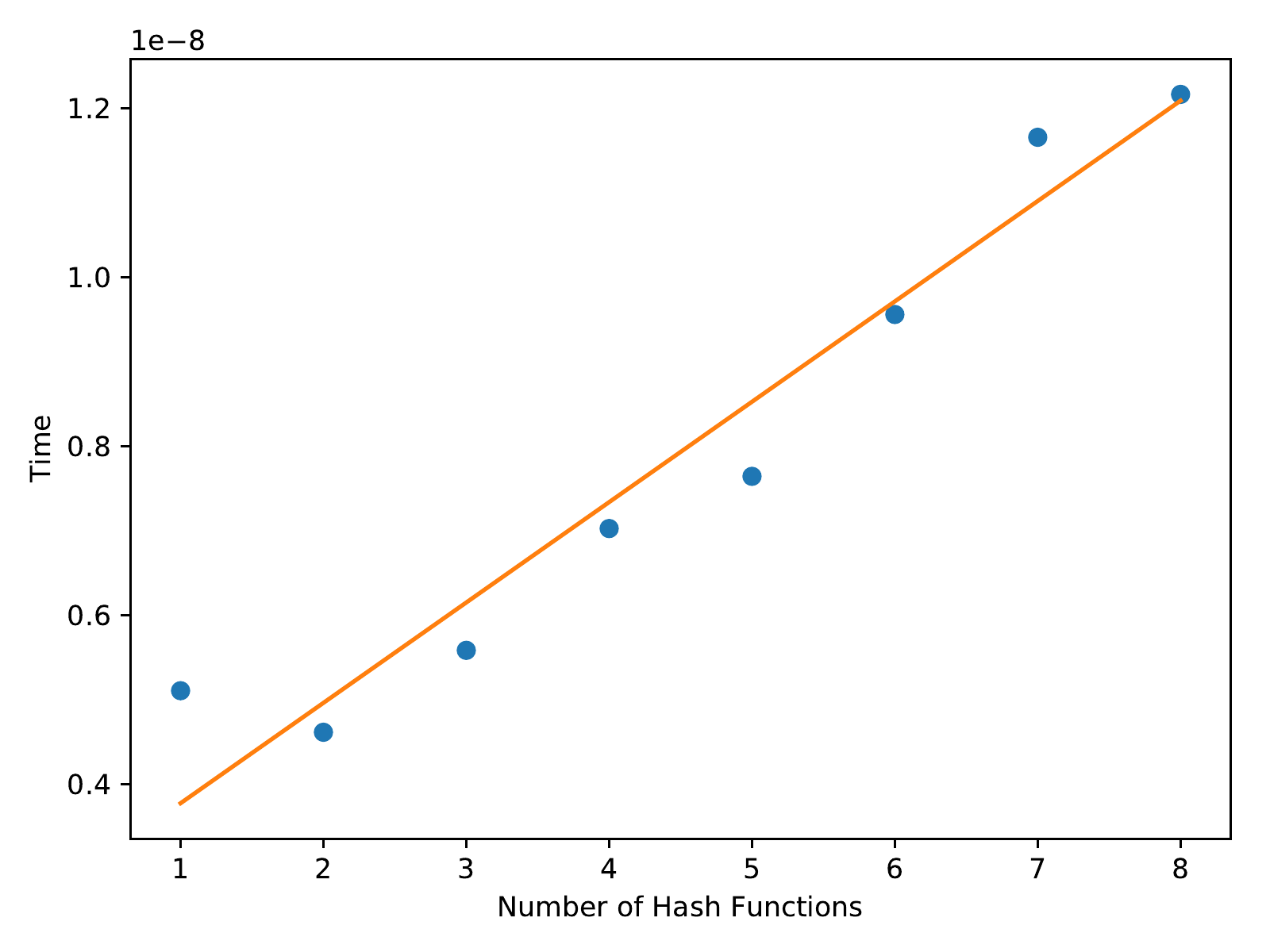}\includegraphics[width=0.45\columnwidth]{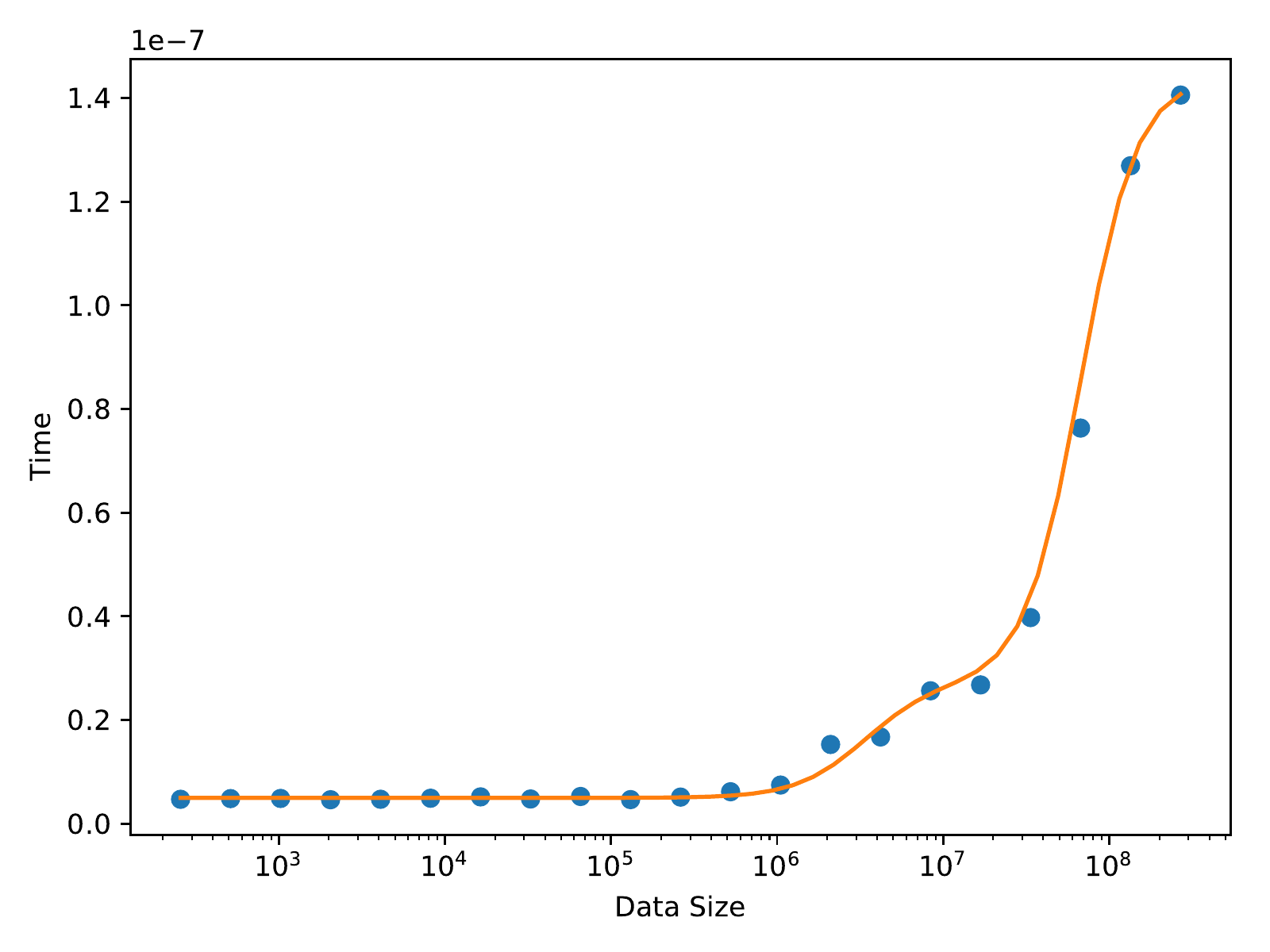}\end{center}
\caption{Example of Benchmark (13). On the left is varying $k$, size constant. On the right is varying size, $k$ constant. }\label{fig:bench13}
\end{figure}

\item \textbf{Bloom Filter Probe (k-wise independent)}\\

\textbf{Description:} This benchmark first builds a bloom filter by computing $k$ hash values for each entry and setting the bits at the computed $k$ hash values to 1.  It then probes the bloom filter using the same $k$ hash functions. In this benchmark, each of the $k$ hash functions is from the multiply shift family. 

There are two inputs the bloom filter benchmark, the size of the filter and the number of hash functions. For each of the $n$ trials, we randomly generated a value to probe for. 

The following describes the multiply shift hash function. 1) The bloom filter size must be a power of 2. 2) If $S$ is the size of the created hash array, let $s = \log_{2} S$. The equation for the multiply shift hash function is $$h(x) = a * x >> (64 - s)$$ where it is assumed that the processor uses 64 bit words. In the equation, $a$ is a randomly drawn odd integer. \\

\textbf{Model:} The model for the bloom filter is the sum of sum of sigmoids model or the weighted nearest neighbor model. The model is trained similarly to the sum of sigmoids model. More details on the model training can be found in the description for the random memory access benchmark. 

{\center Benchmarking PseudoCode}\\
\doublerule

\begin{algorithmic}[1]
\REQUIRE pa: array of $k$ longs, n: numAccesses, s: size
\STATE build bloom filter BF of size s (build is similar to probe shown below)
\STATE fill each slot of pa with random x. 
\STATE count = 0
\FOR{$i = 0, \dots n-1$}
\STATE possiblyInSet = true
\FOR{$j = 0 \dots, k-1$}
\STATE hashBit = [a[j] * pa[i])= + b[j]] \% p
\STATE bitOffset = hashBit \& 7
\STATE notInSet = !((1 << bitOffset) \& BF[hashBit>>3])
\IF{notInSet}
\STATE count += 1
\ENDIF
\ENDFOR
\ENDFOR
\STATE print count
\end{algorithmic}

\hrulefill \\

\textbf{Example of Learned Cost Model}: Figure \ref{fig:bench14} shows slices where the size is kept constant and where the number of hash functions is kept constant. \\

\begin{figure}
\begin{center}\includegraphics[width=0.45\columnwidth]{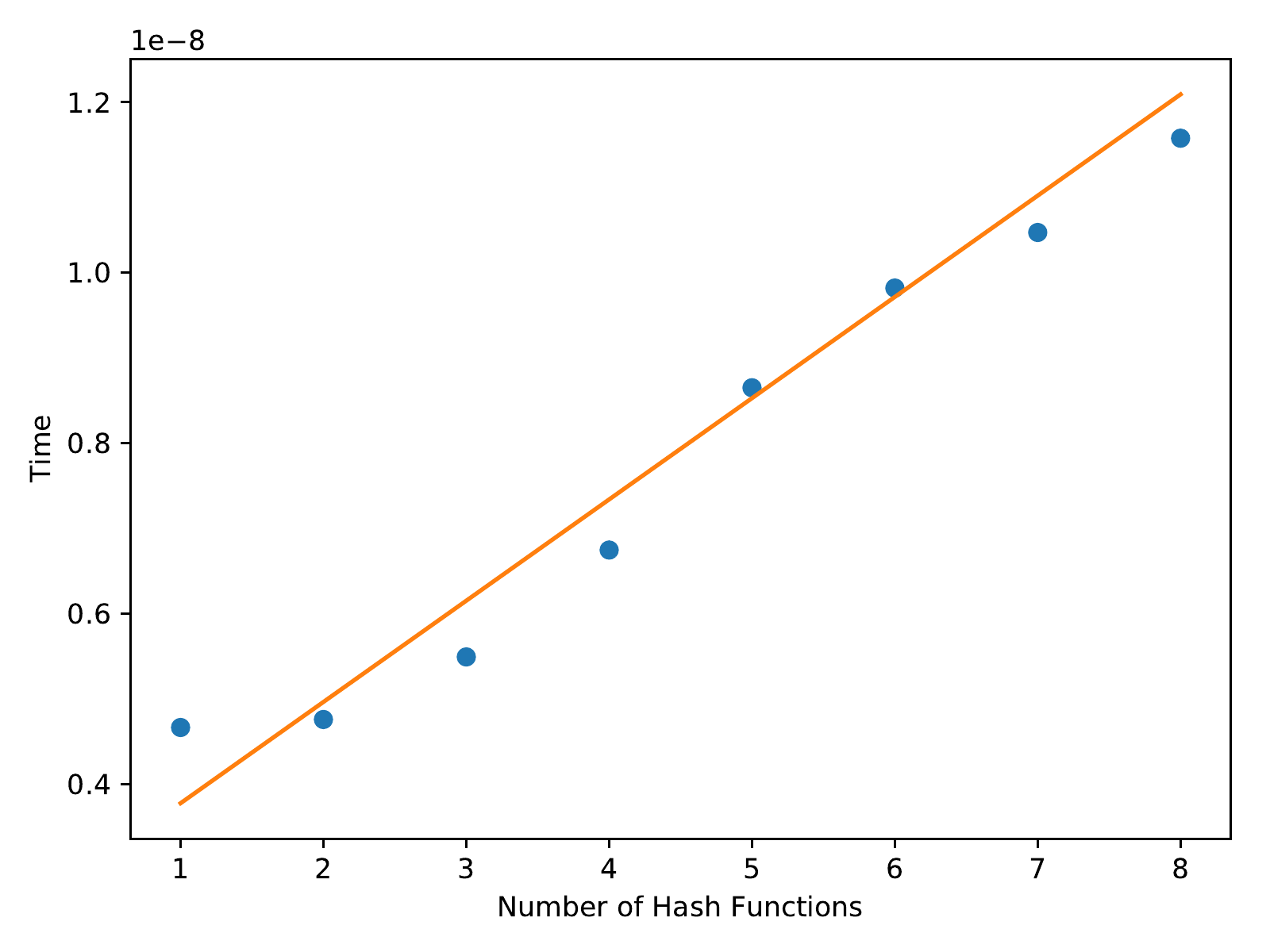}\includegraphics[width=0.45\columnwidth]{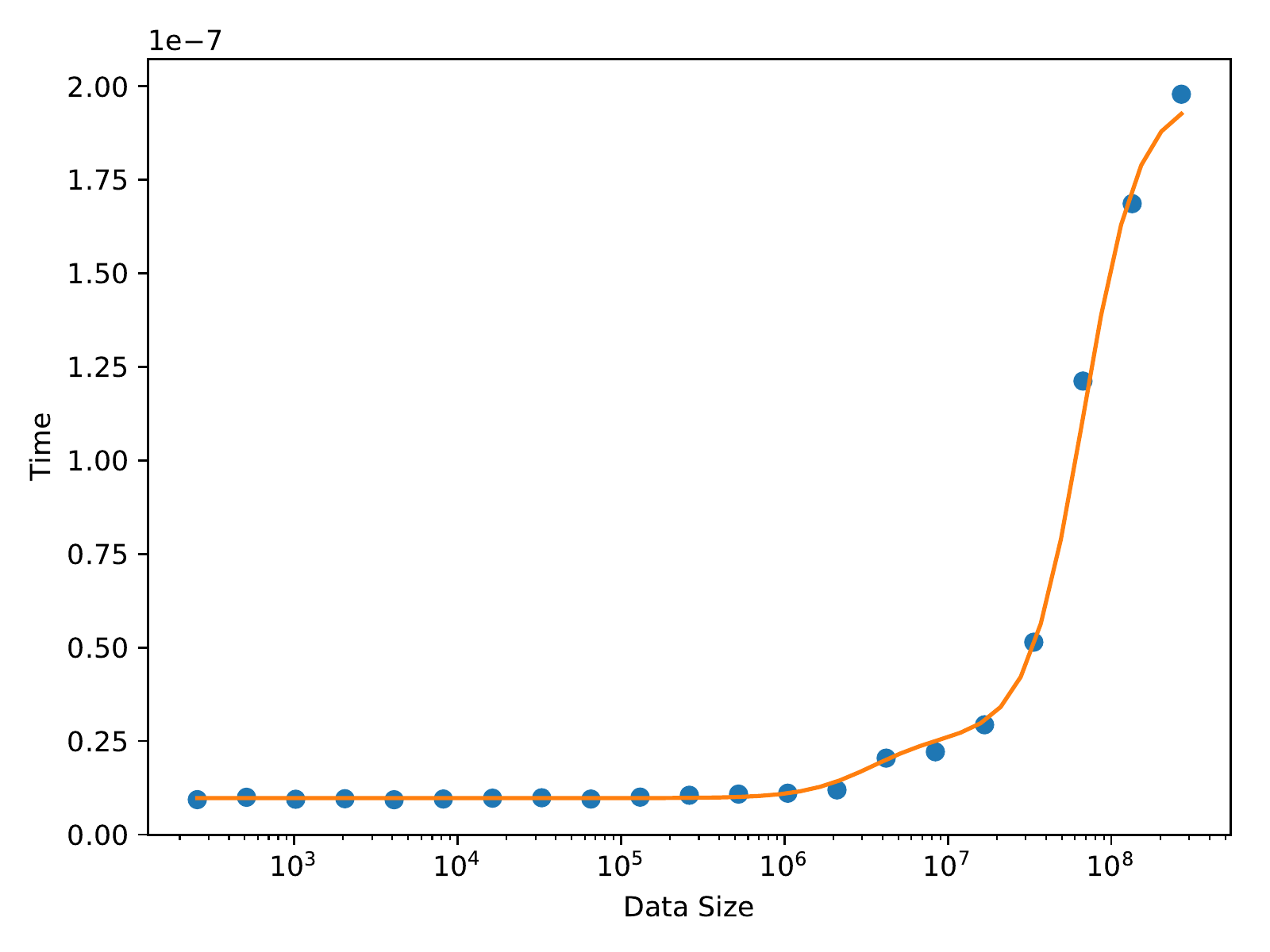}\end{center}
\caption{Example of Benchmark (14). On the left is varying $k$, size constant. On the right is varying size, $k$ constant. }\label{fig:bench14}
\end{figure}

\item \textbf{QuickSort}

\textbf{Description:} This benchmark performs QuickSort to sort a list of keys.  \\

\textbf{Model:} The benchmark is fit by a NLogN model. $f(x) = a x \log x
    + b x + c$ with a,b, and c being learned
    coefficients. \\

{\center Benchmarking PseudoCode}\\
\doublerule \\
QUICKSORT: 
\begin{algorithmic}[1]
\REQUIRE  array of $n$ key-value pairs
\IF {low < high}
\STATE p = partition(A, low, high)
\STATE quicksort (A, low, p - 1)
\STATE quicksort (A, p + 1, high)
\ENDIF
\end{algorithmic}

PARTITION:
\begin{algorithmic}[1]
\REQUIRE  array of $n$ key-value pairs, integers low, high
\STATE pivot = A[high]
\STATE $i = low - 1$ 
\FOR {$j = low, \dots, high - 1$}
\IF {A[j] < pivot}
\STATE i = i + 1
\STATE swap A[i] with A[j]
\ENDIF
\ENDFOR
\STATE swap A[i + 1] with A[hi]
\RETURN i
\end{algorithmic}

\hrulefill \\

\textbf{Example of Learned Cost Model}: Figure \ref{fig:bench15} \\

\begin{figure}
\begin{center}\includegraphics[width=0.9\columnwidth]{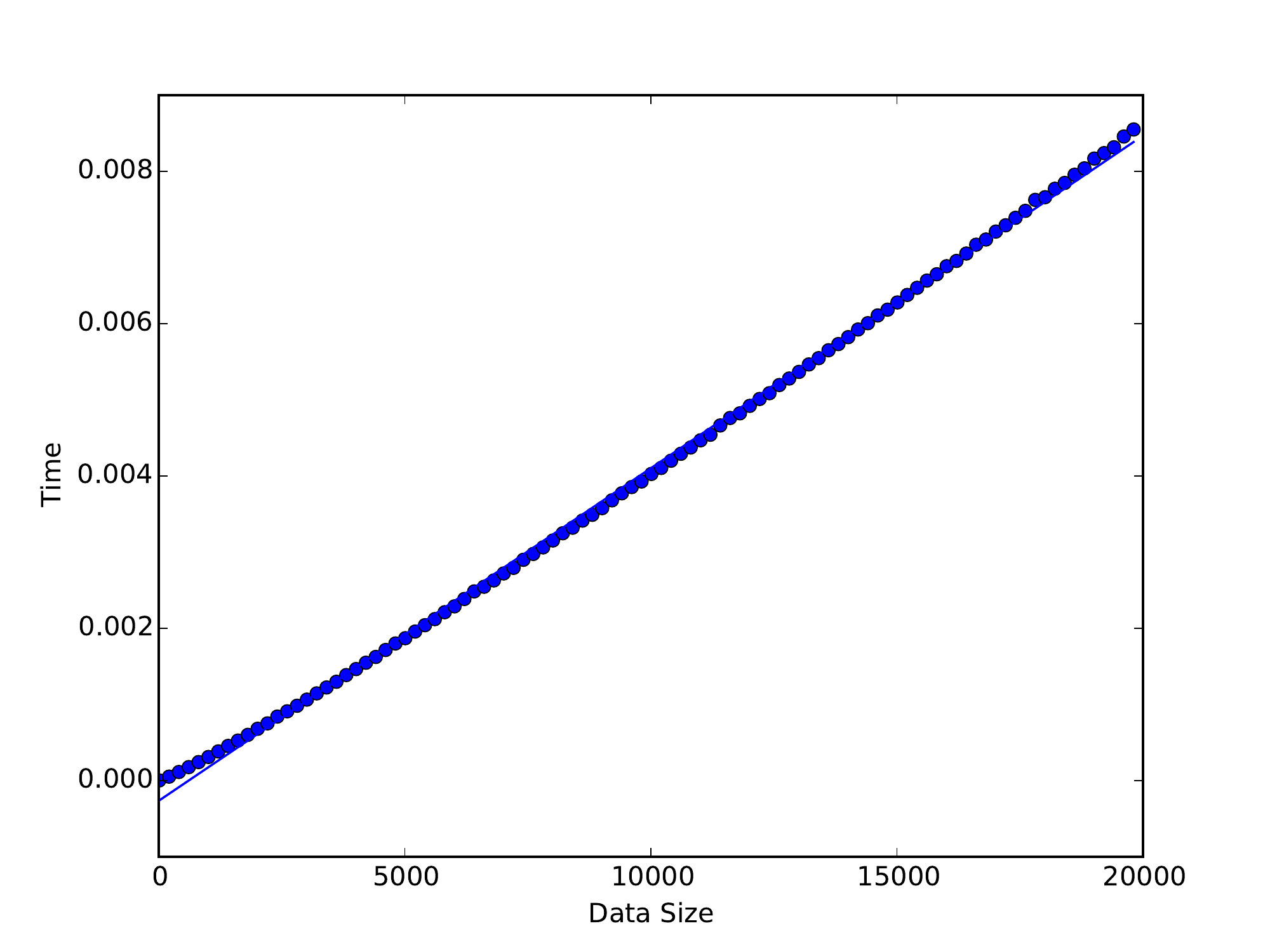}\end{center}
\caption{Example of Benchmark (15)}\label{fig:bench15}
\end{figure}

    \item \textbf{Random Memory Access}\\

    \textbf{Description:} This benchmark performs random memory accesses within
    a region.\\

    To do this, we first generate an array of values of length $k$ = region
    size / sizeof(long). This array will be the ``probe array'', or ``pa'' for
    short. Next, we generate a random number between 0 and $k - 20$ for each
    slot in the array. Additionally, we create a ``scan array'', or ``sa'', of
    $n$ longs which contains random numbers between 0 and $20$. We will be
    performing exactly $i$ random memory accesses, and so this second array is
    generally quite large. \\

    We will initialize a starting pointer value $p$ to be 0. At each round $i =
    0, \dots, n-1$, we will jump to $pa[p] + sa[i]$. At the end of all rounds,
    we will print $p$. This makes the compiler unable to optimize out the
    memory accesses, and is a negligible overhead to the benchmark. The
    combination of two random numbers, one between [0,k-20] and the other
    between [0,20], creates a nearly uniform distribution over [0,k] (the first
    and last 20 slots have slightly lower chances of being picked via a
    combinatorial argument similar to why 7s are the most common dice roll).
    The embedding of one part of the next slot to go to into the probe array pa
    prevents the processor from being able to fetch the next memory location
    before the data arrives. The embedding of a second random part into the
    scan array sa is to prevent cycles in the probing pattern, which would
    occur if we just used the probe array for the next slot.\\

    \textbf{Model:} The benchmark is fit by a sum of sigmoids model: $f(x) =
    \sum_{i=1}^{k} \frac{c_{i}}{1 + e^{-k_{i} (\log x - x_{i})}} +  y_{0}$
    with all coefficients learned as usual. This model is no longer convex in
    its L2 loss and cannot be learned via off the shelf techniques such as
    least squares regression. Instead, we first preprocess the data by
    calculating rates of change for intervals of some size $z$. We find the
    $k$ highest local maxima for this graph of rates of change and set these
    as our initial guesses for the $x_{i}$ values. For the values of the
    $k_{i}, c_{i}$ we initialize them to random positive numbers between $0$
    and $1$. Our initial guess for $y_{0}$ is just the first point. At this
    point, we use scipy's curve fit package to train the parameters and
    observe the results.\\

    The learning of the coefficients for the sum of sigmoid model requires
    good initial conditions for the guesses of the $x_{i}$. It is not
    sensitive to initial guess for the parameters of $k_{i}$ or $c_{i}$. We
    found that for each of $z = 0.1,0.5$, and $1$, the initial guesses for the
    $x_{i}$ were good enough so that the fitted model produces accurate
    results.\\

    Additionally, the user can select a different model for the Random Memory
    Access Benchmark should they so choose. We currently support k nearest
    neighbor regression. In the future, we plan to add support for both locally
    weighted linear regression and Gaussian Process Regression. All these
    techniques are non-parametric and so don't need training, and all three
    techniques produce good predictions given enough data. As a tradeoff, these
    models are significantly slower in producing predictions and are not
    interpretable in any way. This contrasts with the sum of sigmoids model,
    whose values of the $x_{i}$ tell you the cache boundaries and whose values
    $c_{i}$ tell you the difference in memory access latency between each
    subsequent level of cache.\\

{\center Benchmarking PseudoCode}\\
\doublerule

\begin{algorithmic}[1]
\REQUIRE pa: array of $k$ longs, sa: array of $k$ longs, i: numAccesses
\STATE fill each slot of pa with random x=Uniform(0,k-20)
\STATE fill each slot of sa with random x=Uniform(0,20)
\STATE p = 0
\FOR{$i = 0, \dots n-1$}
\STATE p = pa[p] + sa[i]
\ENDFOR
\STATE print p
\end{algorithmic}

\hrulefill \\

\textbf{Example of Learned Cost Model}: Figure \ref{fig:bench16} \\

\begin{figure}
\begin{center}\includegraphics[width=0.9\columnwidth]{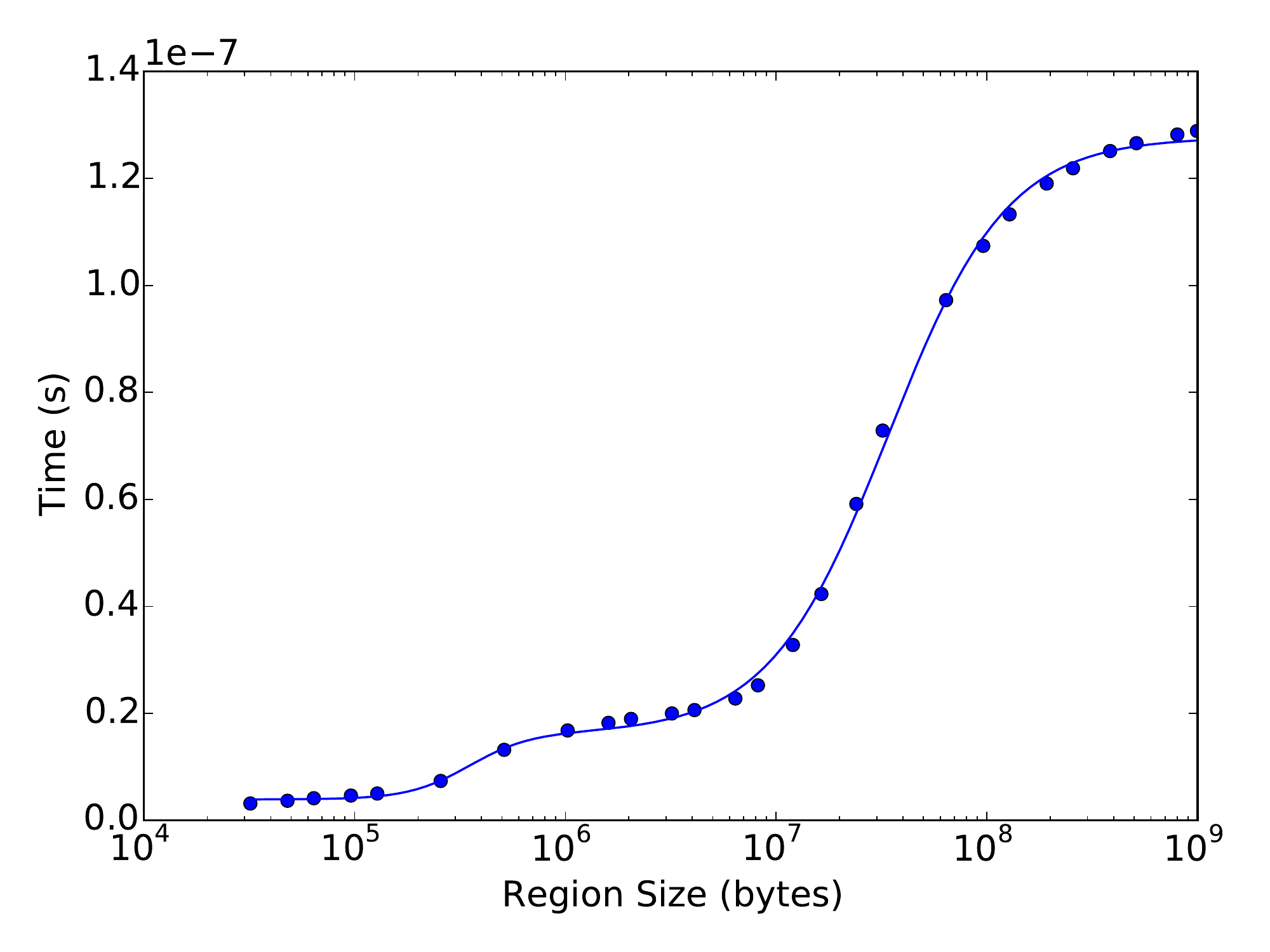}\end{center}
\caption{Example of Benchmark (16)}\label{fig:bench16}
\end{figure}

\item \textbf{Batched Random Memory Access}\\

\textbf{Description: } The batched random memory access benchmark is similar to the random memory access benchmark, except that now there is nothing to prevent the test from batching memory requests. The result is that the time per operation drops dramatically. \\

\textbf{Model: } The batched random memory access benchmark is fit by the weighted nearest neighbors model and the sum of sigmoids model. The training is similar to the random memory access benchmark. 

{\center Benchmarking PseudoCode}\\
\doublerule

\begin{algorithmic}[1]
\REQUIRE pa: array of $k$ longs, sa: array of $k$ longs, i: numAccesses
\STATE fill each slot of pa with random x=Uniform(0,k)
\STATE fill each slot of sa with random x=Uniform(0,k)
\STATE p = 0
\FOR{$i = 0, \dots n-1$}
\STATE p += pa[sa[i]]
\ENDFOR
\STATE print p
\end{algorithmic}

\hrulefill \\

\textbf{Example of Learned Cost Model}: Figure \ref{fig:bench17} \\

\begin{figure}
\begin{center}\includegraphics[width=0.9\columnwidth]{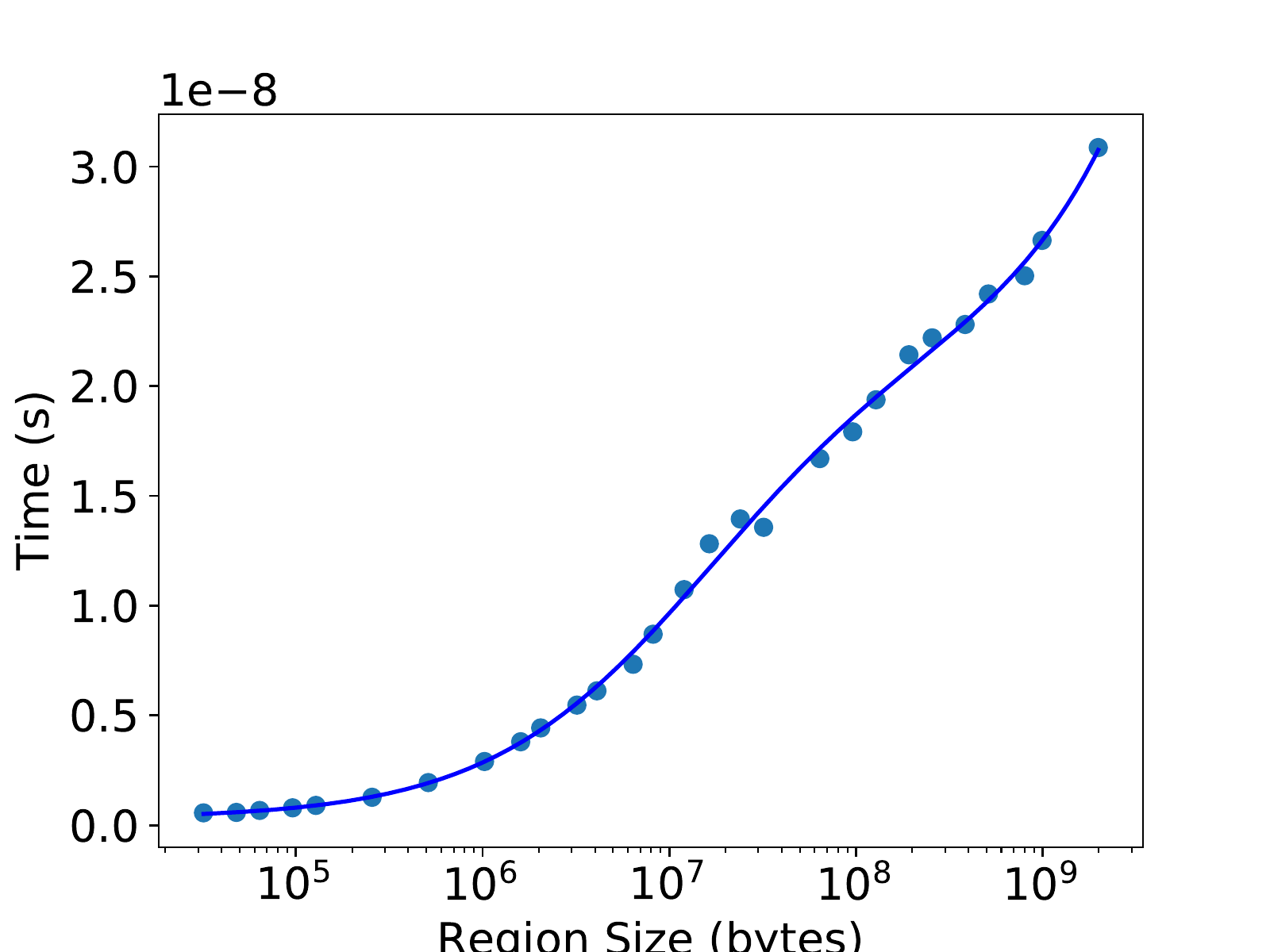}\end{center}
\caption{Example of Benchmark (16)}\label{fig:bench17}
\end{figure}

\textbf{Final Notes: } Currently the technical report contains 17 out of the 24 Level 2 Access
Primitives. We will be adding the remaining 7.

\end{enumerate}

\begin{table*}[t]
 \footnotesize
\begin{tabular}{| x{0.08 cm} | x{3.5 cm} | x{2.43 cm} |  x{4.2 cm} | x{2.5 cm} |}
\hline
\multicolumn{5}{| c |}{\normalsize{\textbf{Data Access Primitives and Fitted Models}}} \tn \hline
& \textbf{Data Access Primitives Level 1} \newline(required parameters ; optional parameters) & \textbf{Model Parameters} & \textbf{Data Access Primitives Layer 2}  & \textbf{Fitted Models} \tn \hline
1 & \textbf{Scan} & Data Size & Scalar Scan (RowStore, Equal) &  Linear Model (1) \tn \cline{1-1}\cline{4-5}
2 & (Element Size, Comparison,& & Scalar Scan (RowStore, Range) & Linear Model (1) \tn \cline{1-1}\cline{4-5}
3 & Data Layout; None) & & Scalar Scan (ColumnStore, Equal) & Linear Model (1) \tn \cline{1-1}\cline{4-5}
4 & & & Scalar Scan (ColumnStore, Range) & Linear Model (1) \tn \cline{1-1}\cline{4-5}
5 & & & SIMD-AVX Scan (ColumnStore, Equal) & Linear Model (1) \tn \cline{1-1}\cline{4-5}
6 &  & & SIMD-AVX Scan (ColumnStore, Range) & Linear Model (1) \tn \hline
7 & \textbf{Sorted Search} & Data Size & Binary Search (RowStore) & Log-Linear Model (2)\tn \cline{1-1}\cline{4-5}
8 & (Element Size, Data Layout; ) & & Binary Search (ColumnStore) & Log-Linear Model (2)\tn \cline{1-1}\cline{4-5}
9 & & & Interpolation Search (RowStore) &  Log $+$ LogLog Model (3) \tn \cline{1-1}\cline{4-5}
10 &  & & Interpolation Search (ColumnStore) & Log $+$ LogLog Model (3) \tn \hline
11 & \textbf{Hash Probe} \newline (; Hash Family) & Structure Size & {\color{black} Linear Probing (Multiply-shift \cite{Dietzfelbinger97})} & Sum of Sigmoids (5), Weighted Nearest Neighbors (7) \tn \cline{1-1}\cline{4-5}
12 & & &{\color{black} Linear Probing (k-wise independent, k=2,3,4,5)} & Sum of Sigmoids (5), Weighted Nearest Neighbors (7)\tn \hline
13 & \textbf{Bloom Filter Probe} \newline(; Hash Family) & Structure Size, Number of Hash Functions & {\color{black}Bloom Filter Probe (Multiply-shift \cite{Dietzfelbinger97})} & Sum of Sum of Sigmoids (6), Weighted Nearest Neighbors (7)\tn \cline{1-1}\cline{4-5}
14 & & & {\color{black}Bloom Filter Probe (k-wise independent, k=2,3,4,5)} & Sum of Sum of Sigmoids (6), Weighted Nearest Neighbors (7) \tn \hline
15 & \textbf{Sort} & Data Size & QuickSort & NLogN Model (4) \tn \cline{1-1}\cline{4-5}
16 & (Element Size; Algorithm) & & MergeSort & NLogN Model (4) \tn \cline{1-1}\cline{4-5}
17 & & & ExternalMergeSort & NLogN Model (4) \tn \hline
18 & \textbf{Random Memory Access} & Region Size & Random Memory Access & Sum of Sigmoids (5), Weighted Nearest Neighbors (7)\tn \hline
19 & \textbf{Batched Random Memory Access} & Region Size & Batched Random Memory Access & Sum of Sigmoids (5), Weighted Nearest Neighbors (7)\tn \hline
20 & \textbf{Unordered Batch Write} & Write Data Size & Contiguous Write (RowStore) & Linear Model (1) \tn \cline{1-1} \cline{4-5}
21 & (Layout; ) &  & Contiguous Write (ColumnStore) & Linear Model (1) \tn \hline
22 & \textbf{Ordered Batch Write} & Write Data Size, & Batch Ordered Write (RowStore) & Linear Model (1) \tn \cline{1-1} \cline{4-5}
23 & (Layout; ) & Data Size & Batch Ordered Write (ColumnStore) & Linear Model (1)\tn \hline
24 & \textbf{Scattered Batch Write} & Number of Elements, Region Size & ScatteredBatchWrite & Sum of Sum of Sigmoids (6), Weighted Nearest Neighbors (7) \tn \hline
\end{tabular}
\begin{tabular}{| x{0.08 cm} | x{2 cm} | x{6.5 cm} | x{4.5 cm} |}
\multicolumn{4}{| c |}{\normalsize{\textbf{Models used for fitting data access primitives}}} \tn \hline
& \textbf{Model} & \textbf{Description} & \textbf{Formula} \tn \hline
1 & Linear & Fits a simple line through data & $f(\mathbf{v}) = \mathbf{w^{\top}\phi(v)} + y_{0}; \phi(v) = (v)$ \tn \hline
2 & Log-Linear & Fits a linear model with a basis composed of the identity and logarithmic functions plus a bias & $f(\textbf{v}) = \mathbf{w^{\top}\phi(v)} + y_{0}; \phi(v) = \begin{pmatrix} v \tn \log v \end{pmatrix}$ \tn \hline
3 & Log + LogLog & Fits a model with $\log, \log \log,$ and linear components & $f(\textbf{v}) = \mathbf{w^{\top}\phi(v)} + y_{0}; \phi(v) = \begin{pmatrix} v \tn \log v \\ \log \log v \end{pmatrix}$\tn \hline
4 & NLogN & Fits a model with primarily an NLogN and linear component & $f(\textbf{v}) = \mathbf{w^{\top}\phi(v)} + y_{0}; \phi(v) = \begin{pmatrix} v\log v \\ v \end{pmatrix}$ \tn \hline
5 & Sum of Sigmoids & Fits a model that has $k$ approximate steps. Seen as sigmoids in $\log x$ as this fits various cache behaviors nicely & $f(x) = \sum_{i=1}^{k} \frac{c_{i}}{1 + e^{-k_{i} (\log x - x_{i})}} +  y_{0}$\tn \hline
6 & Sum of Sum of Sigmoids & Fits a model which has two cost components, both of which have $k$ approximate steps occuring at the same locations. & $f(x, m) = \sum_{i=1}^{k} \frac{c_{i1}}{1 + e^{-k_{i} (\log x - x_{i})}} + \newline(m-1)(\sum_{i=1}^{k} \frac{c_{i2}}{1 + e^{-k_{i} (\log x - x_{i})}} + y_{1}) + y_{0} $\tn \hline
7 & Weighted Nearest Neighbors & Takes the $k$ nearest neighbors under the $l_{2}$ norm and computes a weighted average of their outputs. The input $x$ is allowed to be a vector of any size. & Let $x_{1}, ... x_{k}$ be the nearest neighbors of $x$ with costs $y_{1}, \dots, y_{k}$. Then $f(x) = \frac{1}{\sum_{i=1}^{k} \frac{1}{d(x,x_{i})}} \sum_{i=1}^{k} \frac{1}{d(x,x_{k})} y_{k}$ \tn \hline
\multicolumn{4}{| p{13.08 cm} |}{Notation: $f$ is a function, \textbf{v} is a vector, and x, m are scalars. $\log(\mathbf{v})$ returns a vector with $\log$ applied on an element by element basis to $\mathbf{v}$; i.e. if $v = \begin{pmatrix} v_{1} \\ v_{2} \end{pmatrix}$, then $\log v = \begin{pmatrix} \log v_{1} \\ \log v_{2} \end{pmatrix}$. Finally, if we have vectors $v^{(1)}, v^{(2)}$ of lengths $n,m$ stacked on each other as $\begin{pmatrix} v^{(1)} \tn v^{(2)} \end{pmatrix}$, then this signifies the $n + m$ length vector produced by stacking the entries of $v^{(1)}$ on top of the entries of $v^{(2)}$; i.e. $\begin{pmatrix} v^{(1)} \tn v^{(2)} \end{pmatrix} = \begin{pmatrix} v^{(1)}_{1}, \dots, v^{(1)}_{n}, v^{(2)}_{1}, \dots, v^{(2)}_{m} \end{pmatrix}^{\top}$.} \tn \hline
\end{tabular}
\caption{\ch{Data access primitives and models used for operation cost synthesis.}}
\label{table:accessPrimitives}
\end{table*}


\newpage
\section{Cost Synthesis}
\label{sec:cs}

\begin{figure*}[th!]
    \centering
    \includegraphics[width=0.95\textwidth]{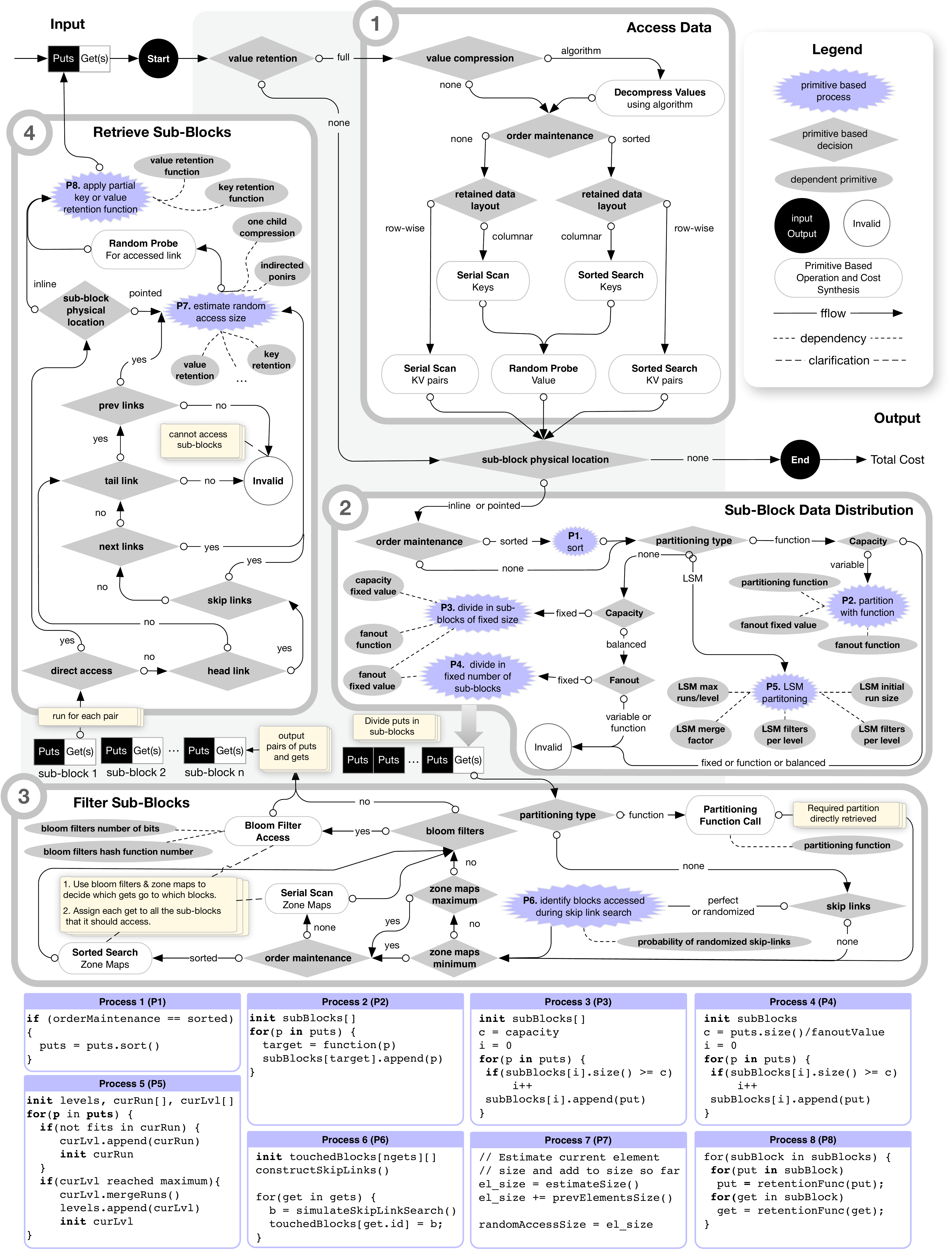}
    \caption{\label{fig:es}Expert System for get cost synthesis}
\end{figure*}
Given a data layout specification and a workload, the Data Calculator uses Level 1 access primitives to synthesize operations and subsequently each Level 1 primitive is translated to the appropriate Level 2 primitive to compute the cost of the overall operation.
This process relies on an expert system which captures fundamental properties of how basic layout principles are expected to behave and which access (high-level) algorithms are expected to best utilize each layout. 
In the main part of the paper, we presented a more high level view of the expert system. Figure~\ref{fig:es} now provides a more detailed look into the internals of the expert system. 
Each empty rounded shape in Figure~\ref{fig:es} represents a Level 1 access primitive call, each diamond a decision, and each dark round shape a data access primitive.
Finally, blue star shaped (marked P1-P8) represent additional processes 
described at the bottom of Figure~\ref{fig:es}. Together they form the complete expert system for algorithm and cost synthesis for the Get operation.

\section{Input Examples}
\label{sec:inex}
Here we provide more examples of data structure input specifications.
The complete set of elements used in the experiments are defined in~Table~\ref{specs}.
Specifically, we list the definition of all 8 elements used in the paper: Unordered Data Page (UDP), Ordered Data Page (ODP), Hash partitioning (Hash),
Range partitioning (Range), Trie node, B+Tree internal node (B+), Linked List (LL), Skip List (SL).

\begin{figure*}[htp]
\centering
\includegraphics[trim={0 5cm 5cm 0},width=1.8\columnwidth]{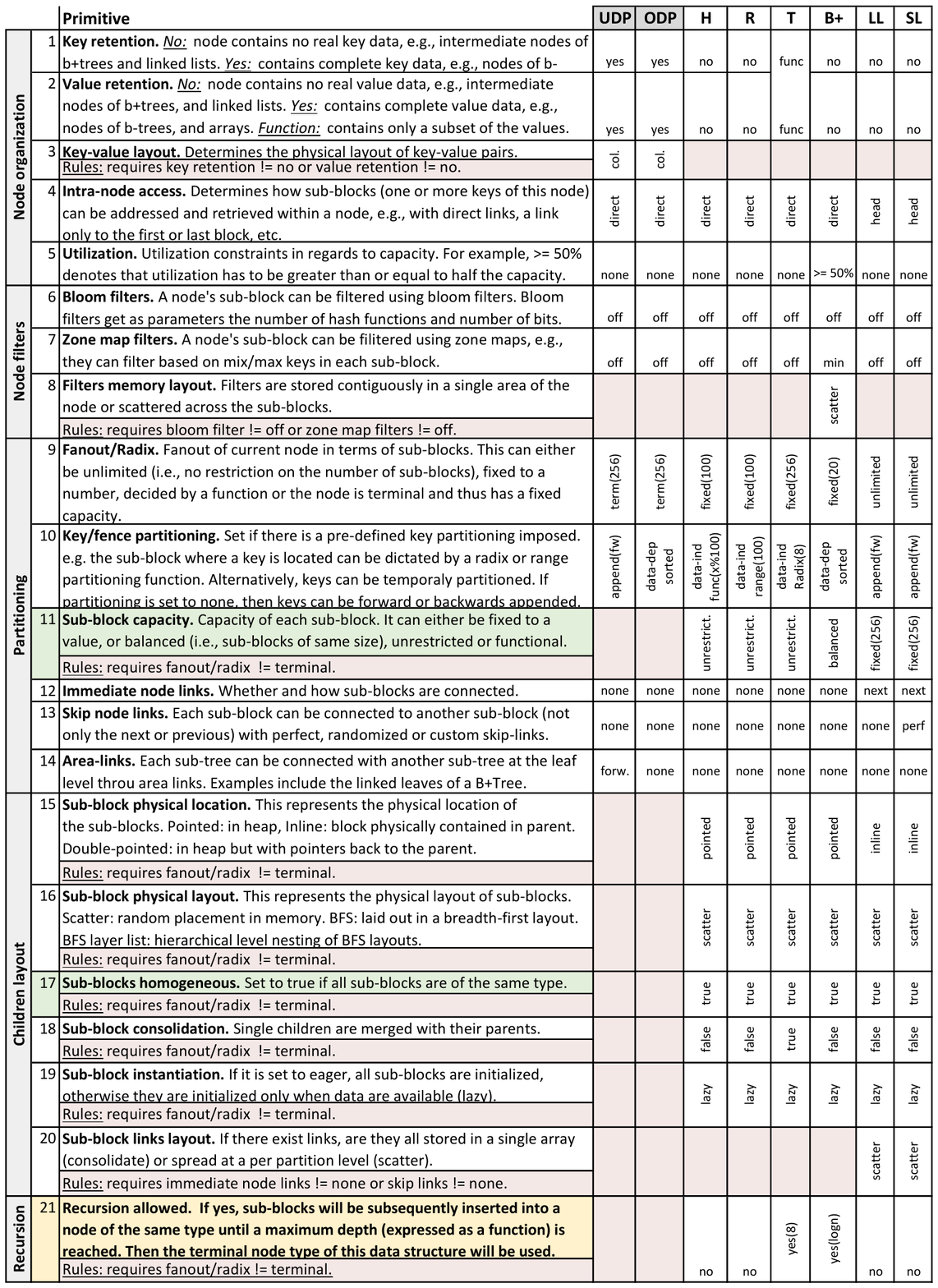}
\caption{Data Layout Specifications\label{specs}}
\end{figure*}

Then full data structures are defined as hierarchies of those elements. We use the following notation: A $\rightarrow$ B, which means that
all sub-blocks of element A are of type B.
Dots represent multiple levels in the hierarchy.
Below we describe all full data structures used in the experiments using this notation. 

\begin{itemize}
    \item \textbf{Linked List: } LL $\rightarrow$ UDP
    \item \textbf{Array: }  UDP with fixed capacity value equal to the number of puts.
    \item \textbf{Range Partitioned Linked List: } Range $\rightarrow$ LL $\rightarrow$ UDP
    \item \textbf{Skip-List: } SL $\rightarrow$ UDP
    \item \textbf{Trie: } Trie $\rightarrow \ldots \rightarrow$ Trie $\rightarrow$ UDP
    \item \textbf{B+Tree: } B+ $\rightarrow \ldots \rightarrow$ B+  $\rightarrow$ ODP
    \item \textbf{Sorted Array: } ODP with fixed capacity value equal to the number of puts.
    \item \textbf{Hash Table: } Hash $\rightarrow$ LL $\rightarrow$ UDP (with fixed capacity value 5)
\end{itemize}

The number of levels an actual instance of a data structure would have depends on the size of the dataset for the case of the B+Tree, and on the size of the keys for the Trie. 

\section{Output Examples}
\label{sec:outex}
In this section, we give examples of the raw output of the Data Calculator. This includes both cases where the output is a calculation of the response time and cases where the output is the design of a new data structure.

\subsection{Output Performance Break Down}
We first present a representative cost breakdown for performing 1 Get query in 100,000 inserts.
These numbers correspond to Figure 6. The cost for each operation is broken down in 3 Level 1 access primitives: $P$ stands for random probe, $S$ for serial scan, and $B$ for binary search.
The input in each function is the size of the data to be accessed.

\begin{itemize}
    \item \textbf{Linked List: } $P(782) + 6P(200974)+ 5S(256) + S(141) + P(256)$   \\  1 probe for getting the head of the linked list, 6 random probes for fetching the first 6 pages (since the answer is found on the 6-th page), 5 full scans for the first 5 pages, 1 scan until 141 (the answer) and 1 probe for getting the value (data is stored in a columnar format).
    \item \textbf{Array: } $ S(1421) + P(100000)$ \\ 1 serial scan of the keys until the record is found (position 1421), and a random probe in the whole array of values (data is stored in a columnar format) to find the values.
    \item \textbf{Range Partitioned Linked List: } $ P(842) + P(216394) + S(185) + P(256) $ \\ 1 probe for the head of the data structure, 1 probe for the linked list head of the given range, one scan on the first page of the linked list and one probe for the value in the first page of the linked list (data is stored in a columnar format).
    \item \textbf{Skip-List: } $ P(1181) + 201P(391) + P(201373) + B(256) + P(256)$ \\ Various probes as we navigate in the skip-list checking the zone maps, one binary search when the correct page is found, followed by a probe for the value.
    \item \textbf{Trie: }  $P(256) + P(512)+ P(768) + P(1024) + P(1280) + P(1536) + P(2048) + P(102144) + P(228628) + P(32608532) + S(185)  +  P(256)$ \\ Multiple probes as we navigate the trie, a serial scan on the target page, followed by a probe for getting the value.
    \item \textbf{B+Tree: } $2B(1400) + P(205640) + P(40) + P(840) + B(256) + P(256)$ \\ Multiple probes as we navigate the tree, followed by a binary search for getting the key, and a probe for getting the value.
    \item \textbf{Sorted Array: } $B(100000) + P(100000)$  \\ Binary search for finding the key, probe for finding the value.
    \item \textbf{Hash Table: } $ P(39718) + P(10207526) + S(2) + P(5)$ \\ Probe for the head, probe for the linked list, serial scan first page of linked list, probe for value.
\end{itemize}

\subsection{Output Design Example }
We now discuss design output. In the experiment of Figure 9, the Data Calculator designed new data structures given a workload. For the first scenario (Figure 9 left side) the Data Calculator computed a design where a hashing element at the upper levels of the hierarchy allows to quickly access data but then data is split between the write and read intensive parts of the domain to simple unsorted pages (like a log) and B+tree -style indexing for the read intensive part. For the second scenario (right side of Figure 9), the Data Calculator produces a design which similarly to the previous one takes care of read and writes separately but this time also distinguishes between range and point gets by allowing the part of the domain that receives point queries to be accessed with hashing and the rest via B+tree style indexing.

In both scenarios the following elements were chosen by the Data Calculator: Hash Partitioning, B+Tree Internal, and Ordered Data Page. Below, we list all specifications as generated by the system in JSON format and blended into a full designs as shown in Figure 9.

\subsubsection{Hash Partitioning Element}
{\scriptsize
\begin{verbatim}
{
  "external.links.next": false,
  "external.links.prev": false,
  "inter-block.blockAccess.direct": true,
  "inter-block.blockAccess.headLink": false,
  "inter-block.blockAccess.tailLink": false,
  "inter-block.fanout.fixedValue": 100,
  "inter-block.fanout.function": "",
  "inter-block.fanout.type": "fixed",
  "inter-block.orderMaintenance.type": "none",
  "inter-block.partitioning.function": "KeyMod(100)",
  "inter-block.partitioning.logStructured.filtersPerLevel": false,
  "inter-block.partitioning.logStructured.filtersPerRun": false,
  "inter-block.partitioning.logStructured.initialRunSize": 0,
  "inter-block.partitioning.logStructured.maxRunsPerLevel": 0,
  "inter-block.partitioning.logStructured.mergeFactor": 0,
  "inter-block.partitioning.type": "function",
  "intra-block.blockProperties.location": "inline",
  "intra-block.blockProperties.layout": "inline",
  "intra-block.blockProperties.homogeneous": true,
  "intra-block.capacity.function": "",
  "intra-block.capacity.type": "variable",
  "intra-block.capacity.value": 0,
  "intra-block.dataRetention.keyRetention.compression": "",
  "intra-block.dataRetention.keyRetention.function": "",
  "intra-block.dataRetention.keyRetention.type": "none",
  "intra-block.dataRetention.retainedDataLayout": "",
  "intra-block.dataRetention.valueRetention.compression": "",
  "intra-block.dataRetention.valueRetention.function": "",
  "intra-block.dataRetention.valueRetention.type": "none",
  "intra-block.filters.bloomFilter.active": false,
  "intra-block.filters.bloomFilter.hashFunctionsNumber": 0,
  "intra-block.filters.bloomFilter.numberOfBits": 0,
  "intra-block.filters.filtersMemoryLayout": "scatter",
  "intra-block.filters.zoneMaps.max": false,
  "intra-block.filters.zoneMaps.min": false,
  "intra-block.filters.zoneMaps.exact": false,
  "intra-block.links.linksMemoryLayout": "scatter",
  "intra-block.links.next": false,
  "intra-block.links.prev": false,
  "intra-block.links.skipLinks.probability": 0,
  "intra-block.links.skipLinks.type": "none",
  "intra-block.utilization.constraint": "none",
  "intra-block.utilization.function": "",
  "layout.oneChildCompression": false,
  "layout.zeroElementNullable": true,
  "layout.indirectedPointers": false
}
\end{verbatim}
}

\subsubsection{Ordered Datapage}
{\scriptsize
\begin{verbatim}
{
  "external.links.next": true,
  "external.links.prev": false,
  "inter-block.blockAccess.direct": true,
  "inter-block.blockAccess.headLink": false,
  "inter-block.blockAccess.tailLink": false,
  "inter-block.fanout.fixedValue": 256,
  "inter-block.fanout.function": "",
  "inter-block.fanout.type": "fixed",
  "inter-block.orderMaintenance.type": "sorted",
  "inter-block.partitioning.function": "",
  "inter-block.partitioning.logStructured.filtersPerLevel": false,
  "inter-block.partitioning.logStructured.filtersPerRun": false,
  "inter-block.partitioning.logStructured.initialRunSize": 0,
  "inter-block.partitioning.logStructured.maxRunsPerLevel": 0,
  "inter-block.partitioning.logStructured.mergeFactor": 0,
  "inter-block.partitioning.type": "none",
  "intra-block.blockProperties.location": "",
  "intra-block.blockProperties.layout": "",
  "intra-block.blockProperties.homogeneous": true,
  "intra-block.capacity.function": "",
  "intra-block.capacity.type": "fixed",
  "intra-block.capacity.value": 1,
  "intra-block.dataRetention.keyRetention.compression": "uncompressed",
  "intra-block.dataRetention.keyRetention.function": "",
  "intra-block.dataRetention.keyRetention.type": "full",
  "intra-block.dataRetention.retainedDataLayout": "columnar",
  "intra-block.dataRetention.valueRetention.compression": "uncompressed",
  "intra-block.dataRetention.valueRetention.function": "",
  "intra-block.dataRetention.valueRetention.type": "full",
  "intra-block.filters.bloomFilter.active": false,
  "intra-block.filters.bloomFilter.hashFunctionsNumber": 0,
  "intra-block.filters.bloomFilter.numberOfBits": 0,
  "intra-block.filters.filtersMemoryLayout": "scatter",
  "intra-block.filters.zoneMaps.max": false,
  "intra-block.filters.zoneMaps.min": false,
  "intra-block.filters.zoneMaps.exact": false,
  "intra-block.links.linksMemoryLayout": "scatter",
  "intra-block.links.next": false,
  "intra-block.links.prev": false,
  "intra-block.links.skipLinks.probability": 0,
  "intra-block.links.skipLinks.type": "none",
  "intra-block.utilization.constraint": "leq_capacity",
  "intra-block.utilization.function": "",
  "layout.oneChildCompression": false,
  "layout.zeroElementNullable": false,
  "layout.indirectedPointers": false
}
\end{verbatim}
}

\subsubsection{B+Tree Internal Node Element}
{\scriptsize
\begin{verbatim}
{
    "external.links.next": false,
    "external.links.prev": false,
    "inter-block.blockAccess.direct": true,
    "inter-block.blockAccess.headLink": false,
    "inter-block.blockAccess.tailLink": false,
    "inter-block.fanout.fixedValue": 20,
    "inter-block.fanout.function": "",
    "inter-block.fanout.type": "fixed",
    "inter-block.orderMaintenance.type": "sorted",
    "inter-block.partitioning.function": "",
    "inter-block.partitioning.logStructured.filtersPerLevel": false,
    "inter-block.partitioning.logStructured.filtersPerRun": false,
    "inter-block.partitioning.logStructured.initialRunSize": 0,
    "inter-block.partitioning.logStructured.maxRunsPerLevel": 0,
    "inter-block.partitioning.logStructured.mergeFactor": 0,
    "inter-block.partitioning.type": "none",
    "intra-block.blockProperties.location": "inline",
    "intra-block.blockProperties.layout": "inline",
    "intra-block.blockProperties.homogeneous": true,
    "intra-block.blockProperties.storage": "pointed",
    "intra-block.capacity.function": "",
    "intra-block.capacity.type": "balanced",
    "intra-block.capacity.value": 0,
    "intra-block.dataRetention.keyRetention.compression": "",
    "intra-block.dataRetention.keyRetention.function": "",
    "intra-block.dataRetention.keyRetention.type": "none",
    "intra-block.dataRetention.retainedDataLayout": "dump",
    "intra-block.dataRetention.valueRetention.compression": "",
    "intra-block.dataRetention.valueRetention.function": "",
    "intra-block.dataRetention.valueRetention.type": "none",
    "intra-block.filters.bloomFilter.active": false,
    "intra-block.filters.bloomFilter.hashFunctionsNumber": 0,
    "intra-block.filters.bloomFilter.numberOfBits": 0,
    "intra-block.filters.filtersMemoryLayout": "scatter",
    "intra-block.filters.zoneMaps.max": false,
    "intra-block.filters.zoneMaps.min": true,
    "intra-block.filters.zoneMaps.exact": false,
    "intra-block.links.linksMemoryLayout": "scatter",
    "intra-block.links.next": false,
    "intra-block.links.prev": false,
    "intra-block.links.skipLinks.probability": 0,
    "intra-block.links.skipLinks.type": "none",
    "intra-block.utilization.constraint": "none",
    "intra-block.utilization.function": "",
    "layout.oneChildCompression": false,
    "layout.zeroElementNullable": true,
    "layout.indirectedPointers": false
    }
\end{verbatim}
}

\subsection{Data Structure Instances}
Finally, we present the raw output of the optimal data structure instances for both scenarios in Figure 9.
For demonstration purposes we converted the raw data structures to graph structures.
Each graph node represents an element and each line represents a connection from a parent element to its sub-blocks.
Since we are plotting an instance of the data structure there is a lot of information.
Zooming in the plots allows readers to read the type of each node.
However, intuitively, very high fanout elements represent hash nodes, low fanout elements represent B+Tree internal nodes and no-fanout (terminal) elements represent data pages.

\noindent\textbf{Scenario 1}. We have a hash-map followed by either shallow B+Trees or large log-style data pages for write intensive parts of the domain (corresponding to the design at the left of Figure 9).

\noindent\textbf{Scenario 2}. We have a B+Tree followed by either hash-maps for point-get intensive parts of the domain or B+Trees for range intensive parts, and log-style big data pages for write intensive parts of the domain (corresponding to the design at the right of Figure 9).

\begin{figure*}[htbp]
    \includegraphics[width=\textwidth]{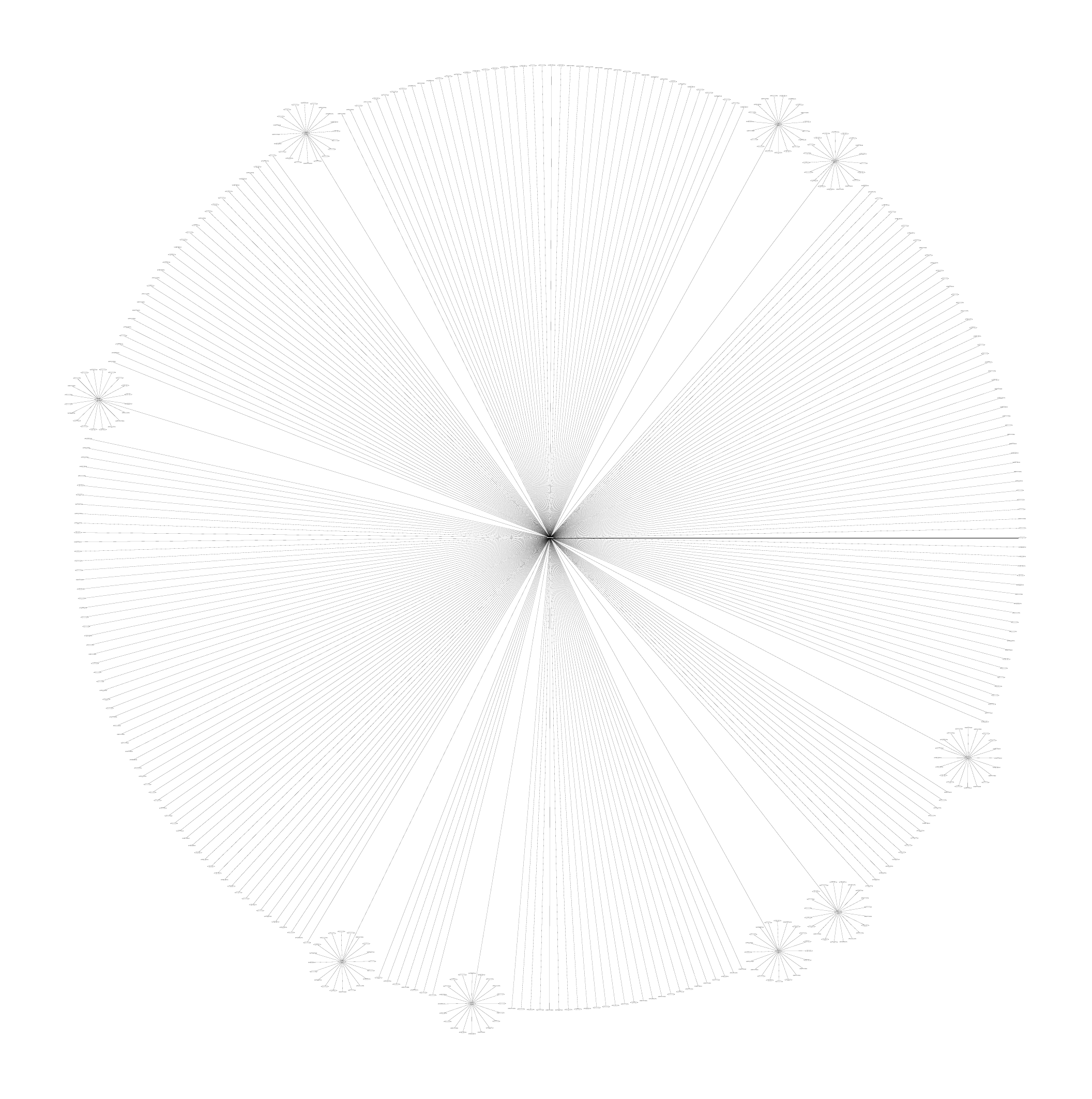}
    \caption{Real output instance (Scenario 1 of Figure 1).}
\end{figure*}

\begin{figure*}[htbp]
    \includegraphics[width=\textwidth]{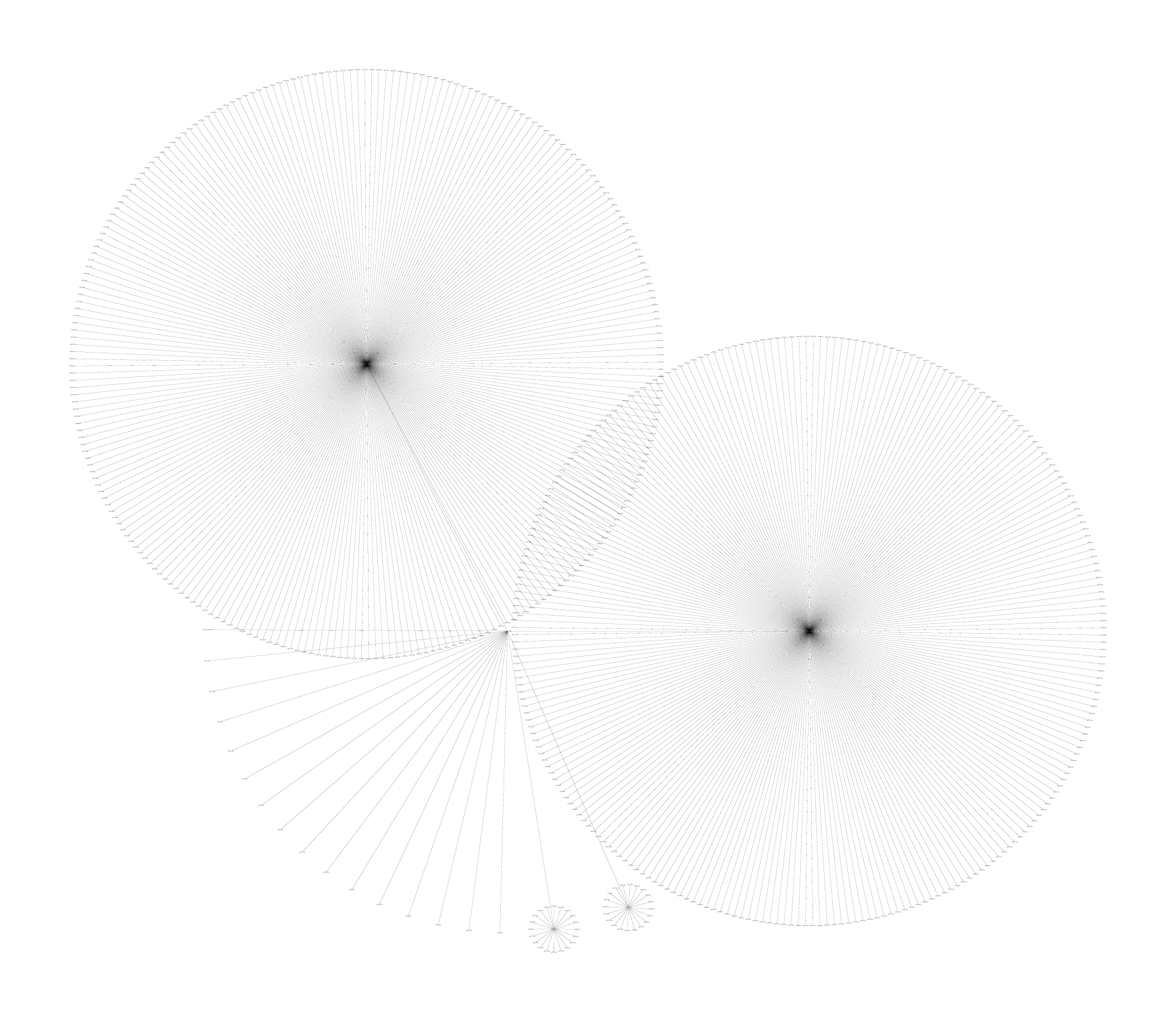}
    \caption{Real output instance (Scenario 2 of Figure 9).}
\end{figure*}

\end{document}